\documentclass[aps,prx,twocolumn,amsmath,amssymb,floatfix,superscriptaddress]{revtex4-2}
\usepackage[latin9,utf8]{inputenc}
\usepackage{stmaryrd}
\usepackage{graphicx}
\usepackage{subfigure}
\usepackage{multirow}
\usepackage{mathtools}
\usepackage{xcolor}
\usepackage{textcomp}
\usepackage{gensymb}
\usepackage{calligra}
\usepackage{braket}
\usepackage{comment}
\usepackage{float}
\usepackage{bbold}

\pdfoutput=1

\usepackage{xcolor,mathrsfs,dsfont}
\definecolor{darkblue}{RGB}{0,0,150}
\definecolor{nightblue}{RGB}{0,0,100}
\definecolor{evergreen}{RGB}{0,120,0}
    
\usepackage{graphicx,mathtools,bm}
\graphicspath{{./Figures/}}
\usepackage[colorlinks = true,
            linkcolor = blue,
            urlcolor  = blue,
            citecolor = blue,
            anchorcolor = blue]{hyperref}

\usepackage[english]{babel}
\usepackage[babel,kerning=true,spacing=true]{microtype}
\usepackage[utf8]{inputenc}

\makeatother

\begin{document}
\title{Topological (in)equivalence encoded in zeros of wavefunction overlaps}
\author{Carolina Paiva}
\affiliation{Raymond and Beverly Sackler School of Physics and Astronomy, Tel Aviv University, Tel Aviv, Israel}
\author{Bruno Mera}
\affiliation{Instituto de Telecomunicações, 1049-001 Lisboa, Portugal}
\affiliation{Department of Mathematics, Instituto Superior Técnico, Universidade de Lisboa, 1049-001 Lisboa,
Portugal}
\affiliation{Advanced Institute for Materials Research (WPI-AIMR), Tohoku University, Sendai 980-8577, Japan}
\author{Roni Ilan}
\email{ronilan@tauex.tau.ac.il}
\affiliation{Raymond and Beverly Sackler School of Physics and Astronomy, Tel Aviv University, Tel Aviv, Israel}
\affiliation{Simons Emmy Noether Fellow, Perimeter Institute for Theoretical Physics, 31 Caroline Street North, Waterloo, ON N2L 2Y5, Canada}
\author{Raquel Queiroz}
\email{raquel.queiroz@columbia.edu}
\affiliation{Department of Physics, Columbia University, New York, NY 10027, USA}
\affiliation{Center for Computational Quantum Physics, Flatiron Institute, New York, New York 10010, USA}
\date{\today}

\begin{abstract}
Topological equivalence between two gapped band structures is determined by whether their occupied states can be continuously deformed into one another without closing the energy gap or breaking the protecting symmetries. We show that this relative topology can be read directly from the overlap matrix between the two occupied subspaces, in a way that treats Chern and $\mathbb{Z}_2$ invariants within a single framework. The determinant of the overlap matrix defines a global section of a line bundle over the Brillouin zone, which vanishes at the momenta where the map between the subspaces fails to be invertible. A singular connection, whose curvature is concentrated at these zeros, assigns each zero an integer charge given by its phase winding. In two-dimensional class A, the total charge equals the difference between the Chern numbers of the two systems. In class AII, time reversal pairs zeros of opposite charge, and the relative $\mathbb{Z}_2$ invariant is the parity of the total charge in half the Brillouin zone; the Kane-Mele Pfaffian is recovered as a special case. The zeros coincide with the gap closings of an interpolation between the two systems, and crystalline symmetries constrain their positions and charges. We illustrate the construction in the Haldane, Kane-Mele, and Qi-Wu-Zhang models, including cases where rotation symmetry alone enforces the zeros.
\end{abstract}

\maketitle

\section{Introduction}

Topological phases are distinguished by an obstruction to continuously deforming one gapped system into another while preserving the symmetries that protect the phase~\cite{Fruchart_2013,bernevig_2013,sun_2016}. Topology is therefore, at its core, a statement about the equivalence of gapped quantum systems. This distinction has observable consequences ranging from quantized responses~\cite{TKNN_1982,Avron_1983,Thouless_1983,Niu_Thouless_Wu_1985,Qi_Hughes_Zhang_2008,Essin_Moore_Vanderbilt_2009} to robust edge states, whose existence is tied to the topology of the bulk through the bulk-boundary correspondence~\cite{Halperin_1982,Hatsugai_1993,Kane_Mele_Graphene_2005}.

In electronic band structures topology is usually diagnosed by assigning a topological invariant to each bulk system separately, rather than by constructing a quantity directly from a pair of systems. The underlying question of topological equivalence is instead intrinsically relative: whether two band structures can be continuously deformed into one another without closing the gap or breaking the protecting symmetries. The most familiar example is the Chern insulator. The occupied Bloch states define a Berry connection and Berry curvature~\cite{Berry_1984}, and in a two-dimensional Chern insulator the integral of the Berry curvature over the Brillouin zone gives the Chern number, which determines the quantized Hall response~\cite{TKNN_1982,Haldane_1988,bernevig_2013}. For time-reversal-invariant insulators, this Chern-number diagnosis becomes trivial. Time-reversal symmetry forces the total Chern number of the occupied bands to vanish, even though the occupied states can remain topologically distinct from those of a trivial insulator. Kane and Mele showed that the two phases are instead distinguished by a $\mathbb Z_2$ invariant~\cite{Kane_Mele_2005}. In their original formulation, this invariant is expressed through the zeros of a Pfaffian constructed from the occupied states. The same index can be expressed through time-reversal polarization~\cite{Fu_Kane_2006}, or as an obstruction to choosing a smooth periodic Bloch frame that respects time reversal~\cite{Roy_2009_QSH,Fiorenza_Monaco_Panati_2016}. The Chern and $\mathbb Z_2$ invariants thus admit descriptions in terms of apparently different geometric quantities: the Berry curvature in the former case and the zeros of a Pfaffian in the latter.

Zeros of a carefully chosen quantity can provide an alternative description of topology also in the case of Chern insulators. Kohmoto related the Chern number to the vorticities of zeros of a Bloch-state component in the magnetic Brillouin zone~\cite{Kohmoto_1985}. For an occupied multiplet, Hatsugai used a reference frame to fix the gauge and recovered the Chern number from the vortices at the singularities of that gauge~\cite{Hatsugai_2004,Hatsugai_2005}, which are unavoidable because a nonzero Chern number obstructs a smooth periodic gauge~\cite{Brouder_2007}. More recently, the total vorticity of the overlap between a Chern band and a trial Wannier state was related to its Chern number~\cite{eslam_2024}, and the zero of the overlap between a topological band and the Bloch states of a local orbital was shown to protect in-gap states bound to strong impurities~\cite{Queiroz_2024_rings}. Ref.~\cite{Mera_2021} formulated the relation between zeros and topology geometrically in terms of a singular connection, which requires the theory of currents~\cite{Harvey_Lawson_1993}. In this construction, one chooses a section of a line bundle whose zeros capture the topology, and uses this section to define a connection whose curvature is concentrated at those zeros. The Chern number is then recovered from the total winding of the phase around the zeros, or equivalently, from the total flux contributed by the singular curvature. Zeros can also reveal topological information in real space, where crystalline symmetries can constrain the zeros of Bloch wavefunctions~\cite{Ingham_2026}, and have proved useful in other settings, including excitons, superconductors, and magnons~\cite{paiva_2024,schindler_2026,Yong_Baek_2026,Chen_2026}. In a Chern band, the leading momentum-space envelope of a local orbital must likewise vanish, with total vorticity fixed by the Chern number~\cite{Verma_et_al_2026_CNO,Mahalingam_et_al_2026_CNO}.

We develop a framework that relies on defining global sections of appropriate line bundles and the singular connections associated with their zeros. Bloch bands naturally define vector bundles over the Brillouin zone, whose fibers are the occupied subspaces at each momentum~\cite{Fruchart_2013,Nakahara_2003}. This approach can be formulated for the case of a single or several bands, where for the latter, taking the antisymmetric product of an occupied basis produces a one-dimensional space associated with each subspace~\cite{Slager_2023}. These spaces form a line bundle, whose topology can be detected through the zeros of a global section~\cite{Bott_tu_1982,Rui_loja_24} and their phase windings, captured geometrically by singular connections~\cite{Mera_2021}.

For the purpose of comparing the topology of two band structures with the same number of bands, we construct such a section from the matrix of overlaps between their occupied Bloch bases. As we show, the determinant of the matrix then naturally defines a global section of a relative line bundle, whose sections are linear maps between the two determinant lines. We call this section the ``comparison section''. Its zeros occur precisely at momenta where the overlap matrix loses rank, and its determinant vanishes. Applying the singular-connection construction to the comparison section defined from this determinant, we assign a winding to each isolated zero. 

We further show that comparing two systems belonging to class A~\cite{Altland_Zirnbauer_1997}, the total winding number is equal to the relative Chern number, namely, the difference between the Chern numbers of the two systems. For two systems belonging to class AII, time-reversal symmetry constrains the zeros of the section and their windings, and the parity of the total winding in half the Brillouin zone, with one momentum from each pair $\{\mathbf k,-\mathbf k\}$, gives the relative $\mathbb Z_2$ invariant. 

The comparison section can also be directly linked to the notion of a spectral flow. We show this by constructing a Hamiltonian that interpolates between the two occupied subspaces, where the zeros of the comparison section appear as gap closings. 

We further discuss how crystal symmetries can impose constraints on the comparison section, relating zeros and their windings at symmetry-related momenta. At high-symmetry points, a mismatch between the symmetry representations of the two occupied subspaces can also force the comparison section to vanish. This can occur even when both Chern numbers vanish. In the Haldane model with the amplitude of the purely imaginary next-nearest-neighbor hopping set to zero, for example, reversing the staggered potential forces a zero at each valley, and the two zeros carry opposite windings. We illustrate these constraints for threefold rotation symmetry in the Haldane~\cite{Haldane_1988} and Kane-Mele~\cite{Kane_Mele_2005} models, and for fourfold rotation symmetry in the Qi-Wu-Zhang (QWZ) model~\cite{Qi_Wu_Zhang_2006}.

The question of topological equivalence has been addressed before directly by asking whether two band structures can be connected by a gapped, symmetry-preserving adiabatic path. In Ref.~\cite{sun_2016}, the matrix of overlaps between their occupied states was used to provide a sufficient criterion for the existence of such a path. If this matrix is nonsingular throughout the Brillouin zone, a gapped, symmetry-preserving adiabatic path between the two systems can be constructed. Conversely, topologically inequivalent band structures necessarily lead to a singular overlap matrix at some momentum, and hence its determinant must vanish. These zeros identify where the sufficient criterion for adiabatic equivalence fails, although their presence alone does not imply topological inequivalence. For two Bloch bands with different Chern numbers the overlap must have zeros~\cite{Yang_Bahramy_Nagaosa_2013,Huang_Balatsky_2016}, and their number is bounded below by the absolute difference of the Chern numbers~\cite{Huang_Balatsky_2016}. Singular connections provide a geometric way of assigning topological charges to zeros, while overlap matrices allow two band structures to be related directly. Here we combine the two. The reference is an arbitrary band structure rather than a trivial one, the same determinant gives the relative $\mathbb Z_2$ invariant as a winding parity, and its zeros are the gap closings of an interpolation between the two systems.

The paper is organized as follows. In Sec.~\ref{sec:singular_connections}, we introduce the global comparison section and the singular connection that associates a winding to its zeros. In Sec.~\ref{subsec:absence_symmetries}, we apply this construction to class A, relate the total winding of the zeros of the comparison section to the relative Chern number, and develop its spectral interpretation and Haldane-model realization. In Sec.~\ref{sec:time_reversal_singular_connection}, we extend the construction to class AII, connect it to the Pfaffian formulation of the $\mathbb Z_2$ invariant, formulate the relative $\mathbb Z_2$ obstruction, and illustrate the result in the Kane-Mele model. In Sec.~\ref{sec:crystalline_symmetries}, we study how crystalline symmetries constrain the positions and winding charges of the zeros of the comparison section and can enforce zeros at symmetry-invariant momenta.

\section{Singular connection and topological equivalence}
\label{sec:singular_connections}

In this section we develop the mathematical framework that is used throughout the paper to
characterize the relative topology of two sets of occupied
states. These may belong to two different
Hamiltonians, or to two parameter regimes of the same Hamiltonian. We
assume that both systems are gapped and have the same number of occupied bands $N_{\mathrm{occ}}$~\footnote{The notion of topological equivalence considered here is based on isomorphisms between vector bundles, which require them to have the same rank. Systems with different numbers of occupied bands can nevertheless still be compared by adding topologically trivial occupied bands until the ranks of the two vector bundles agree. The construction can then be applied to these enlarged bundles.}. We refer to the bands being compared as occupied, although the construction applies to any set of bands separated from the rest of the spectrum by a gap. Their Brillouin zones need not arise from the same lattice geometries~\cite{sun_2016}. Rather, we assume a fixed smooth identification between the two Brillouin-zone tori, compatible with the momentum action of any symmetry being preserved. After this identification, we regard both systems as defined over a common Brillouin zone $\mathrm{BZ}^2$. We further assume that their Bloch states are
represented in the same Hilbert space $\mathbb C^n$, with $n$ the total number of bands. At each momentum their occupied states span two
$N_{\mathrm{occ}}$-dimensional subspaces,
$\mathcal E^\alpha_{\mathrm{occ},\mathbf{k}}$ and
$\mathcal E^\sigma_{\mathrm{occ},\mathbf{k}}$. To characterize their relative topology, we consider a particular map between these subspaces and investigate where it becomes singular. Although the singularities of this map do not necessarily imply topological inequivalence, we will show that the winding charges associated with the zeros of its determinant encode the relative topology of the two systems.

\subsection{Comparison section from band structure overlaps}

To formulate the problem globally, we use the vector-bundle
description of Bloch bands
~\cite{Nakahara_2003,Fruchart_2013,Kobayashi_2014,Rui_loja_24}\footnote{A vector bundle may be viewed as a family of vector spaces attached
smoothly to the points of a base space. In band theory the base space is
the Brillouin zone, and the vector space attached to each momentum is
the space spanned by the corresponding Bloch states.}.
As $\mathbf{k}$ varies over the Brillouin zone, the occupied
subspaces of each system $\mu\in \{\alpha,\sigma\}$ form a rank-$N_{\mathrm{occ}}$ complex vector bundle
\begin{equation}
\mathcal E^\mu_{\mathrm{occ}} \longrightarrow \mathrm{BZ}^2,
\qquad
\mu\in\{\alpha,\sigma\},
\end{equation}
whose fiber at $\mathbf{k}$ is precisely
$\mathcal E^\mu_{\mathrm{occ},\mathbf{k}}$. A local choice of occupied
states
$\{\ket{\psi_i^\mu(\mathbf{k})}\}_{i=1}^{N_{\mathrm{occ}}}$ provides a basis for this fiber.  

A section assigns to each momentum a vector in the corresponding fiber. It is global if this assignment is smooth throughout the Brillouin zone, and local if defined only on a region. Two occupied bundles are equivalent when there exists a smooth, everywhere-invertible bundle map between them (commuting with the symmetries of the system). Such maps are global sections of the homomorphism bundle
\begin{equation}
\operatorname{Hom}\left(
\mathcal E^\sigma_{\mathrm{occ}},
\mathcal E^\alpha_{\mathrm{occ}}
\right),
\end{equation}
whose fiber at each momentum $\mathbf{k}$ is the vector space of all linear maps from $\mathcal E^\sigma_{\mathrm{occ},\mathbf{k}}$ to
$\mathcal E^\alpha_{\mathrm{occ},\mathbf{k}}$. A global section thus specifies one such linear map smoothly at each momentum. If the two bundles are topologically inequivalent, every symmetry-compatible global section of this bundle must fail to be invertible at some momentum. The converse, however, is not necessarily true: a particular section may fail to be invertible even when the bundles are equivalent. Rather than searching for an everywhere-invertible global section, we therefore consider a particular global section and investigate what information about equivalence can be extracted from the points where it loses rank. Since the two subspaces have the same dimension, these points correspond precisely to zeros of the determinant of the map.

The determinant of a linear map between the bundles naturally defines a linear map between their determinant line bundles. Introducing the line bundles 
\begin{equation}
\mathcal L_\mu
=
\det \mathcal E^\mu_{\mathrm{occ}}
:=
\bigwedge^{N_{\mathrm{occ}}}
\mathcal E^\mu_{\mathrm{occ}},
\qquad
\mu\in\{\alpha,\sigma\},
\end{equation}
the fiber of $\mathcal L_\mu$ at each momentum is the one-dimensional complex vector space spanned by the antisymmetrized state of all the occupied single-particle states given by
\begin{equation}
\ket{\Psi^\mu(\mathbf{k})}
=
\ket{\psi^\mu_1(\mathbf{k})}
\wedge\cdots\wedge
\ket{\psi^\mu_{N_{\mathrm{occ}}}(\mathbf{k})}.
\label{eq:local_basis_wedge}
\end{equation}
Here, $\{\ket{\psi^\mu_n(\mathbf{k})}\}$ denotes a locally defined
basis of occupied states. The determinant of a map between occupied
fibers is thus a map from $\mathcal L_{\sigma,\mathbf{k}}$ to
$\mathcal L_{\alpha,\mathbf{k}}$, represented locally by a single
complex scalar.

As $\mathbf{k}$ varies, the determinants of smooth global bundle maps define global sections of the line bundle 
\begin{equation}
\mathcal L_r
=
\operatorname{Hom}
\left(
\mathcal L_\sigma,\mathcal L_\alpha
\right)
=
\mathcal L_\alpha\otimes\mathcal L_\sigma^*,
\label{eq:relative_line_bundle}
\end{equation}
which we term the ``relative line bundle''. A complex line bundle
admits a nowhere-vanishing global section if and only if it is
trivial~\cite{Nakahara_2003,Bott_tu_1982}. Thus, a nontrivial
relative line bundle forces every global section to have zeros.
For a generic section on the two-dimensional Brillouin zone,
the zeros are isolated and have integer phase windings.

A global section of $\mathcal L_r$ can be constructed by
projecting one occupied subspace onto the other. At each momentum, this
defines the map
\begin{equation}
T_{\alpha\sigma}(\mathbf{k}):
\mathcal E^\sigma_{\mathrm{occ},\mathbf{k}}
\longrightarrow
\mathcal E^\alpha_{\mathrm{occ},\mathbf{k}},
\qquad
\ket{\psi^\sigma}
\longmapsto
P^\alpha(\mathbf{k})\ket{\psi^\sigma},
\end{equation}
where $P^\alpha(\mathbf{k})$ is the projector onto the occupied subspace
of $\alpha$.
If
$\{\ket{\psi^{\alpha}_{i}(\mathbf{k})}\}_{i=1}^{N_{\mathrm{occ}}}$
and
$\{\ket{\psi^{\sigma}_{j}(\mathbf{k})}\}_{j=1}^{N_{\mathrm{occ}}}$
are local orthonormal bases of the two subspaces, the map
$T_{\alpha\sigma}(\mathbf{k})$ is represented in these bases by the
matrix of overlaps~\footnote{Overlap matrices of the same form, but between occupied states of a single system at neighboring momenta, appear in Wannier and Wilson-loop constructions~\cite{Marzari_Vanderbilt_1997,Dai_2011}.} 
\begin{equation}
S_{ij}(\mathbf{k})
=
\braket{\psi^{\alpha}_{i}(\mathbf{k})|
\psi^{\sigma}_{j}(\mathbf{k})}.
\label{eq:overlap_matrix}
\end{equation}
This projection between subspaces is also the map that enters the index of a pair of projections~\cite{Avron_1994}.
At a given momentum $\mathbf{k}$, this matrix loses rank when some nonzero occupied state of $\sigma$ is orthogonal to every occupied state of $\alpha$ at the same momentum. Thus,
\begin{equation}
\det S(\mathbf{k})=0
\qquad\Longleftrightarrow\qquad
T_{\alpha\sigma}(\mathbf{k})
\ \text{is not invertible}.
\end{equation}
This makes the determinant of $T_{\alpha\sigma}$ a good choice for a global section of
the relative line bundle $\mathcal L_r$. To write this section explicitly
we use the bases $\ket{\Psi^\alpha(\mathbf{k})}$ and
$\ket{\Psi^\sigma(\mathbf{k})}$ introduced above as the local
basis of $\mathcal L_r$,
\begin{equation}
e_r(\mathbf{k})
=
\ket{\Psi^\alpha(\mathbf{k})}
\bra{\Psi^\sigma(\mathbf{k})}.
\label{eq:relative_line_basis}
\end{equation}
In this basis, the section induced by $\det T_{\alpha\sigma}$ is
\begin{equation}
a_S(\mathbf{k})
=
\det S(\mathbf{k})\,e_r(\mathbf{k}),
\label{eq:relative_overlap_section}
\end{equation}
which we call the ``comparison section''. 

\begin{figure}[t]
    \centering
    \includegraphics[width=1\linewidth]{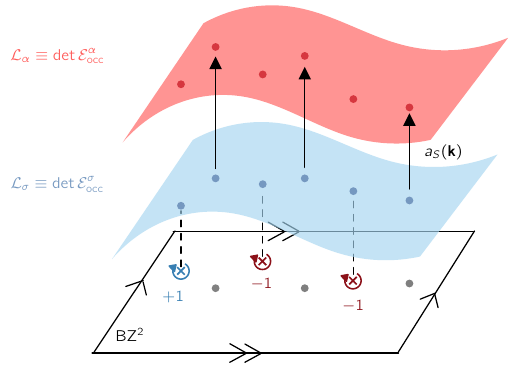}
    \caption{
    Schematic representation of the relative line bundle
    $\mathcal L_r=\operatorname{Hom}(\mathcal L_\sigma,\mathcal L_\alpha)$.
    At each momentum, its fiber is the one-dimensional vector space of
    linear maps from the determinant line of $\sigma$ to that of
    $\alpha$. A section selects one such map at each momentum. For the
section $a_S$ constructed from the overlap matrix, its zeros occur
precisely where the projection map $T_{\alpha\sigma}$ fails to be
invertible.
    }
    \label{fig:homomorphism_bundle}
\end{figure}

We note that although $\det S(\mathbf{k})$ depends on the choice of bases,
this dependence is exactly compensated by the transformation of
$e_r(\mathbf{k})$. The section $a_S(\mathbf{k})$ is therefore
gauge invariant, as shown explicitly in
Appendix~\ref{app:gauge_invariance_comparison_section}.
Since $e_r(\mathbf{k})$ is nonvanishing, zeros of the section are in one-to-one correspondence with the zeros of $\det S(\mathbf{k})$, namely 
\begin{equation}
a_S(\mathbf{k})=0
\qquad\Longleftrightarrow\qquad
\det S(\mathbf{k})=0.
\end{equation}
Figure~\ref{fig:homomorphism_bundle} illustrates the relative line bundle and its comparison section.

It is important to note that 
$T_{\alpha\sigma}$ is a particular choice for a bundle map, and the existence of zeros in this map does not necessarily imply a topological obstruction. Meaning, even if the bundle is trivial, it is still possible that zeros may appear. It is the global information about the total winding around the zeros of the determinant that indicates whether an obstruction is present or not, and hence if the topology is nontrivial or not. We now extract this information using the singular connection associated with $a_S$.

\subsection{Topological characterization via singular connections}
\label{subsec:singular_connection_general}

To characterize the topology via the zeros of the global section, we must assign them with a charge. This is done by defining a singular connection~\cite{Mera_2021}. Consider a global
section $a(\mathbf{k})$ of a complex line bundle, such as
$\mathcal L_r$ or the line bundles of Sec.~\ref{sec:time_reversal_singular_connection}, not necessarily the comparison section $a_S$ of Eq.~\eqref{eq:relative_overlap_section}. In a local basis
$e(\mathbf{k})$, the section can be written as
\begin{equation}
a(\mathbf{k})=s(\mathbf{k})\,e(\mathbf{k}),
\end{equation}
where $s(\mathbf{k})$ is a locally defined complex function
that transforms inversely to the local basis $e(\mathbf{k})$,
so that the section $a(\mathbf{k})$ is independent of the
choice of basis.

The singular connection is defined by requiring the
section to be covariantly constant away from its zeros,
\begin{equation}
\nabla a=0.
\label{eq:singular_parallel_section}
\end{equation}
With the convention
$\nabla e=i\mathcal A_{\mathrm{sing}}e$,
this condition gives
\begin{equation}
\mathcal A_{\mathrm{sing}}
=
i\,\mathrm d\log s.
\label{eq:singular_connection_general}
\end{equation}
Under a change of local basis $e'=h\,e$,
the coefficient transforms as $s'=h^{-1}\,s$,
so that
\begin{equation}
\mathcal A'_{\mathrm{sing}}
=
\mathcal A_{\mathrm{sing}}-i h^{-1}\,\mathrm dh,
\end{equation}
as required for a connection. The singular connection
generally takes complex values and is flat away from the zeros of $a$.

At a zero, $\log s$ becomes singular, and
the phase of $s$ may wind. For an isolated
zero at $\mathbf{k}_*$, we define a local charge as
\begin{equation}
\nu_{\mathbf{k}_*}
=
\frac{1}{2\pi i}
\oint_{\partial C_{\mathbf{k}_*}}
\mathrm d\log s(\mathbf{k}),
\label{eq:zero_winding_general}
\end{equation}
where $\partial C_{\mathbf{k}_*}$ is a small positively
oriented loop surrounding the zero, and the local
basis is chosen smoothly over the enclosed region.
The integer $\nu_{\mathbf{k}_*}$ counts the phase
winding around the zero. With our convention,
the curvature $\mathrm d\mathcal A_{\mathrm{sing}}$
has a contribution $-2\pi\nu_{\mathbf{k}_*}$
concentrated at each isolated zero. The singular connection thus associates integer
charges with the zeros of a global section. We can immediately apply these principles to
characterize the relative topology of occupied
bundles in classes A and AII.

\section{Class A: relative Chern number}
\label{subsec:absence_symmetries}

We now show that the relative line bundle of Sec.~\ref{sec:singular_connections} captures the difference in Chern numbers $C_\alpha-C_\sigma$ for two-dimensional systems belonging to class A. To do so, we first address the equivalence problem using the standard approach, i.e. using Berry connections, and then argue that the same information can be obtained simply by picking an alternative connection that extracts this information directly from the zeros of the section.   

For each occupied bundle, the determinant line bundle
$\mathcal L_\mu=\det\mathcal E^\mu_{\mathrm{occ}}$ has a well-defined Chern number~\cite{Nakahara_2003,Bott_tu_1982},
\begin{equation}
c_1\!\left(\mathcal E^\mu_{\mathrm{occ}}\right)
=
c_1(\mathcal L_\mu),
\qquad
\mu\in\{\alpha,\sigma\}.
\end{equation}
In addition, one can define an Abelian Berry connection from the
non-Abelian connection of the occupied bands. Writing
\begin{equation}
\left(\mathcal A^\mu_{\mathrm{NA}}\right)_{ij}
=
-i\braket{\psi_i^\mu(\mathbf{k})|
\mathrm d\psi_j^\mu(\mathbf{k})},
\end{equation}
the Abelian connection is obtained by differentiating the wedge product of the occupied states, which gives the trace of the non-Abelian Berry connection~\cite{Slager_2023}
\begin{equation}
\mathcal A^\mu_{\det}
=
\operatorname{Tr}\mathcal A^\mu_{\mathrm{NA}}
=
-i\braket{\Psi^\mu(\mathbf{k})|
\mathrm d\Psi^\mu(\mathbf{k})}.
\label{eq:det_berry_connection}
\end{equation}
Thus the determinant line retains the $U(1)$ contribution that determines the first Chern number~\cite{Slager_2023}.

The relative line bundle, defined in
Eq.~\eqref{eq:relative_line_bundle}, inherits the corresponding relative Berry connection. Since
$\mathcal L_\sigma^*$ is the dual line bundle, its connection enters with the opposite sign, and therefore
\begin{equation}
\mathcal A_r
=
\mathcal A^\alpha_{\det}
-
\mathcal A^\sigma_{\det}.
\label{eq:relative_berry_connection}
\end{equation}
The corresponding Berry curvature is
\begin{equation}
\Omega_r
=
\mathrm d\mathcal A_r
=
\Omega^\alpha_{\det}
-
\Omega^\sigma_{\det},
\label{eq:relative_berry_curvature}
\end{equation}
where $\Omega^\mu_{\det}=\mathrm d\mathcal A^\mu_{\det}$.
With $\mathcal A=-i\braket{\psi|\mathrm d\psi}$ and $C=-\frac{1}{2\pi}\int_{\mathrm{BZ}^2}\mathrm d\mathcal A$, integration gives
\begin{equation}
C_r
\equiv
C(\mathcal L_r)
=
C_\alpha-C_\sigma.
\label{eq:relative_chern_number}
\end{equation}
This shows that the relative line bundle contains information required to distinguish between the topology of two class A bundles. Since complex vector bundles of equal rank over a two-dimensional base are classified by their first Chern class, $\mathcal E^\alpha_{\mathrm{occ}}\simeq\mathcal E^\sigma_{\mathrm{occ}}$ if and only if $C_r=0$.

The relative Berry connection measures this invariant through smooth curvature. The singular connection of Eq.~\eqref{eq:singular_connection_general} must represent the same Chern class through a different distribution of curvature concentrated at the isolated zeros of $a_S$~\cite{Mera_2021}. Since both connections transform as $\mathcal A\to\mathcal A-i h^{-1}\,\mathrm dh$ under a change of local basis, their difference is a globally defined one-form on $\mathrm{BZ}^2\setminus\mathcal V$, where $\mathcal V$ is the finite zero set of the section. Since $\mathcal A_{\mathrm{sing}}$ is flat away from the zeros, we apply Stokes' theorem to the Brillouin zone after excluding a small disk around each zero. Taking the radii of these disks to zero, the resulting boundary integrals give the winding numbers associated with the zeros, $\int_{\mathrm{BZ}^2}\mathrm d\mathcal A_r=\sum_{\mathbf k_*}\oint_{\partial C_{\mathbf k_*}}\mathcal A_{\mathrm{sing}}=-2\pi\sum_{\mathbf k_*}\nu_{\mathbf k_*}$. Hence, for any global section with isolated zeros, and in particular the section $a_S$,
\begin{equation}
\sum_{\mathbf{k}_*\in \mathcal{V}}
\nu_{\mathbf{k}_*}
=
C_r
=
C_\alpha-C_\sigma,
\label{eq:classA_main_result}
\end{equation}
where $\mathcal{V}$ denotes the zero set of
$\det S(\mathbf{k})$. Equation~\eqref{eq:classA_main_result} establishes the relation between the zeros of the comparison section and the relative Chern number, or difference in Chern numbers, of two systems in class A.

The simplest relative invariant compares a single band $\alpha$ with a momentum-independent reference state, for which $C_\sigma=0$ and the windings of the zeros of the overlap sum to $C_\alpha$~\cite{Mera_2021}. For a position eigenstate at $\mathbf r$, the overlap is the complex conjugate Bloch wavefunction, $S(\mathbf k)=(\psi^\alpha_{\mathbf k}(\mathbf r))^{*}$ up to a smooth phase, so its zeros are the momenta at which $\psi^\alpha_{\mathbf k}$ vanishes in real space~\cite{Kohmoto_1985}. The winding numbers of these zeros are opposite to those of the wavefunction itself. Such zeros are well known in bands with a holomorphic structure, such as the lowest Landau level~\cite{Haldane_Rezayi_1985,Arovas_1988}, the chiral flat bands of twisted bilayer graphene~\cite{Tarnopolsky_2019}, and ideal Chern bands~\cite{Mera_Ozawa_2021_Kahler,Ledwith_Vishwanath_Parker_2023}, where all zeros wind in the same direction and their number is exactly $|C_\alpha|$.

For a single band in each system, this relation was previously obtained in Ref.~\cite{Huang_Balatsky_2016}, where the local charges are given by the winding of the wavefunction overlap around its zeros. Our determinant construction extends this result to subspaces of arbitrary equal rank. Physically, $|\det S(\mathbf k)|^2$ represents the momentum-resolved fidelity between the occupied states of the two systems~\cite{Amin_Mera_2018}. In a two-band model, its zeros correspond to momenta where a sudden quench from $\alpha$ to $\sigma$ completely inverts the band population: a state initially occupying the valence band of $\alpha$ lies entirely in the conduction band of $\sigma$. Such zeros can also provide the conditions for robust dynamical quantum phase transitions~\cite{Heyl_Polkovnikov_Kehrein_2013,Huang_Balatsky_2016}, and the resulting dynamical vortices have been observed in cold atoms~\cite{Flaschner_2018}.

When $\sigma$ is an atomic band constructed from a localized orbital $\ket{\phi^\sigma}$, the overlap reduces to the orbital weight $|S(\mathbf k)|^2=|\langle\phi^\sigma_{\mathbf k}|\psi^\alpha_{\mathbf k}\rangle|^2$ studied in Ref.~\cite{Queiroz_2024_rings}. Zeros of this weight in a topological band are associated with impurity-bound ring states. Since $C_\sigma=0$, Eq.~\eqref{eq:classA_main_result} shows that the total winding of these zeros equals $C_\alpha$.

Note that neither the positions nor the number of individual zeros of a section is a topological invariant. The zeros may move under smooth
deformations, and pairs with opposite winding may be created or annihilated throughout the process.  Their total winding is, however, fixed by the relative Chern number $C_\alpha-C_\sigma$. The next subsection gives these zeros a spectral interpretation. 

\subsection{Spectral flow and the zeros of the comparison section
}
\label{subsec:spectral_flow}
\begin{figure*}[t]
    \centering
    \includegraphics[width=\linewidth]{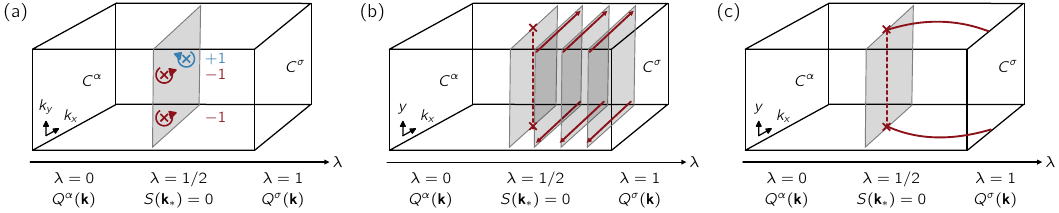}
    \caption{
    (a) Spectral flow interpretation of the overlap zeros. The interpolation parameter $\lambda$ extends the two-dimensional Brillouin zone to the auxiliary parameter space $(k_x,k_y,\lambda)$, with $Q^\alpha$ at $\lambda=0$ and $Q^\sigma$ at $\lambda=1$. A zero of the overlap section produces a gap closing at $(\mathbf{k}_*,\lambda=1/2)$, which is locally an auxiliary Weyl point. The Chern number of the fixed-$\lambda$ slices can change only when such a crossing is passed.
    (b,c) Boundary interpretation of the auxiliary Weyl point. Opening the
    system along $y$ leaves $(k_x,\lambda)$ as the boundary parameter
    space. A bulk zero of the comparison section at
    $((k_{x})_{*},(k_{y})_{*},\lambda=1/2)$ projects to
    $((k_{x})_{*},\lambda=1/2)$ at the boundary.
    (b) The change in Chern number across the auxiliary Weyl point
is accompanied by chiral boundary spectral flow, illustrated
by the oppositely propagating modes on the two edges.
(c) The chiral boundary states form Fermi-arc-like curves
in $(k_x,\lambda)$. Since $\lambda$ is an interpolation
parameter rather than a periodic crystal momentum,
these curves need not connect two projected Weyl points.
    }
    \label{fig:fermi_arc_slab_interpretation}
    \label{fig:bulk_spectral_flow}
\end{figure*}

Zeros of the comparison section can be linked to bulk gap closings in the spectrum of an auxiliary Hamiltonian that interpolates between the occupied subspaces. This establishes an interpretation of these zeros in terms of a bulk spectral flow.  

Consider the simplest case of a two-band model with a single occupied
band, $N_{\mathrm{occ}}=1$. In this case the matrix of overlaps between the two ground states, represented by $\ket{\psi^\alpha(\mathbf{k})}$ and $\ket{\psi^\sigma(\mathbf{k})}$ reduces to a scalar, equal to its determinant and to the overlap between the two occupied
states $\braket{\psi^\alpha(\mathbf{k})|\psi^\sigma(\mathbf{k})}
$. 
Let
\begin{equation}
P^\mu(\mathbf{k})
=
\ket{\psi^\mu(\mathbf{k})}
\bra{\psi^\mu(\mathbf{k})},
\qquad
Q^\mu(\mathbf{k})
=
2P^\mu(\mathbf{k})-\mathbb 1,
\end{equation}
with $\mu=\alpha,\sigma$, be the projectors onto the occupied bands and their spectrally flattened
version, respectively. With this sign the occupied states have eigenvalue $+1$, we follow the positive-energy band throughout. Define the
auxiliary interpolation Hamiltonian
\begin{equation}
H(\lambda,\mathbf{k})
=
(1-\lambda)Q^\alpha(\mathbf{k})
+
\lambda Q^\sigma(\mathbf{k}),
\label{eq:flattened_interpolation}
\end{equation}
where $\lambda\in[0,1]$. The endpoints of the interpolation at $\lambda=0,1$, correspond to the flattened projectors onto $\alpha$ and $\sigma$. 
The two eigenvalues of $H(\lambda,\mathbf{k})$ are given by (see Appendix~\ref{app:spectral_flow_details})
\begin{equation}
E_\pm(\lambda,\mathbf{k})
=
\pm
\sqrt{
(1-2\lambda)^2
+
4\lambda(1-\lambda)
\left|\det S(\mathbf{k})\right|^2
}.
\label{eq:spectral_flow_eigenvalues}
\end{equation}
Since both terms under the square
root are non-negative for $\lambda\in[0,1]$, the gap closes if and only
if
\begin{equation}
\lambda=\frac{1}{2},
\qquad
\det S(\mathbf{k})=0.
\label{eq:gap_closing_overlap_zero}
\end{equation}
Thus, the zeros of the comparison
section occur precisely at the momenta where the interpolating
Hamiltonian $H(\lambda,\mathbf{k})$ closes its gap.
In other words, the same points that carry the local charges of the singular
connection therefore identify the spectral events through
which topology can change between the two endpoint phases.

In fact, the interpolating Hamiltonian is a Weyl Hamiltonian in the three-dimensional parameter space defined by $\mathbf{k},\lambda$. For a simple zero $\mathbf{k}_*$, we write locally
\begin{equation}
\det S(\mathbf{k})
=
S_1(\mathbf{k})+iS_2(\mathbf{k}),
\qquad
\eta=\lambda-\frac{1}{2}.
\end{equation}
To linear order around $(\mathbf{k}_*,1/2)$, the interpolating
Hamiltonian can be brought to the form
\begin{equation}
H_{\mathrm{eff}}
\simeq
S_1(\mathbf{k})\,\sigma_1
-
S_2(\mathbf{k})\,\sigma_2
-
2\eta\,\sigma_3.
\label{eq:weyl_expansion_overlap}
\end{equation}
A simple zero of $S(\mathbf{k})$ therefore becomes an isolated Weyl point in $(k_x,k_y,\lambda)$, with chirality that is determined by the winding $\nu_{\mathbf{k}_*}$ (see Appendix~\ref{app:spectral_flow_details}). As for any isolated band crossing, this charge determines the change of the Chern number across it~\cite{Bellissard_1995}.

Away from $\lambda=1/2$, the positive-energy band of each fixed-$\lambda$ slice has a well-defined Chern number which is equal to $C_\alpha$ for $\lambda<1/2$ and $C_\sigma$ for $\lambda>1/2$. The Weyl point at  $\lambda=1/2$ compensates for the difference in Chern numbers between these two ground states, equal to minus its chirality (see Appendix~\ref{app:spectral_flow_details}), so that $C_\sigma-C_\alpha=-\sum_{\mathbf k_*}\nu_{\mathbf k_*}$. The spectral flow thus reproduces Eq.~\eqref{eq:classA_main_result} independently of the singular-connection argument. Figure~\ref{fig:bulk_spectral_flow}(a) illustrates this interpretation.

The existence of an auxiliary Weyl Hamiltonian picture raises the question whether there is a meaningful interpretation of
its boundary modes. Suppose that one physical direction, say $y$, is terminated, while $k_x$ remains a good quantum number. The boundary spectrum of $H(\lambda,\mathbf{k})$
then depends on $(k_x,\lambda)$, and a bulk Weyl
point at $((k_{x})_{*},(k_{y})_{*},1/2)$ projects onto $((k_{x})_{*},1/2)$. For each gapped fixed-$\lambda$ slice, the
bulk-boundary correspondence relates its Chern number to the net
chirality of its boundary modes. As $\lambda$ passes through a Weyl
point and the Chern number of a $\lambda$ slice changes, a chiral boundary mode is created.

The zero-energy locus of this boundary branch traces a
Fermi-arc-like curve in the $(k_x,\lambda)$ plane, that terminates at the
projection of the bulk Weyl point, as illustrated in
Fig.~\ref{fig:fermi_arc_slab_interpretation}(b,c). This is analogous to the
surface Fermi arc of a Weyl semimetal, with the important distinction
that $\lambda$ is an interpolation parameter rather than a physical
crystal momentum, and is not periodic. Note that the surfaces of the three-dimensional Hamiltonian $H(\lambda,\mathbf{k})$ at the end points $\lambda=0,1$ represent the two insulators whose relative topology is being studied. If these insulators have different Chern numbers, then the Fermi arc extends into one of the phases and translates into a boundary mode of the insulator of the endpoint. This allows for the model to have a single bulk Weyl point, as opposed to the three-dimensional genuine lattice Weyl semimetal, where Weyl points must appear in pairs of opposite chiralities, and Fermi arcs terminate at another projection of a Weyl point of opposite bulk chirality.

The above construction extends to several occupied bands. The scalar overlap
of the single-band case is then replaced by the matrix of overlaps of
Eq.~\eqref{eq:overlap_matrix}. As shown in
Appendix~\ref{app:z2_spectral_flow_details}, the singular-value
decomposition of the overlap matrix allows the interpolation of
Eq.~\eqref{eq:flattened_interpolation}, with
$Q^\mu=2P^\mu-\mathbb 1$ now built from rank-$N_{\mathrm{occ}}$
projectors, to be reduced to independent two-level sectors. In the appendix, we work directly with the projectors $P^\mu$ rather than $Q^\mu$. The two interpolations differ only by a shift and rescaling of their eigenvalues, so their gap-closing conditions are equivalent.
The interpolation becomes gapless at $\lambda=1/2$ when one of
the singular values of $S(\mathbf{k})$ vanishes, equivalently when
\begin{equation}
\det S(\mathbf{k})=0.
\end{equation}
The same zeros of the comparison section that carry the relative Chern number therefore identify the bulk gap closings of the interpolation also in the multiband case.

\subsection{Application: Haldane model}
\label{subsec:chern_insulator_application}

We illustrate the class A construction using the Haldane model on the
honeycomb lattice~\cite{Haldane_1988}. In the sublattice basis, its Bloch
Hamiltonian can be written as
\begin{equation}
H(\mathbf{k})
=
\mathbf{d}(\mathbf{k})\cdot\boldsymbol{\sigma},
\label{eq:Haldane_Hamiltonian}
\end{equation}
the explicit components of $\mathbf d(\mathbf{k})$ are given in Appendix~\ref{app:haldane_model}, where $t_1$ is the nearest-neighbor hopping, $t_2$ the amplitude of the purely imaginary next-nearest-neighbor hopping (we take the Haldane flux $|\phi|=\pi/2$), and $M$ the staggered sublattice potential.
Taking the valence band to be occupied, we study the relative topology of two such bands at two parameter regimes of the model.

\begin{figure}[t]
    \centering
    \includegraphics[width=1\linewidth]{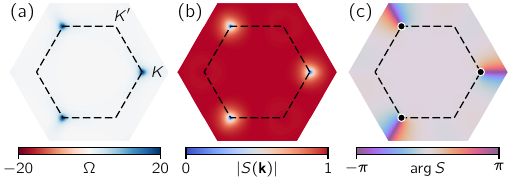}
    \caption{     Relative topology in the Haldane model.     (a) Berry curvature of the topological valence band with     $C_\alpha=-1$, for $M=0.8$, $t_2^\alpha=0.2$, and $t_1=1$.     (b) Magnitude of the overlap with the trivial valence band,     obtained for $t_2^\sigma=0.02$ at the same $M$ and $t_1$.     The apparent copies of the zero in the extended-zone     representation correspond to a single point on the     Brillouin-zone torus. Its winding is $-1$, in agreement with     $C_\alpha-C_\sigma=-1$. The dashed hexagon bounds the first Brillouin zone, and $\Omega$ in (a) is the Berry curvature in our convention, $C=-\frac{1}{2\pi}\int\Omega$. (c) Phase of the overlap, in a frame that is smooth away from the zero at $K$, around which it winds by $-2\pi$. Black dots mark this zero and its copies.}
    \label{fig:overlap_haldane}
\end{figure}

\begin{figure}[t]
    \centering
    \includegraphics[width=\linewidth]{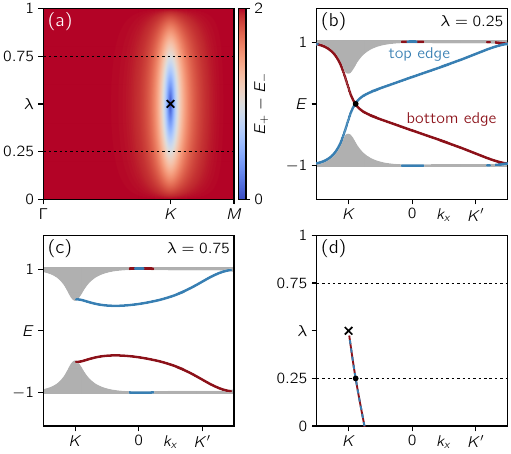}
    \caption{Spectral flow for the pair of Fig.~\ref{fig:overlap_haldane}. (a) Gap $E_+-E_-$ of the interpolation Hamiltonian of Eq.~\eqref{eq:flattened_interpolation}, obtained by numerical diagonalization, along $\Gamma$-$K$-$M$ as a function of $\lambda$. It closes only at $K$ and $\lambda=1/2$ (cross), where the overlap vanishes. The dotted lines mark the values of $\lambda$ used in (b) and (c). (b) Spectrum of a zigzag ribbon at $\lambda=0.25$. Bulk states are grey, and each edge carries one chiral mode. The two modes cross zero energy at the same $k_x$ (dot). (c) Same at $\lambda=0.75$, where no state crosses the gap. (d) Zero-energy edge states in the $(k_x,\lambda)$ plane. The curve passes through the dot of (b) and ends at the projection of the gap closing (cross). Its other end is at $\lambda=0$, and not at a second Weyl point. $K$ and $K'$ label the projections of the valleys.}
    \label{fig:haldane_spectral_flow}
\end{figure}

Figure~\ref{fig:overlap_haldane} shows the Berry curvature of a valence band with $C_\alpha=-1$, as well as the overlap with a trivial valence band.
As seen in Fig.~\ref{fig:overlap_haldane}(b), the comparison section
vanishes at a single independent point of the Brillouin-zone torus.
The phase of the overlap winds once with negative orientation around
this zero (see Fig.~\ref{fig:overlap_haldane}(c)), reproducing the relative Chern number $C_r=-1$. 
Figure~\ref{fig:haldane_spectral_flow} shows the spectral flow of Sec.~\ref{subsec:spectral_flow} for the same pair. The gap of the interpolation Hamiltonian closes only at $K$ and $\lambda=1/2$, and on a ribbon the chiral edge modes present for $\lambda<1/2$ end at the projection of this point.

For this model the zero lies at $K$. Section~\ref{sec:crystalline_symmetries} explains how threefold rotation constrains its position.

\section{Class AII: relative $\mathbb{Z}_2$ invariant}
\label{sec:time_reversal_singular_connection}

We now extend the construction to two-dimensional
time-reversal-symmetric insulators in class AII.
Time-reversal symmetry forces the first Chern number
of the occupied bundle, and hence the total winding of the zeros of any global section
over the full Brillouin zone, to vanish.
Nevertheless, these systems can carry a nontrivial
$\mathbb Z_2$ invariant
~\cite{Kane_Mele_2005,Kane_Fu_2007}.
The $\mathbb Z_2$ invariant was originally formulated via the vorticity of a Pfaffian
of a matrix constructed from matrix elements of the time reversal operator between occupied states.
As we show below, this Pfaffian can be associated with the definition of a global section of an
appropriate line bundle. We then extend the
construction to the comparison of occupied bundles and show that
the parity of the total winding around zeros over a time-reversal fundamental domain gives their relative
$\mathbb Z_2$ invariant.

For simplicity, we consider a single occupied Kramers pair of bands, whose bundle $\mathcal E$ has rank two. The construction extends to higher even ranks (see Appendix~\ref{app:AII_details}). The time-reversal operator $\Theta$ is antiunitary and acts on the fibers as
\begin{equation}
\Theta^2=-1,
\qquad
\Theta:
\mathcal E_{\mathbf{k}}
\longrightarrow
\mathcal E_{-\mathbf{k}}.
\label{eq:AII_TR_action}
\end{equation}
The properties define what is mathematically called a Quaternionic structure on the bundle: an anti-linear map relating the fibers at $\mathbf{k}$ and $-\mathbf{k}$ that squares to $-1$. Since time reversal relates the states at $\mathbf{k}$ and $-\mathbf{k}$, we work with a time-reversal fundamental domain $\mathrm{BZ}^2_+$.
Its interior contains one representative of each generic pair $\{\mathbf{k},-\mathbf{k}\}$, while its boundary is mapped into itself by time reversal.
For a local basis
\begin{equation}
U(\mathbf{k})
=
\bigl(
\ket{\psi_1(\mathbf{k})},
\ket{\psi_2(\mathbf{k})}
\bigr),
\end{equation}
the determinant line
\begin{equation}
\mathcal L
=
\det\mathcal E
=
\wedge^2\mathcal E
\end{equation}
has local basis vector
\begin{equation}
e_{\mathcal E}(\mathbf{k})
=
\ket{\psi_1(\mathbf{k})}
\wedge
\ket{\psi_2(\mathbf{k})}.
\end{equation}
The new ingredient here beyond the class A treatment is the nontrivial action of time reversal on $e_{\mathcal E}(\mathbf{k})$
\begin{equation}
\Theta_{\det}
\left(
\ket{\psi_1}\wedge\ket{\psi_2}
\right)
=
\Theta\ket{\psi_1}
\wedge
\Theta\ket{\psi_2}.
\label{eq:induced_TR_action}
\end{equation}
Although $\Theta^2=-1$ on each state, the two minus signs
cancel on the wedge product, giving
\begin{equation}
\Theta_{\det}^2=+1.
\label{eq:AII_det_real_structure}
\end{equation}
Thus, the determinant line bundle has a Real structure, i.e. an anti-linear action between fibers at $\mathbf{k}$ and $-\mathbf{k}$ that squares to $+1$.

Our goal is to detect the $\mathbb Z_2$ obstruction through the zeros of a global section. However, requiring the section to respect time reversal is not sufficient: two time-reversal-compatible sections of the same bundle can have different winding parities in half the Brillouin zone. This ambiguity arises because time reversal does not fix the signs of the sections at the time-reversal-invariant momenta (TRIMs). To ensure that the winding parity reflects the topology of the occupied bundle rather than the choice of section, we impose two conditions. First, the section must respect time reversal,
\begin{equation}
a(-\mathbf{k})
=
\Theta_{\det}a(\mathbf{k}).
\label{eq:AII_section_TR}
\end{equation}
Second, we require the section to be nonzero and to have a fixed sign relative to the occupied Kramers pair at every TRIM. We now explain how this sign can be defined independently of the choice of occupied states.

We use $\Gamma_i$ as a generic label for the four TRIMs. In a specific lattice we refer to them by their conventional names: $\{\Gamma,M,M',M''\}$ in the hexagonal Brillouin zone and $\{\Gamma,X,Y,M\}$ in the square one.
At a TRIM $\Gamma_i$, with $i=0,1,2,3$,
$-\Gamma_i=\Gamma_i$ modulo a reciprocal lattice vector, and
\begin{equation}
\Theta:
\mathcal E_{\Gamma_i}
\longrightarrow
\mathcal E_{\Gamma_i},
\qquad
\Theta^2=-1.
\end{equation}
For any normalized state $\ket{u}\in\mathcal E_{\Gamma_i}$, the states $\ket{u}$ and $\Theta\ket{u}$ form an orthonormal Kramers pair, and hence define the nonzero determinant vector
\begin{equation}
e_{K}(\Gamma_i)
=
\ket{u}\wedge\Theta\ket{u}.
\label{eq:AII_canonical_TRIM_vector}
\end{equation}
Importantly, this vector is independent of the choice of normalized $\ket{u}$, since different Kramers bases are related by an $SU(2)$ transformation, which leaves their antisymmetric product unchanged (see Appendix~\ref{app:AII_details}). It also satisfies
\begin{equation}
\Theta_{\det}e_{K}(\Gamma_i)
=
e_{K}(\Gamma_i).
\end{equation}
Thus, time reversal provides a preferred nonzero determinant vector at every TRIM, independent of the choice of basis. We can use this vector to define the sign of a section at these momenta.

Let local determinant-line bases be chosen in compatibility with time reversal,
\begin{equation}
e(-\mathbf{k})
=
\Theta_{\det}e(\mathbf{k}).
\end{equation}
A basis change $e(\mathbf{k})\mapsto h(\mathbf{k})e(\mathbf{k})$ preserving this condition must obey
\begin{equation}
h(-\mathbf{k})
=
(h(\mathbf{k}))^{*}.
\label{eq:AII_real_gauge_condition}
\end{equation}
At a TRIM $\Gamma_i$,
a time-reversal-compatible gauge transformation
$h\in U(1)$ is restricted to $h=\pm1$.
Thus, time reversal restricts the phase freedom at the TRIMs to a sign, but does not determine that sign.
The determinant vector
$e_{K}(\Gamma_i)$ defined in
Eq.~\eqref{eq:AII_canonical_TRIM_vector}
provides a reference for distinguishing these
two possibilities. As we show in the next subsection, fixing the signs relative to these vectors is essential for extracting a well-defined $\mathbb Z_2$ invariant from the zeros of a global section.
The derivation of
Eq.~\eqref{eq:AII_real_gauge_condition}
is given in Appendix~\ref{app:AII_details}.

\subsection{Class AII $\mathbb Z_2$ invariant from singular connections}
\label{subsec:AII_absolute}

We first obtain the invariant of a single occupied Kramers pair of bands by comparing it with a fixed, trivial Kramers pair. This is the time-reversal analogue of the singular connection approach to characterizing Chern bands of Ref.~\cite{Mera_2021}, where a fixed reference state is projected onto a single band. Here, the reference must instead consist of two states related by time reversal.

Let $R$ be the reference bundle generated by a fixed, normalized Kramers pair
\begin{equation}
\ket{R_1},
\qquad
\ket{R_2}=\Theta\ket{R_1}.
\label{eq:AII_reference_pair}
\end{equation}
This defines a trivial rank-two Quaternionic bundle with a globally defined, nowhere-vanishing section of its determinant line,
\begin{equation}
e_R
=
\ket{R_1}\wedge\ket{R_2}.
\end{equation}
By construction, $e_R$ satisfies the two conditions discussed above. First, it is time-reversal compatible, since $\Theta_{\det}e_R=e_R$. Second, at every TRIM, it coincides with the determinant vector defined by its Kramers pair, $e_R=\ket{R_1}\wedge\Theta\ket{R_1}$, and therefore has a fixed positive sign.

Following the relative construction of
Sec.~\ref{sec:singular_connections}, we construct a section relating the determinant line of the occupied bundle to that of the reference through
\begin{equation}
\mathcal L_{\mathcal E,R}
=
\operatorname{Hom}
(\det R,\det\mathcal E)
=
\det\mathcal E\otimes(\det R)^*.
\label{eq:AII_reference_relative_line}
\end{equation}
Because $\det R$ has the global nonvanishing basis $e_R$,
$\mathcal L_{\mathcal E,R}$ is isomorphic to
$\det\mathcal E$ as a complex line bundle. We nevertheless keep the reference explicit, since
it makes clear that the section constructed below
encodes the topology of the occupied Kramers pair
relative to a fixed time-reversal-compatible reference.

For a local basis
$\{\ket{\psi_1(\mathbf{k})},\ket{\psi_2(\mathbf{k})}\}$, we define
the matrix of overlaps
\begin{equation}
O_{ij}(\mathbf{k})
=
\braket{\psi_i(\mathbf{k})|R_j},
\qquad
i,j=1,2.
\label{eq:AII_reference_overlap}
\end{equation}
The corresponding comparison section~\footnote{With the $R_j$ taken as trial orbitals, $O_{ij}$ is the projection matrix used to initialize a Wannier construction~\cite{Marzari_Vanderbilt_1997}. Soluyanov and Vanderbilt exhibited zeros of its determinant for time-reversal-paired trial orbitals in the quantum spin Hall phase~\cite{Soluyanov_Vanderbilt_2011} and subsequently constructed a smooth gauge by decomposing the occupied subspace into bands of opposite Chern number~\cite{Soluyanov_Vanderbilt_2012}.} is then given by 
\begin{equation}
a_O(\mathbf{k})
=
\det O(\mathbf{k})\,
e_{\mathcal E}(\mathbf{k})\otimes e_R^*.
\label{eq:TR_reference_section}
\end{equation}
It is globally defined and satisfies our first requirement, i.e. it is time-reversal compatible
\begin{equation}
\det O(-\mathbf{k})
=
(\det O(\mathbf{k}))^{*}.
\label{eq:AII_reference_TR_constraint}
\end{equation}
The derivation is given in Appendix~\ref{app:AII_details}.

We now show that $a_O$ also satisfies our second requirement, namely that it is nonzero and has a fixed positive sign relative to the occupied Kramers pair at every TRIM. At a TRIM $\Gamma_i$, we can choose the occupied basis such that $\ket{\psi_2(\Gamma_i)}=\Theta\ket{\psi_1(\Gamma_i)}$, giving $e_{\mathcal E}(\Gamma_i)=e_{K}(\Gamma_i)$. Since the reference states also form a Kramers pair, their overlap satisfies
\begin{equation}
\det O(\Gamma_i)
=
|\braket{\psi_1(\Gamma_i)|R_1}|^2
+
|\braket{\psi_1(\Gamma_i)|R_2}|^2
\geq 0.
\end{equation}
Thus, whenever it is nonzero, $a_O(\Gamma_i)$ is a positive multiple of $e_{K}(\Gamma_i)\otimes e_R^*$, as required. For the zero-counting construction, we choose the reference $R$ generically so that $\det O(\Gamma_i)>0$ at every TRIM. This choice is justified in Appendix~\ref{app:AII_generic_reference}. The section $a_O$ is therefore nonvanishing at these points.

The comparison section $a_O$ can also be obtained directly
by projecting the two reference states onto the occupied
subspace. Denoting by $P_{\mathcal E}(\mathbf{k})$ the
projector onto the occupied subspace, their wedge product gives (see derivation in Appendix~\ref{app:AII_details})
\begin{equation}
P_{\mathcal E}(\mathbf{k})\ket{R_1}
\wedge
P_{\mathcal E}(\mathbf{k})\ket{R_2}
=
\det O(\mathbf{k})\,
e_{\mathcal E}(\mathbf{k}).
\label{eq:AII_projected_area}
\end{equation}
Geometrically, the wedge product represents the area element
spanned by the two projected reference states in the occupied
Hilbert space, with $\det O(\mathbf{k})$ giving its coefficient
relative to the local area element
$e_{\mathcal E}(\mathbf{k})
=\ket{\psi_1(\mathbf{k})}\wedge\ket{\psi_2(\mathbf{k})}$.
When $\det O(\mathbf{k})=0$, the projected states become
linearly dependent and fail to span the occupied subspace. As in class A, these zeros define the singularities of the connection associated with $a_O$, and each isolated zero carries a winding charge.

Because the comparison section respects time reversal, a zero at $\mathbf{k}_*$ is accompanied by a zero at $-\mathbf{k}_*$. We denote the phase winding around a zero by $\nu^{\mathcal E}_{\mathbf{k}_*}$. Since $\mathbf k\mapsto-\mathbf k$ preserves the orientation of the two-dimensional Brillouin zone, while complex conjugation reverses the phase winding, we have
\begin{equation}
\nu^{\mathcal E}_{-\mathbf{k}_*}
=
-\nu^{\mathcal E}_{\mathbf{k}_*}.
\label{eq:AII_reference_winding_pair}
\end{equation}
The signed sum of the zero charges over the complete
Brillouin zone therefore vanishes. However, the sum
over a time-reversal fundamental domain, such as
$\mathrm{BZ}^2_+$, containing one representative of
each generic pair $\{\mathbf{k},-\mathbf{k}\}$,
need not vanish.

Assuming zeros are isolated, we choose the fundamental
domain so that its boundary contains no zeros, and let
\begin{equation}
Z_O^+
=
\left\{
\mathbf{k}\in\mathrm{BZ}^2_+:
\det O(\mathbf{k})=0
\right\},
\end{equation}
denote the zeros contained in this domain. We then define the corresponding half-Brillouin-zone winding
\begin{equation}
N_{\mathcal E}
=
\sum_{\mathbf{k}_*\in Z_O^+}
\nu^{\mathcal E}_{\mathbf{k}_*}
\label{eq:AII_half_BZ_integer}
\end{equation}
that depends on which member of each
time-reversal pair is selected by the fundamental domain. Different choices can result in exchanging a zero of winding $\nu$ with its time-reversed partner, which changes the sum by $-2\nu$, but leaves the parity of the total winding to be the same. Hence we can define the topological invariant 
\begin{equation}
\nu_{\mathbb Z_2}(\mathcal E)
=
\sum_{\mathbf{k}_*\in Z_O^+}
\nu^{\mathcal E}_{\mathbf{k}_*}
\quad \mathrm{mod}\;2.
\label{eq:Z2_invariant}
\end{equation}

To support this definition of the invariant, we must also establish that the parity does not depend on the choice of global section, provided it satisfies the two conditions discussed above. To see this, consider how the zeros can move under continuous deformations that preserve time reversal and keep the sections nonzero and positive relative to the Kramers-pair determinants at the TRIMs. Two zeros of opposite winding may meet and annihilate inside $\mathrm{BZ}^2_+$, with the time-reversed annihilation occurring simultaneously in the complementary half of the Brillouin zone. Such annihilations remove two time-reversal pairs and therefore do not change the parity. A single time-reversal pair, however, cannot
annihilate by itself away from the TRIMs. Its two
members can meet only at a TRIM, which would require the section to vanish
there, violating our second requirement. Moreover, any two global sections satisfying these conditions can be continuously interpolated while preserving time reversal and their positive signs relative to the Kramers-pair determinants at the TRIMs. The parity is therefore independent of the choice of section and measures the obstruction to finding a globally nonvanishing, time-reversal-compatible section satisfying the required sign condition at every TRIM.

In the next subsection, we show that the Pfaffian
introduced by Kane and Mele also defines a global,
time-reversal-compatible section satisfying the same sign condition at the TRIMs, with an appropriate overall sign convention. Its half-Brillouin-zone winding parity therefore coincides with the
$\mathbb Z_2$ invariant defined in
Eq.~\eqref{eq:Z2_invariant}.

\subsection{Relation to the Kane-Mele invariant}
\label{subsec:AII_pfaffian}

To connect the above approach to the standard formulation of the $\mathbb Z_2$ invariant, we recall the Pfaffian construction of Kane and Mele~\cite{Kane_Mele_2005,Kane_Fu_2007}. For a local basis
$\{\ket{\psi_1(\mathbf{k})},\ket{\psi_2(\mathbf{k})}\}$,
we introduce the antisymmetric matrix
\begin{equation}
M_{ij}(\mathbf{k})
=
\braket{
\psi_i(\mathbf{k})
|
\Theta\psi_j(\mathbf{k})
}.
\label{eq:AII_pfaffian_matrix}
\end{equation}
For a single occupied Kramers pair, the Pfaffian of $M$ is a complex scalar,
\begin{equation}
p(\mathbf{k})
=
\operatorname{Pf}M(\mathbf{k}).
\end{equation}
Under a change of basis, $p$ transforms with the inverse determinant, canceling the transformation of $e_{\mathcal E}$. Hence
\begin{equation}
a_{\mathrm{Pf}}(\mathbf{k})
=
p(\mathbf{k})\,
e_{\mathcal E}(\mathbf{k})\otimes e_R^*
\label{eq:pfaffian_global_section}
\end{equation}
defines a global, time-reversal-compatible section of the same line bundle
$\mathcal L_{\mathcal E,R}$ introduced in
Eq.~\eqref{eq:AII_reference_relative_line}. The sections $a_O$ and $a_{\mathrm{Pf}}$ are therefore two different global sections satisfying our first requirement. 
At each TRIM, choosing a Kramers basis $\ket{\psi_2}=\Theta\ket{\psi_1}$ gives $p(\Gamma_i)=-1$. Thus, $-a_{\mathrm{Pf}}$ has a positive sign relative to the Kramers-pair determinant at every TRIM, just like $a_O$, and therefore also satisfies our second requirement.
As shown in Appendix~\ref{app:AII_details},
the two sections can therefore be related through
time-reversal-compatible deformations without forcing
a zero through a TRIM, hence their half-Brillouin-zone
zero-counting parities must agree. Fukui and Hatsugai related the vorticity of the Pfaffian in half the Brillouin zone to the $\mathbb Z_2$ index and gave a lattice formulation~\cite{Fukui_Hatsugai_2007}. Having recovered the usual class AII invariant using a fixed trivial Kramers reference, we now extend the construction to the comparison of two arbitrary time-reversal-symmetric occupied bundles.

\subsection{Relative $\mathbb{Z}_2$ invariant}
\label{subsec:AII_relative}

Let $\mathcal E^\alpha$ and $\mathcal E^\sigma$ be rank-two occupied
bundles over the same Brillouin zone, represented in the same Bloch
Hilbert space and carrying the same time-reversal symmetry
$\Theta^2=-1$. We use the relative determinant line introduced in
Sec.~\ref{sec:singular_connections}, which coincides with $\mathcal L_r$ of Eq.~\eqref{eq:relative_line_bundle} and which we denote here by $\mathcal L_{\alpha\sigma}$,
\begin{equation}
\mathcal L_{\alpha\sigma}
=
\det\mathcal E^\alpha
\otimes
(\det\mathcal E^\sigma)^*
=
\operatorname{Hom}
\left(
\det\mathcal E^\sigma,
\det\mathcal E^\alpha
\right).
\label{eq:AII_relative_line}
\end{equation}
Time reversal acts on this line through its action on the determinant lines of the two bundles. At each TRIM, the Kramers pairs of the two systems also provide a reference vector
$e^{K}_\alpha(\Gamma_i)\otimes\bigl(e^{K}_\sigma(\Gamma_i)\bigr)^*$, which allows us to define the sign of the comparison section at these momenta.

For local bases
\begin{equation}
U^\mu(\mathbf{k})
=
\bigl(
\ket{\psi_1^\mu(\mathbf{k})},
\ket{\psi_2^\mu(\mathbf{k})}
\bigr),
\qquad
\mu\in\{\alpha,\sigma\},
\end{equation}
we form the matrix of overlaps
$S(\mathbf{k})$ of
Eq.~\eqref{eq:overlap_matrix}. Its determinant is the local coefficient
of the global comparison section $a_S$. 
We first verify that $a_S$ satisfies the two requirements discussed above. As in the construction with a trivial reference, the comparison section respects time reversal. In time-reversal-compatible determinant-line bases, this gives
\begin{equation}
\det S(-\mathbf{k})
=
(\det S(\mathbf{k}))^{*}.
\label{eq:AII_relative_TR_constraint}
\end{equation}
At a TRIM $\Gamma_i$,
the overlap between two Kramers bases satisfies
\begin{equation}
\det S(\Gamma_i)\geq 0.
\end{equation}
Thus, whenever it is nonzero, $a_S(\Gamma_i)$ is a
positive multiple of
$e^{K}_\alpha(\Gamma_i)\otimes
\bigl(e^{K}_\sigma(\Gamma_i)\bigr)^*$,
as required by our second condition. For the zero-counting construction,
we assume that $\det S(\Gamma_i)>0$ at every TRIM,
as it holds generically. Although a zero may occur at a TRIM, a small time-reversal-preserving deformation can make the overlap nonzero at that momentum without changing the topological class of either system, as shown in Appendix~\ref{app:AII_details}. Such a deformation does not remove the global topological obstruction, which remains encoded in the winding parity of the zeros away from the TRIMs. If an additional crystalline symmetry enforces a zero at a TRIM, that symmetry must be broken for the zero to be removed from that momentum. The comparison section therefore satisfies both requirements of our construction.

A zero of $\det S$ therefore has the same meaning as in class A: the matrix of overlaps loses rank, and the projection map between the two subspaces fails to be invertible at that momentum. Its phase winding can therefore be computed using the singular connection introduced in Sec.~\ref{sec:singular_connections}. We choose a fundamental domain whose boundary contains no zeros, so that the half-Brillouin-zone construction of Sec.~\ref{subsec:AII_absolute} applies directly to $a_S$.

It remains to identify this parity with the usual class AII invariants of the two systems. Using the same
trivial reference $R$ as in Sec.~\ref{subsec:AII_absolute}, the
relative line can be written as
\begin{equation}
\mathcal L_{\alpha\sigma}
\simeq
\mathcal L_{\mathcal E^\alpha,R}
\otimes
\mathcal L_{\mathcal E^\sigma,R}^{\,*}.
\label{eq:AII_relative_reference_factorization}
\end{equation}
The winding parity is additive modulo two under this tensor product, with the dual factor contributing with the opposite sign. The parity associated with $\mathcal L_{\alpha\sigma}$ is therefore the difference between the absolute class AII invariants of the two systems. Hence,
\begin{equation}
\nu_{\mathbb Z_2}(\mathcal E^\alpha)
-
\nu_{\mathbb Z_2}(\mathcal E^\sigma)
=
\sum_{\mathbf{k}_*\in Z_{\alpha\sigma}^+}
\nu_{\mathbf{k}_*}
\quad
\mathrm{mod}\;2.
\label{eq:relative_Z2_det_overlap}
\end{equation}
Here $Z_{\alpha\sigma}^+$ denotes the zeros of the comparison section contained in the chosen time-reversal fundamental domain. The subtraction on the left-hand side is, of course, understood modulo two. The detailed equivalence, including the treatment of boundary zeros, is given in Appendix~\ref{app:AII_details}. Equation~\eqref{eq:relative_Z2_det_overlap} is the class AII analogue of the relative Chern-number formula Eq.~\eqref{eq:classA_main_result}. 

If the two systems also share an inversion symmetry $I$ with $[I,\Theta]=0$, the antiunitary symmetry $I\Theta$ fixes each momentum and squares to $-1$. In local $I\Theta$-Kramers bases, the overlap determinant takes the form $\det S=|S_{11}|^2+|S_{12}|^2\geq0$ and therefore cannot wind around any zero, whether at a TRIM or elsewhere. Consequently, the comparison section is not generic, and Eq.~\eqref{eq:relative_Z2_det_overlap} cannot be applied directly to it. To recover the winding formulation, one must construct a different, generic time-reversal-compatible section. In the presence of inversion symmetry, however, the relative $\mathbb Z_2$ invariant can be determined more simply from the inversion eigenvalues using the Fu-Kane parity formula.
At each TRIM $\Gamma_i$, both occupied states of system $\mu$ have the same inversion eigenvalue $\xi_I^\mu(\Gamma_i)$, and $S(\Gamma_i)=0$ whenever $\xi_I^\alpha(\Gamma_i)\neq\xi_I^\sigma(\Gamma_i)$. Applying the Fu-Kane parity formula~\cite{Kane_Fu_2007} to each system gives
\begin{equation}
(-1)^{\nu_{\mathbb Z_2}(\mathcal E^\alpha)-\nu_{\mathbb Z_2}(\mathcal E^\sigma)}
=\prod_{i=0}^{3}\xi_I^\alpha(\Gamma_i)\,\xi_I^\sigma(\Gamma_i),
\label{eq:relative_fu_kane}
\end{equation}
so the relative $\mathbb Z_2$ invariant is the parity of the number of TRIMs at which $S$ vanishes. In the interpolation of Sec.~\ref{subsec:spectral_flow_Z2}, each such TRIM corresponds to a fourfold crossing at $\lambda=1/2$ rather than a pair of Weyl points. A winding formulation can nevertheless be recovered by comparing each system with a trivial reference Kramers pair that is not an inversion eigenstate, as in Sec.~\ref{subsec:AII_absolute}, or by introducing a time-reversal-preserving deformation that breaks inversion and allows the zeros to move away from the TRIMs.

\begin{figure*}[t]
    \centering
    \includegraphics[width=\linewidth]{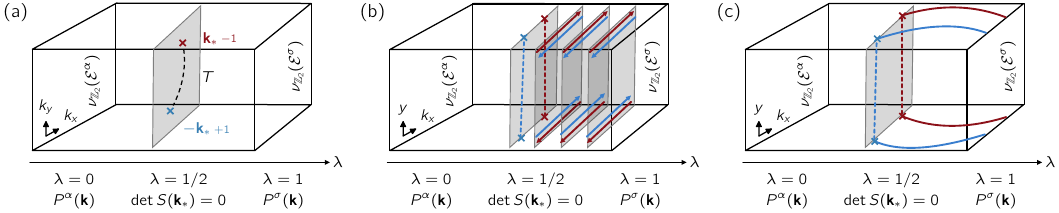}
    \caption{
    Spectral flow interpretation of the relative
    $\mathbb Z_2$ invariant.
    \textbf{(a)} Zeros of the comparison section
    correspond to auxiliary Weyl crossings at
    $\lambda=1/2$, occurring in time-reversed
    pairs with opposite winding.
    \textbf{(b)} A nontrivial fixed-$\lambda$
    slice supports a Kramers pair of
    counterpropagating boundary modes.
    \textbf{(c)} Opening the system along $y$
    projects the bulk crossings onto the boundary
    parameter space $(k_x,\lambda)$, where they
    are accompanied by time-reversal-related
    boundary spectral flow.
    }
    \label{fig:z2_spectral_flow}
\end{figure*}

By the same deformation argument as in
Sec.~\ref{subsec:AII_absolute}, opposite-winding
zeros of the comparison section may annihilate
in time-reversal-related pairs. A single
time-reversal pair, however, cannot annihilate by itself away from a TRIM, while annihilation at a TRIM is excluded by our requirement that the comparison section remain nonzero there. An odd parity therefore implies that every global section satisfying our two conditions must have zeros. Consequently, no smooth, everywhere-invertible, time-reversal-compatible map can exist between the two bundles, since the determinant of such a map would define a section satisfying both conditions without zeros. Conversely, since two-dimensional class AII bundles are classified by their $\mathbb Z_2$ invariant~\cite{DeNittis_Gomi_2015}, an even parity implies that the two bundles are topologically equivalent in class AII, even though the particular comparison section $a_S$ may still have zeros. Additional crystalline symmetries may distinguish the two systems further. We return to these additional constraints in Sec.~\ref{sec:crystalline_symmetries}.

\subsection{Spectral flow interpretation}
\label{subsec:spectral_flow_Z2}

As in class A, the zeros of the comparison section
correspond to gap closings of an auxiliary interpolation.
For the rank-two occupied spaces considered here,
we interpolate between the occupied projectors,
\begin{equation}
H(\lambda,\mathbf{k})
=
(1-\lambda)P^\alpha(\mathbf{k})
+
\lambda P^\sigma(\mathbf{k}),
\qquad
\lambda\in[0,1].
\label{eq:z2_projector_interpolation}
\end{equation}
As shown in Appendix~\ref{app:z2_spectral_flow_details},
the gap at energy $1/2$ closes precisely when
\begin{equation}
\lambda=\frac{1}{2},
\qquad
\det S(\mathbf{k})=0.
\label{eq:rank_two_gap_closing_detS}
\end{equation}
Near a simple zero, one singular value of $S$
vanishes while the other remains nonzero,
leaving an effective two-level Weyl crossing
whose charge equals the winding of the zero,
as in Sec.~\ref{subsec:spectral_flow}. 

Time reversal relates a crossing at $\mathbf{k}_*$
to one at $-\mathbf{k}_*$ with opposite charge. Such paired gap closings at generic momenta also occur at the transition between an ordinary insulator and a quantum spin Hall insulator without inversion symmetry~\cite{Murakami_et_al_2007}.
Their total charge therefore vanishes, but the
parity of the crossings in a time-reversal
fundamental domain can remain nontrivial.
By Eq.~\eqref{eq:relative_Z2_det_overlap},
this parity gives the relative $\mathbb Z_2$
invariant of the two occupied bundles.

The spectral interpretation follows directly.
For $\lambda<1/2$, the rank-two eigenbundle
above energy $1/2$ is adiabatically connected
to $\mathcal E^\alpha$, while for $\lambda>1/2$
it is connected to $\mathcal E^\sigma$.
Consequently, a nontrivial relative $\mathbb Z_2$
invariant requires an odd number of time-reversal
pairs of simple gap closings in the auxiliary
interpolation, as illustrated in
Fig.~\ref{fig:z2_spectral_flow}.

The same picture has a boundary manifestation,
shown schematically in
Fig.~\ref{fig:z2_spectral_flow}(b,c).
Opening the system along $y$ leaves
$(k_x,\lambda)$ as the boundary parameter space.
A nontrivial fixed-$\lambda$ slice supports
a Kramers pair of counterpropagating boundary
modes. As $\lambda$ passes through the auxiliary
bulk crossings, the boundary spectrum undergoes
spectral flow. The parity of the helical boundary spectral flow
reflects the relative $\mathbb Z_2$ invariant,
providing the boundary counterpart of the
bulk zero-counting construction.
Further details are given in
Appendix~\ref{app:z2_spectral_flow_details}.

\subsection{Application: Kane-Mele model}
\label{subsec:kane_mele_application}

We illustrate the class AII construction using the Kane-Mele model on
the honeycomb lattice~\cite{Kane_Mele_2005}. At half filling, the two
lowest-energy bands form a rank-two occupied bundle with
time-reversal symmetry $\Theta^2=-1$. The explicit Hamiltonian and our
lattice conventions are given in
Appendix~\ref{app:kane_mele_model}.

We first consider the absolute construction of
Sec.~\ref{subsec:AII_absolute}. For
\begin{equation}
\frac{\lambda_v}{t}=0.1,
\qquad
\frac{\lambda_R}{t}=0.05,
\qquad
\frac{\lambda_{\mathrm{SO}}}{t}=0.06,
\label{eq:KM_absolute_parameters}
\end{equation}
the occupied bundle is in the nontrivial class AII phase. As the
trivial reference, we choose the momentum-independent Kramers pair
\begin{equation}
\ket{R_1}=\ket{B,\downarrow},
\qquad
\ket{R_2}=\ket{B,\uparrow}.
\end{equation}
Using this reference, we evaluate the determinant-overlap section
$a_O$ of Eq.~\eqref{eq:TR_reference_section}. We also evaluate the
Pfaffian section $a_{\mathrm{Pf}}$ introduced in
Sec.~\ref{subsec:AII_pfaffian}.

\begin{figure}[t]
    \centering
    \includegraphics[width=\linewidth]{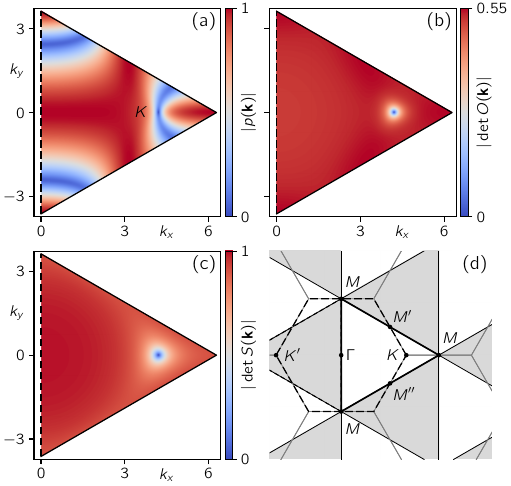}
    \caption{
    Kane-Mele illustration of the class AII construction in a chosen
    time-reversal fundamental domain $\mathrm{BZ}^2_+$.
    \textbf{(a)} Magnitude of the Pfaffian coefficient
    $|p(\mathbf{k})|$.
    \textbf{(b)} Magnitude of the determinant-overlap coefficient
    $|\det O(\mathbf{k})|$ obtained from the trivial reference Kramers
    pair. Both sections contain an odd-winding zero and therefore give
    $\nu_{\mathbb Z_2}=1$.
    \textbf{(c)} Magnitude of the relative determinant overlap
    $|\det S(\mathbf{k})|$ between a topological and a
    trivial occupied bundle. Its odd-winding zero gives a nontrivial
    relative $\mathbb Z_2$ obstruction.
    The pale arcs in (a) are a valley along which $|p|\geq0.05$; the only zero of $p$ in $\mathrm{BZ}^2_+$ is at $K$. \textbf{(d)} Tiling of reciprocal space by hexagonal Brillouin zones. The triangle with bold edges is $\mathrm{BZ}^2_+$, the region shown in (a)-(c). Its copies under reciprocal-lattice translations are white and the copies of its image under $\mathbf k\to-\mathbf k$ are grey, so that one white and one grey triangle together cover the Brillouin zone (dashed hexagon) once. Dots mark $\Gamma$, the valleys $K$ and $K'$, and the TRIMs $M$, $M'$ and $M''$. Figure~\ref{fig:km_pair_annihilation} is drawn around the $M$ at the tip of the triangle.
    }
    \label{fig:kane_mele_z2_overlap}
\end{figure}

As shown in Fig.~\ref{fig:kane_mele_z2_overlap}(a,b), each section has one zero of odd winding in the chosen time-reversal fundamental domain, giving $\nu_{\mathbb Z_2}=1$. For the present parameters and
reference states, both zeros occur at the $K$ point. Their coincidence
is not required by the class AII topology: the two sections are
different, and only the parity of their winding is fixed by the
$\mathbb Z_2$ invariant. Their location at a high-symmetry momentum is instead enforced by threefold rotation, as discussed in Sec.~\ref{subsec:c3_haldane}.

We next apply the direct relative construction of
Sec.~\ref{subsec:AII_relative}. Keeping
$\lambda_R/t=0.05$ and $\lambda_{\mathrm{SO}}/t=0.06$ fixed, we study the relative topology between the parameter regimes
\begin{equation}
\frac{\lambda_v^\alpha}{t}=-0.1,
\qquad
\frac{\lambda_v^\sigma}{t}=0.6.
\label{eq:KM_relative_parameters}
\end{equation}
For these parameters,
$\nu_{\mathbb Z_2}(\mathcal E^\alpha)=1$ while
$\nu_{\mathbb Z_2}(\mathcal E^\sigma)=0$, so
Eq.~\eqref{eq:relative_Z2_det_overlap} predicts an odd total winding of
the zeros of the comparison section in $\mathrm{BZ}^2_+$.

Figure~\ref{fig:kane_mele_z2_overlap}(c) shows precisely this behavior:
the relative determinant overlap has a single odd-winding zero in the
chosen fundamental domain. Its winding parity is therefore nontrivial,
in agreement with the difference of the $\mathbb Z_2$ indices of the
two endpoint bundles.

To separate the time-reversal constraint from the valley pinning, we change two of the three real nearest-neighbor hoppings in system $\sigma$ to $t(1+\eta)$ and $t(1+0.55\eta)$, keeping the third at $t$ (see Appendix~\ref{app:km_winding_numerics}). We keep $\alpha$ isotropic and set $\lambda_{\rm SO}/t=0.06$ and $\lambda_R/t=0.05$ in both systems. With $\lambda_v^\sigma/t=0.6$, the choices $\lambda_v^\alpha/t=-0.1$ and $-0.6$ give relative indices one and zero, respectively, as determined from the Wilson loops of the individual occupied subspaces~\cite{Dai_2011,Soluyanov_Vanderbilt_2011_Wilson}. The two cases distinguish an odd winding from an even, nonzero winding of the same overlap determinant.

\begin{figure}[t]
    \centering
    \includegraphics[width=\linewidth]{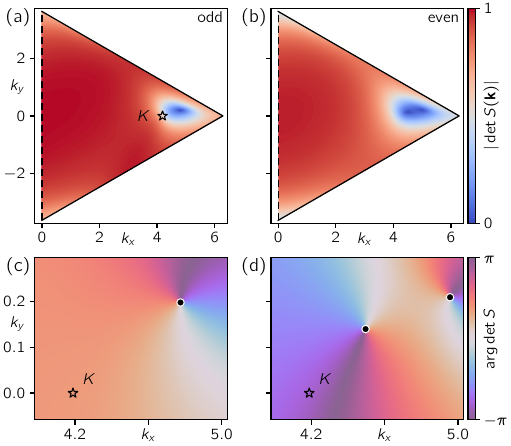}
    \caption{Relative overlap under a time-reversal-preserving hopping anisotropy, $\eta=1.55$. (a,b) $|\det S|$ in the same half Brillouin zone as Fig.~\ref{fig:kane_mele_z2_overlap}, for the pairs with odd and even relative index. (c,d) Phase of $\det S$ near the valley, in a frame that is smooth in the region shown. The star marks the valley $K$, from which the zeros move away as $\eta$ increases. Black dots mark zeros of winding $-1$.}
    \label{fig:km_anisotropic_test}
\end{figure}

Figure~\ref{fig:km_anisotropic_test} shows the result for $\eta=1.55$. As $\eta$ increases from zero, the zeros move continuously away from the valley $K$. In the half Brillouin zone with $k_x>0$ the pair with odd relative index has a single zero, of winding $-1$, while the pair with even relative index has two zeros, each of winding $-1$. The half-Brillouin-zone sums are therefore $-1$ and $-2$, respectively. Every zero has a partner at the opposite momentum with winding $+1$, so the full-zone sums vanish. We locate the zeros using occupied-frame determinant links, following Ref.~\cite{Fukui_Hatsugai_Suzuki_2005}, and measure each winding in a smooth frame on a small enclosing circle (see Appendix~\ref{app:km_winding_numerics}).


\begin{figure}[t]
    \centering
    \includegraphics[width=\linewidth]{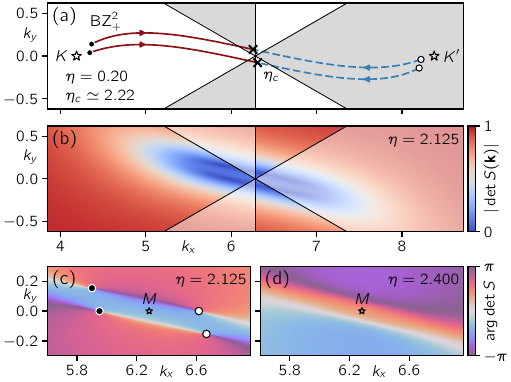}
    \caption{Annihilation of the zeros for the pair with even relative index. (a) Trajectories of the four zeros as the anisotropy $\eta$, applied to both systems, increases from $0.20$. The zeros of winding $-1$ (black dots, solid lines) start near the valley $K$ and their time-reversed partners of winding $+1$ (white dots, dashed lines) near $K'$ (stars). They annihilate at the crosses, at $\eta_c\simeq2.22$. The shading is that of Fig.~\ref{fig:kane_mele_z2_overlap}(d): the white region containing $K$ is the half Brillouin zone $\mathrm{BZ}^2_+$ in which the windings are counted, and the grey regions are copies of its image under $\mathbf k\to-\mathbf k$. The lines are the boundaries between them and meet at the TRIM $M$. (b) $|\det S|$ at $\eta=2.125$, shortly before the annihilation. (c) Phase of $\det S$ at $\eta=2.125$, with its four zeros. (d) Phase of $\det S$ at $\eta=2.40$ in the same window and frame. No zeros remain, and $|\det S|\geq0.07$ in the whole half Brillouin zone. The small star in (c,d) marks the TRIM $M$.}
    \label{fig:km_pair_annihilation}
\end{figure}

Only the parity of the half-Brillouin-zone winding is protected. To illustrate this, we apply the same anisotropy $\eta$ to both systems of the pair with even relative index, with $\lambda_R/t=0.1$, and increase $\eta$, as shown in Fig.~\ref{fig:km_pair_annihilation}. Both systems remain gapped and trivial along this path, with an indirect gap larger than $0.65t$, and $|\det S|\geq0.1$ at the TRIMs. The two zeros of winding $-1$ move away from $K$, while their time-reversed partners of winding $+1$ move away from $K'$. At $\eta_c\simeq2.22$, each zero of winding $-1$ meets the time-reversed partner of the other one and the two annihilate, at a generic momentum close to the TRIM $M$. For $\eta>\eta_c$ the comparison section has no zeros. Before the annihilation, zeros also cross the boundary of the fundamental domain away from the TRIMs, which changes the half-Brillouin-zone winding by two without changing its parity. In contrast, for the pair with odd relative index the single zero in the fundamental domain persists along the same path for as long as both gaps remain open.

The same construction thus yields both the absolute invariant, relative to a trivial Kramers reference, and the relative obstruction between two physical bundles.

\section{Crystalline symmetry constraints on zeros of the section}
\label{sec:crystalline_symmetries}

Crystalline
symmetries can impose constraints on the zeros of the comparison section. The two key points of the discussion are that symmetry imposes a covariance condition that relates
zeros at symmetry-related momenta and constraints their
winding numbers. Additionally, at symmetry-invariant momenta, a
mismatch between the symmetry representations of the
two occupied subspaces can also force the comparison
section to vanish.

\subsection{Unitary crystalline symmetry}
\label{subsec:crystalline_comparison_section}

Let $g$ be a crystalline symmetry acting on momentum as
$\mathbf{k}\mapsto g\mathbf{k}$ and represented on the
Bloch Hilbert space by the unitary map
\begin{equation}
\hat g(\mathbf{k}):
\mathcal H_{\mathbf{k}}
\longrightarrow
\mathcal H_{g\mathbf{k}}.
\end{equation}
We assume that both systems preserve the same microscopic symmetry, i.e
\begin{equation}
\hat g(\mathbf{k})
H^\mu(\mathbf{k})
\hat g(\mathbf{k})^\dagger
=
H^\mu(g\mathbf{k}),
\qquad
\mu\in\{\alpha,\sigma\}.
\end{equation}
Symmetry therefore maps their occupied subspaces according to
\begin{equation}
\hat g(\mathbf{k}):
\mathcal E^\mu_{\mathrm{occ},\mathbf{k}}
\longrightarrow
\mathcal E^\mu_{\mathrm{occ},g\mathbf{k}}.
\end{equation}
This means that for any choice of a local frame
\begin{equation}
U^\mu(\mathbf{k})
=
\bigl(
\ket{\psi_1^\mu(\mathbf{k})},
\ldots,
\ket{\psi_{N_{\mathrm{occ}}}^\mu(\mathbf{k})}
\bigr),
\end{equation}
the symmetry-transformed frame and $U^\mu(g\mathbf{k})$ span the same subspace at $g\mathbf{k}$.
 
These bases are therefore
related by a unitary sewing matrix
\begin{equation}
\hat g(\mathbf{k})U^\mu(\mathbf{k})
=
U^\mu(g\mathbf{k})B_g^\mu(\mathbf{k}),
\qquad
B_g^\mu(\mathbf{k})\in U(N_{\mathrm{occ}}).
\label{eq:crystalline_sewing_matrix}
\end{equation}
We now examine how the sewing matrices constrain the comparison section. In particular, the matrix of overlaps of Eq.~\eqref{eq:overlap_matrix} transforms under crystalline symmetry as
\begin{equation}
S(g\mathbf{k})
=
B_g^\alpha(\mathbf{k})
S(\mathbf{k})
B_g^\sigma(\mathbf{k})^\dagger.
\label{eq:crystalline_overlap_covariance}
\end{equation}
Taking the determinant of both sides of the equation gives the gauge-consistent condition
\begin{equation}
\det S(g\mathbf{k})
=
\det B_g^\alpha(\mathbf{k})
\left[
\det B_g^\sigma(\mathbf{k})
\right]^*
\det S(\mathbf{k}),
\label{eq:crystalline_det_overlap_covariance}
\end{equation}
(see Appendix~\ref{app:crystalline_details}).
The unitarity of the sewing matrix implies that 
\begin{equation}
\det S(\mathbf{k}_*)=0
\quad\Longrightarrow\quad
\det S(g\mathbf{k}_*)=0.
\label{eq:crystalline_zero_mapping}
\end{equation}
Denoting the crystalline symmetry group as $\mathcal{G}$, the zeros of the comparison section therefore occur in
symmetry orbits
\begin{equation}
\mathcal{G}\mathbf{k}_*
=
\left\{
g\mathbf{k}_*:g\in G
\right\}.
\end{equation}
Generic momenta belong to full orbits, while momenta left invariant by a nontrivial subgroup of $G$ belong to shorter ones.

So far we obtained a constraint on the position of the zeros. But symmetry also constrains the local winding carried by each zero.
Writing the action of $g$ on momentum space as
$g\mathbf{k}=R_g\mathbf{k}$ modulo reciprocal lattice
vectors, where $R_g$ is the $2\times 2$ matrix representing this action, we find
\begin{equation}
\nu_{g\mathbf{k}_*}
=
\operatorname{sgn}\!\left[\det R_g\right]
\nu_{\mathbf{k}_*},
\label{eq:crystalline_winding_transformation}
\end{equation}
for a full derivation see Appendix~\ref{app:crystalline_details}. The sewing matrix determinant factor in
Eq.~\eqref{eq:crystalline_det_overlap_covariance} contributes no local
winding because it is smooth and nonvanishing near the zero. The sign in Eq.~\eqref{eq:crystalline_winding_transformation} is therefore determined by whether $R_g$ preserves or reverses the orientation of momentum space (see Appendix~\ref{app:crystalline_details}). Proper rotations therefore relate zeros with the same winding, whereas
orientation-reversing symmetries, such as reflections, relate zeros
with opposite winding. 

\subsection{Symmetry-enforced zeros at invariant momenta}
\label{subsec:symmetry_enforced_zeros}

A stronger constraint arises at momenta left invariant by a subgroup of the crystalline symmetry group. We denote symmetry-invariant momenta by $\mathbf k_\star$, to distinguish them from the zeros $\mathbf k_*$ of the comparison section. Let $\mathbf{k}_\star$ be such a momentum and define
\begin{equation}
\mathcal{G}_{\mathbf{k}_\star}
=
\left\{
g\in G:
g\mathbf{k}_\star
=
\mathbf{k}_\star+\mathbf{G}
\right\},
\end{equation}
where $\mathbf{G}$ is a reciprocal lattice vector. With the reciprocal-lattice identification understood, the matrices $B_g^\mu(\mathbf{k}_\star)$  describe how the occupied subspace
transforms under the crystalline symmetries that leave
$\mathbf{k}_\star$ invariant.

At $\mathbf{k}_\star$,
Eq.~\eqref{eq:crystalline_overlap_covariance} becomes
\begin{equation}
S(\mathbf{k}_\star)
B_g^\sigma(\mathbf{k}_\star)
=
B_g^\alpha(\mathbf{k}_\star)
S(\mathbf{k}_\star),
\qquad
g\in \mathcal{G}_{\mathbf{k}_\star}.
\label{eq:crystalline_fixed_point_constraint}
\end{equation}
This equation shows that the overlap matrix relates the symmetry representations of the two occupied subspaces at $\mathbf{k}_\star$.

If $S(\mathbf{k}_\star)$ were invertible,
Eq.~\eqref{eq:crystalline_fixed_point_constraint} would imply
\begin{equation}
B_g^\alpha(\mathbf{k}_\star)
=
S(\mathbf{k}_\star)
B_g^\sigma(\mathbf{k}_\star)
[S(\mathbf{k}_\star)]^{-1},
\label{eq:crystalline_rep_equivalence}
\end{equation}
for every $g\in \mathcal{G}_{\mathbf{k}_\star}$. The two occupied symmetry
representations at $\mathbf{k}_\star$ would therefore be equivalent.
Conversely, if the two occupied subspaces transform according to
inequivalent symmetry representations at $\mathbf{k}_\star$, no
invertible map satisfying
Eq.~\eqref{eq:crystalline_fixed_point_constraint} can exist, and hence
\begin{equation}
\det
S(\mathbf{k}_\star)
=
0.
\label{eq:crystalline_representation_zero}
\end{equation}
A mismatch between the symmetry representations therefore forces the comparison section to vanish at $\mathbf{k}_\star$, and this zero cannot be removed without changing the symmetry representations.

The intertwining condition also provides information about the rank of the overlap matrix at a symmetry-invariant momentum $\mathbf k_\star$. Since $S(\mathbf k_\star)$ can only connect states transforming under the same symmetry representation, a mismatch between the representations of the two systems forces some directions of the overlap matrix to vanish. To quantify this, let $n_\rho^\mu$ denote the number of times an irreducible representation $\rho$, of dimension $d_\rho$, appears among the occupied states of system $\mu=\alpha,\sigma$. By Schur's lemma, the contribution of each representation to the rank of $S$ is at most $d_\rho\min(n_\rho^\alpha,n_\rho^\sigma)$, since only representations shared by both systems can have nonzero overlap. Using the fact that both systems have $N_{\mathrm{occ}}=\sum_\rho d_\rho n_\rho^\mu$ states, we obtain
\begin{equation}
\dim\ker S(\mathbf k_\star)
\geq
\frac12\sum_\rho d_\rho
|n_\rho^\alpha-n_\rho^\sigma|.
\label{eq:symmetry_rank_bound}
\end{equation}
Thus, the mismatch between the occupied symmetry representations determines a lower bound on the number of independent directions with vanishing overlap.
The rank of the overlap matrix therefore contains symmetry information that is not captured by the winding of its determinant. In the Kane-Mele comparisons of Sec.~\ref{subsec:kane_mele_application}, the mismatch of $C_3$ eigenvalues at $K$ and $K'$ requires at least one vanishing overlap direction. This bound is saturated, with $\operatorname{rank}S=1$ at both valleys.

For a single band, the condition becomes particularly simple.
If
\begin{equation}
\hat g(\mathbf{k}_\star)
\ket{\psi^\mu(\mathbf{k}_\star)}
=
\xi_g^\mu(\mathbf{k}_\star)
\ket{\psi^\mu(\mathbf{k}_\star)},
\end{equation}
then the sewing matrix $B_g^\mu(\mathbf{k}_\star)$ reduces to the scalar $\xi_g^\mu(\mathbf{k}_\star)$. Equation~\eqref{eq:crystalline_fixed_point_constraint}
therefore gives
\begin{equation}
\left[ \xi_g^\alpha(\mathbf{k}_\star)
- \xi_g^\sigma(\mathbf{k}_\star) \right] S(\mathbf{k}_\star) = 0.
\end{equation}
Hence a mismatch of symmetry eigenvalues for any $g\in \mathcal{G}_{\mathbf{k}_\star}$ forces
\begin{equation}
S(\mathbf{k}_\star)=0.
\label{eq:crystalline_eigenvalue_zero}
\end{equation}

Finally, it is important to note that crystalline symmetry can enforce zeros of the comparison section even when the strong relative invariant is trivial.

\subsection{Rotation symmetry in the Haldane, Kane-Mele, and QWZ models}
\label{subsec:c3_haldane}

We now return to the Haldane model
and examine the particular case of constraints imposed by threefold rotation
symmetry. At a generic momentum, the three points
\begin{equation}
\mathbf{k},
\qquad
C_3\mathbf{k},
\qquad
C_3^2\mathbf{k}
\end{equation}
form a symmetry orbit. Since $C_3$ preserves the orientation of
momentum space,
Eq.~\eqref{eq:crystalline_winding_transformation} implies that all
three zeros in such an orbit carry the same winding. A generic orbit
therefore contributes a multiple of three to the total winding.
The relative Chern number can consequently be decomposed as
\begin{equation}
C_\alpha-C_\sigma
=
3\sum_{\mathcal O_{\rm gen}}
\nu_{\mathcal O}^{\alpha\sigma}
+
\nu_{\Gamma}^{\alpha\sigma}
+
\nu_{K}^{\alpha\sigma}
+
\nu_{K'}^{\alpha\sigma},
\label{eq:c3_relative_chern_decomposition}
\end{equation}
where the first sum runs over generic $C_3$ orbits and the remaining
terms denote the winding carried by zeros at the three
$C_3$-invariant momenta: the Brillouin-zone center $\Gamma$
and the two valleys $K$ and $K'$, provided such zeros exist. Hence
\begin{equation}
C_\alpha-C_\sigma
=
\nu_{\Gamma}^{\alpha\sigma}
+
\nu_{K}^{\alpha\sigma}
+
\nu_{K'}^{\alpha\sigma}
\quad
\mathrm{mod}\;3.
\label{eq:c3_relative_chern_mod3}
\end{equation}
Thus the windings at rotation-invariant momenta
determine the relative Chern number modulo three.
The winding of an isolated zero at such a momentum
is itself constrained by the rotation eigenvalues
of the two occupied subspaces. For a signed rotation
angle $\phi_g$, taken positive for counterclockwise
rotations, this relation takes the form
\begin{equation}
e^{i\phi_g\nu_{\mathbf{k}_\star}}
=
\det B_g^\alpha(\mathbf{k}_\star)
\left[
\det B_g^\sigma(\mathbf{k}_\star)
\right]^*.
\label{eq:crystalline_winding_eigenvalues}
\end{equation}
Together with
Eq.~\eqref{eq:c3_relative_chern_mod3},
this relates the relative Chern number modulo three
to the rotation eigenvalues
~\cite{Fang_Gilbert_Bernevig_2012}.
The derivation is given in
Appendix~\ref{app:crystalline_details}. The same argument applies to any $n$-fold rotation with signed angle $\phi_n$. Zeros in the same symmetry orbit share the same winding. Let $\mathcal G_{\mathbf k_\star}$ denote the stabilizer of $\mathbf k_\star$, namely the subgroup of rotations that leave this momentum invariant (up to a reciprocal lattice vector). The orbit of $\mathbf k_\star$ then contains $n/|\mathcal G_{\mathbf k_\star}|$ points, where $|\mathcal G_{\mathbf k_\star}|$ is the number of elements in the stabilizer. Applying Eq.~\eqref{eq:crystalline_winding_eigenvalues} to the generator $g_\star$ of each nontrivial stabilizer gives, for isolated zeros,
\begin{equation}
e^{i\phi_n(C_\alpha-C_\sigma)}
=
\prod_{[\mathbf k_\star]}
\det B_{g_\star}^\alpha(\mathbf k_\star)
\left[\det B_{g_\star}^\sigma(\mathbf k_\star)\right]^*,
\label{eq:Cn_localized}
\end{equation}
where the product runs over one representative of each orbit of momenta invariant under a nontrivial rotation. This is the relative form of the rotation-eigenvalue formulas of Ref.~\cite{Fang_Gilbert_Bernevig_2012}, with each factor attached to the zero that carries it (see also Ref.~\cite{Hwang_Davenport_Schindler_2026}).
Note that for a position-eigenstate reference at $\mathbf r_\star$, with $\mathbf r_\star$ a Wyckoff position, Eq.~\eqref{eq:crystalline_eigenvalue_zero} reproduces the symmetry-enforced real-space zeros of Bloch wavefunctions at high-symmetry momenta of Ref.~\cite{Ingham_2026}. The present construction relates the winding of these zeros through Eq.~\eqref{eq:Cn_localized} to the Chern number of $\alpha$.

An overlap between two valence bands belonging to the Haldane model at two different phases can be written as
\begin{equation}
\left|
S(\mathbf{k})
\right|^2
=
\frac{
1+
\hat{\mathbf d}^\alpha(\mathbf{k})
\cdot
\hat{\mathbf d}^\sigma(\mathbf{k})
}{2}.
\label{eq:haldane_overlap_bloch_vectors}
\end{equation}
It therefore vanishes precisely when the two Bloch vectors are
antiparallel (see
Appendix~\ref{app:crystalline_details} for more details). We use this form to reproduce the result of Fig.~\ref{fig:overlap_haldane} from symmetry considerations. From the explicit components of $\mathbf d(\mathbf{k})$ given in Appendix~\ref{app:haldane_model}, $d_1$ and $d_2$ depend only on the nearest-neighbor hopping $t_1$. Since the two Hamiltonians considered in Fig.~\ref{fig:overlap_haldane} have the same $t_1$, their $d_1$ and $d_2$ components are identical. They differ through the $t_2$ dependence of $d_3$. Antiparallel Bloch vectors can therefore occur only where
\begin{equation}
d_1(\mathbf{k})=d_2(\mathbf{k})=0,
\end{equation}
which, for the honeycomb lattice, restricts the possible zeros of the comparison section to the two valleys $K$ and $K'$.

To determine in which of the valleys a zero occurs, consider the Dirac masses at the valleys which are given by
\begin{equation}
d_3^\mu(K)
=
M-3\sqrt3\,t_2^\mu,
\qquad
d_3^\mu(K')
=
M+3\sqrt3\,t_2^\mu.
\label{eq:haldane_dirac_masses_comparison}
\end{equation}
For the parameters used in
Fig.~\ref{fig:overlap_haldane},
$d_3^\alpha(K)$ and $d_3^\sigma(K)$ have opposite signs, whereas
$d_3^\alpha(K')$ and $d_3^\sigma(K')$ have the same sign. Consequently,
\begin{equation}
S(K)=0,
\qquad
\left|S(K')\right|=1.
\end{equation}
At the zone center $\Gamma$ the two Bloch vectors also coincide, so
$|S(\Gamma)|=1$.

The zero in $K$ can also be understood from the symmetry considerations of Sec.~\ref{subsec:symmetry_enforced_zeros}.
At the valleys, $d_1=d_2=0$, so the Hamiltonian reduces to
\begin{equation}
H^\mu(K)
=
d_3^\mu(K)\sigma_3.
\end{equation}
In the sublattice basis, with
$\sigma_3\ket{A}=+\ket{A}$ and
$\sigma_3\ket{B}=-\ket{B}$, the valence state
belongs to sublattice $A$ when $d_3^\mu(K)<0$
and to sublattice $B$ when $d_3^\mu(K)>0$.
Since $d_3^\alpha(K)<0$ and $d_3^\sigma(K)>0$,
the two occupied states at $K$ belong to
different sublattices, and the corresponding spinors are therefore orthogonal.
The sublattice Bloch states at $K$ are eigenstates
of the threefold rotation operator with different
eigenvalues, $\xi_A(K)\neq\xi_B(K)$.
The occupied states of the two Hamiltonians therefore
carry different $C_3$ eigenvalues, and
Eq.~\eqref{eq:crystalline_eigenvalue_zero}
forces their overlap to vanish at $K$.
At $K'$, by contrast, the valence states belong
to the same sublattice and carry the same rotation
eigenvalue, so symmetry does not enforce a zero.

Rotation eigenvalues constrain the winding
of the zero at $K$. Using Eq.~\eqref{eq:crystalline_winding_eigenvalues} with $\phi_{C_3}=-2\pi/3$ for the clockwise rotation, we obtain $\nu_K=-1$ modulo three.
Since $K$ contains the only zero of the comparison
section, its winding accounts for the full relative
Chern number,
\begin{equation}
\nu_K
=
C_\alpha-C_\sigma
=
-1.
\end{equation}
Note that symmetry alone fixes $\nu_K$ only modulo three. The fact  that $K$ is the only zero follows from the two systems sharing the same $d_1$ and $d_2$, and the value $\nu_K=-1$ then follows from Eq.~\eqref{eq:classA_main_result}.

\begin{figure}[t]
    \centering
    \includegraphics[width=\linewidth]{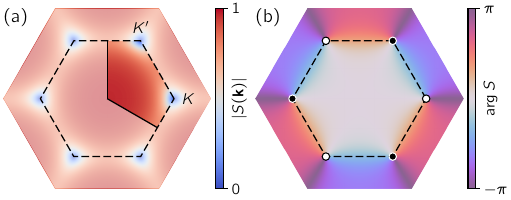}
    \caption{Zeros of the overlap enforced by symmetry alone. The two bands are valence bands of the Haldane model with $t_2=0$, $t_1=1$, and opposite staggered potentials $M^\alpha=-M^\sigma=0.8$. Both have $C=0$, so the total winding of the zeros must vanish, but the two bands are different atomic limits, with the electron on opposite sublattices, and threefold rotation distinguishes them.
(a) $|S(\mathbf k)|$ vanishes at $K$ and $K'$, where the rotation eigenvalues of the two occupied states differ. These zeros cannot be removed while the symmetry is preserved.
(b) Phase of $S(\mathbf k)$. It winds by $+2\pi$ around $K$ (white dots) and by $-2\pi$ around $K'$ (black dots), so the total winding is zero while each zero is individually protected. The dashed hexagon bounds the first Brillouin zone, and the unshaded region in (a) is one fundamental domain of the rotation.}
    \label{fig:haldane_opposite_mass}
\end{figure}

We now set $t_2=0$ in both systems and compare staggered potentials $M$ and $-M$, with $M\ne0$. Both systems have vanishing Chern numbers, since they can be continuously connected to the atomic limit which corresponds to $t_1=0$ without closing the gap. For finite nearest-neighbor hopping, i.e. $t_1\neq0$, Eq.~\eqref{eq:haldane_overlap_bloch_vectors} gives
\begin{equation}
|S(\mathbf k)|^2=\frac{d_1^2+d_2^2}{d_1^2+d_2^2+M^2},
\end{equation}
which vanishes only at $K$ and $K'$, where the nearest-neighbor hopping vanishes linearly. At each valley the occupied states lie on opposite sublattices and carry distinct rotation eigenvalues, so both zeros are symmetry enforced. Their windings are opposite, as required by $C_\alpha-C_\sigma=0$: for $M^\alpha=-M^\sigma>0$ we find $\nu_K=+1$ and $\nu_{K'}=-1$. Figure~\ref{fig:haldane_opposite_mass} shows the magnitude and the phase of this overlap.

Since the Kane-Mele model is based in essence on the Haldane model, threefold rotation similarly constrains the zeros of its relative section.
For the reference and relative determinant overlaps,
$a_O$ and $a_S$, the $C_3$ representations of the two
spaces entering each overlap are inequivalent at $K$
and $K'$. Equation~\eqref{eq:crystalline_representation_zero}
therefore forces zeros at both valleys.
For the section based on the Pfaffian construction of Kane and Mele, however, the mechanism is different.
The occupied $C_3$ eigenspaces at each valley fix an
occupied subspace orthogonal to its time-reversed image,
so the Pfaffian vanishes there. The determinant
of the occupied sewing matrix alone does not provide
a general criterion for zeros of the Pfaffian.
These results are derived in
Appendix~\ref{app:crystalline_details} and summarized in Table~\ref{tab:kane_mele_C3_summary}.

\begin{figure}[t]
    \centering
    \includegraphics[width=\linewidth]{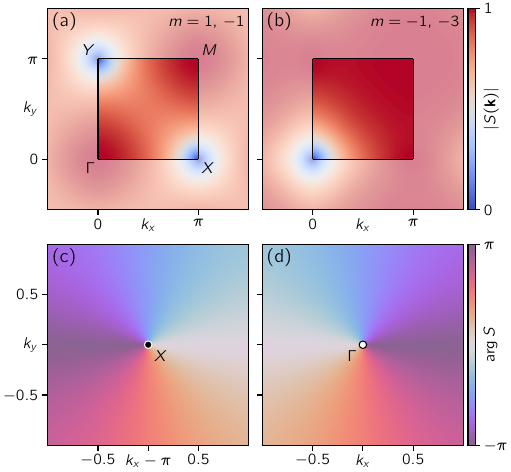}
    \caption{Zeros of the overlap enforced by fourfold rotation in the square-lattice model, for (a,c) $m^\alpha=1$, $m^\sigma=-1$ and (b,d) $m^\alpha=-1$, $m^\sigma=-3$. (a,b) $|S(\mathbf k)|$ vanishes only at rotation-invariant momenta, at $X$ and $Y$ in (a) and at $\Gamma$ in (b). (c,d) Phase of $S(\mathbf k)$ near $X$ and near $\Gamma$, in a frame that is smooth near the zero. It winds by $-2\pi$ around $X$ (black dot) and by $+2\pi$ around $\Gamma$ (white dot). The zeros at $X$ and $Y$ are related by the rotation and have the same winding, so that the total windings are $-2$ and $+1$, equal to $C_\alpha-C_\sigma$. The unshaded region in (a,b) is one fundamental domain of the fourfold rotation.}
    \label{fig:c4_overlap}
\end{figure}

As an illustration with a fourfold rotation symmetry, consider the QWZ model~\cite{Qi_Wu_Zhang_2006} $H(\mathbf k)=\sin k_x\,\sigma_x+\sin k_y\,\sigma_y+(m+\cos k_x+\cos k_y)\,\sigma_z$ on the square lattice, with the lower band occupied. Since two systems with different $m$ share the first two components of the Bloch vector, zeros of their overlap can occur only at the rotation-invariant momenta $\Gamma$, $M$, $X$, and $Y$. For $m^\alpha=1$ and $m^\sigma=-1$ we find $C_\alpha-C_\sigma=-2$ and two zeros, at $X$ and at $Y$, each of winding $-1$, as shown in Fig.~\ref{fig:c4_overlap}. These two momenta are related by fourfold rotation and are individually invariant under twofold rotation. Since the occupied states of the two systems have different $C_2$ eigenvalues at these points, Eq.~\eqref{eq:crystalline_winding_eigenvalues} then requires an odd winding. For $m^\alpha=-1$ and $m^\sigma=-3$, $C_\alpha-C_\sigma=1$ and there is a single zero, at $\Gamma$, of winding $+1$. In both cases Eq.~\eqref{eq:Cn_localized} is satisfied.

\section{Conclusion}

We have developed a geometric framework for characterizing
the relative topology of two sets of Bloch bands directly. The determinant of the matrix of overlaps
provides the local coefficient of a global comparison
section, whose zeros occur precisely where the
projection map between the occupied subspaces fails
to be invertible. These zeros acquire a topological
charge through the singular connection associated
with the comparison section.

For class A systems, the total winding of the zeros
of the comparison section is the difference between
the Chern numbers of the two systems. For class AII
systems, time reversal pairs zeros with opposite
winding, so that the integer charge cancels over the
full Brillouin zone, while its parity in a
time-reversal fundamental domain remains.
With the canonical normalization imposed at the
time-reversal-invariant momenta, this parity gives
the relative $\mathbb Z_2$ invariant.
In this way, the Chern and $\mathbb Z_2$ cases can
be understood through the same geometric construction,
despite their different conventional formulations.

The zeros of the comparison section also have a
direct interpretation in terms of a spectral flow. They correspond to bulk gap closings in the spectrum of a Hamiltonian interpolating between the two subspaces. Conversely, when the comparison section has no zeros, this interpolation remains gapped for all $\lambda$ (Appendix~\ref{app:z2_spectral_flow_details}) and is itself an adiabatic path between the two systems, preserving every symmetry they share. A nontrivial total winding or parity
therefore signals an obstruction to continuously
deforming one occupied subspace into the other while
preserving the gap and the relevant symmetries.
This relative distinction also has a boundary
manifestation through bulk-boundary correspondence
and boundary spectral flow.

Crystalline symmetry adds further structure to
this picture. While the strong invariant fixes the
total winding, or its parity, it does not in general
fix where the zeros occur. Symmetry relates zeros
at different momenta and can enforce zeros at
symmetry-invariant points when the two occupied
subspaces carry different symmetry representations.
We illustrated this explicitly in the Haldane and
Kane-Mele models, and in a square-lattice model with fourfold rotation.

A central message of this work is that the singular connection places the Chern and $\mathbb Z_2$ invariants on the same footing. Conventionally these two invariants are computed from different objects: the Chern number is calculated from the integral of a smooth Berry curvature distributed over the Brillouin zone, while the $\mathbb Z_2$ invariant is computed from the zeros of a Pfaffian of a matrix or other related constructions that have no curvature analog in class A. Our framework provides a unified geometric interpretation of these results. By constructing a singular connection from a global section of a determinant line bundle, we express the topological invariant in terms of the winding charges at the zeros of the section. Within this framework, the Kane–Mele Pfaffian is one possible choice of section, alongside the determinant of the overlap with a trivial Kramers reference. Moreover, symmetry constrains the location and winding charges of these zeros, providing a geometric connection between topological invariants and lattice symmetries. 
Finally, because a section can be built from the overlap of two occupied subspaces, the same procedure yields relative invariants directly, in either class, without needing to compute the invariant of each system separately.

More broadly, the comparison section and its associated singular connection recast topological equivalence as a question about a single gauge-invariant object built only from overlaps of occupied states. Because it requires neither a smooth gauge nor the Berry curvature of either system, it is well suited to numerical band structures: locating the zeros of the determinant of the overlap matrix and counting their windings yields the relative invariant, together with the momenta at which the projector interpolation closes its gap. 
Since the construction depends only on the occupied subspaces rather than the Hamiltonians that produce them, it suggests a possible extension to interacting or disordered ground states, although whether an analogous construction can be developed in these settings remains an open question. In this view, topology is a property of the map between two band structures, and the zeros of that map are where their distinction becomes visible.

Our results apply to two-dimensional, noninteracting, translation-invariant systems in classes A and AII, with isolated zeros of the overlap determinant. When additional symmetries constrain the determinant to be real or enforce a mismatch of symmetry representations along a line, its zeros may instead form lines, for which the winding formulation developed here does not directly apply. In three dimensions, the zeros of a complex overlap determinant generically form loops rather than isolated points. Extending the construction to these settings, as well as investigating whether overlap zeros between distinct atomic limits can characterize crystalline and fragile~\cite{Po_Watanabe_Vishwanath_2018_fragile} topological distinctions, remains an interesting direction for future work.

\emph{Note added.} Upon completion of this manuscript, we became aware of Ref.~\cite{Hwang_Davenport_Schindler_2026}, which also relates the zeros of the overlap between a Bloch band and a reference state to the Chern number and to the rotation eigenvalues, in agreement with our results for class A and with Sec.~\ref{sec:crystalline_symmetries}, and connects these zeros to shifts of the band dispersion. The relative $\mathbb Z_2$ invariant of class AII is not discussed there.

\section{Acknowledgements} R.Q. is supported by the National Science Foundation under Award No. DMR-2340394 and by the Alfred P. Sloan Foundation (FG-2025-24714).  R.I. and C.P. are supported by the Israel Science Foundation (ISF) grant No. 2307/24.
R.I. is grateful for the hospitality of Perimeter Institute, where part of this work was carried out. Research
at Perimeter Institute is supported in part by the Government of Canada through the Department of
Innovation, Science and Economic Development and by the Province of Ontario through the Ministry of
Colleges and Universities. This work was supported by a grant from the Simons Foundation (1034867,
Dittrich). 
The Flatiron Institute is a division of the Simons Foundation. C.P. wants to thank the Alexander Zaks scholarship for its support.
B.M. acknowledges support from the Security and Quantum Information Group (SQIG) in Instituto de Telecomunica\c{c}\~{o}es, Lisbon. This work is funded by FCT/MECI through national funds and when applicable co-funded by EU funds under UID/50008: Instituto de Telecomunicações (IT). B.M. further acknowledges the Scientific Employment Stimulus --- Individual Call (CEEC Individual) --- 2022.05522.CEECIND/CP1716/CT0001, with DOI 10.54499/2022.05522.CEECIND/CP1716/CT0001.

\bibliography{SingularConnections}

\clearpage

\widetext

\appendix

\section{Gauge invariance of global section}
\label{app:gauge_invariance_comparison_section}

Under independent changes of occupied basis
$W^\alpha(\mathbf{k}),W^\sigma(\mathbf{k})\in U(N_{\mathrm{occ}})$,
the overlap matrix transforms as
\begin{equation}
S
\longrightarrow
(W^\alpha)^\dagger S W^\sigma,
\end{equation}
and therefore
\begin{equation}
\det S
\longrightarrow
[\det W^\alpha]^*
\det W^\sigma\,
\det S.
\end{equation}
At the same time, the local basis of the relative line transforms as
\begin{equation}
e_r
\longrightarrow
\det W^\alpha
[\det W^\sigma]^*
e_r.
\end{equation}
The two transformations therefore cancel in
\begin{equation}
a_S(\mathbf{k})
=
\det S(\mathbf{k})\,e_r(\mathbf{k}),
\end{equation}
showing that $a_S(\mathbf{k})$ is independent of the choice of occupied
bases.

\section{Projector interpolation and spectral flow}
\label{app:spectral_flow_details}

In this appendix we provide the details underlying the spectral flow
interpretation of Sec.~\ref{subsec:spectral_flow}. We first derive the spectrum of the interpolation for a two-band system,
then show that a simple zero of the comparison section gives an auxiliary Weyl point
and relate its chirality to the transfer of Chern number. We subsequently
extend the construction to several occupied bands.

\subsection{Two-band interpolation}

Consider a two-band system with a single occupied band. At fixed
$\mathbf{k}$, we use the occupied state of $\alpha$ and its orthogonal
complement as a basis,
\begin{equation}
\left\{
\ket{\psi^\alpha(\mathbf{k})},
\ket{\chi^\alpha(\mathbf{k})}
\right\},
\end{equation}
with
$\braket{\chi^\alpha(\mathbf{k})|\psi^\alpha(\mathbf{k})}=0$.
Since the matrix of overlaps of Eq.~\eqref{eq:overlap_matrix} is a scalar
in this case, the occupied state of $\sigma$ can be expanded as
\begin{equation}
\ket{\psi^\sigma(\mathbf{k})}
=
S(\mathbf{k})
\ket{\psi^\alpha(\mathbf{k})}
+
T(\mathbf{k})
\ket{\chi^\alpha(\mathbf{k})},
\label{eq:app_state_expansion}
\end{equation}
where
\begin{equation}
T(\mathbf{k})
=
\braket{\chi^\alpha(\mathbf{k})|
\psi^\sigma(\mathbf{k})}.
\end{equation}
Normalization gives
\begin{equation}
|S(\mathbf{k})|^2+|T(\mathbf{k})|^2=1.
\label{eq:app_st_normalization}
\end{equation}

In this basis,
\begin{equation}
P^\alpha
=
\begin{pmatrix}
1&0\\
0&0
\end{pmatrix},
\qquad
P^\sigma
=
\begin{pmatrix}
|S|^2 & S T^*\\
T S^* & |T|^2
\end{pmatrix},
\end{equation}
where the momentum dependence is suppressed for compactness. Their
flattened representatives $Q^\mu=2P^\mu-\mathbb 1$, with $\mu\in\{\alpha,\sigma\}$, are therefore
\begin{equation}
Q^\alpha
=
\begin{pmatrix}
1&0\\
0&-1
\end{pmatrix},
\qquad
Q^\sigma
=
\begin{pmatrix}
2|S|^2-1 & 2S T^*\\
2T S^* & 2|T|^2-1
\end{pmatrix}.
\end{equation}

Substituting these expressions into the interpolation
Eq.~\eqref{eq:flattened_interpolation} gives
\begin{equation}
H(\lambda,\mathbf{k})
=
\begin{pmatrix}
1-2\lambda+2\lambda|S|^2
&
2\lambda S T^*
\\
2\lambda T S^*
&
-1+2\lambda-2\lambda|S|^2
\end{pmatrix}.
\label{eq:app_interpolating_matrix}
\end{equation}
The matrix is traceless and, using
Eq.~\eqref{eq:app_st_normalization}, has determinant
\begin{equation}
\det H
=
-
\left[
(1-2\lambda)^2
+
4\lambda(1-\lambda)|S|^2
\right].
\end{equation}
Its eigenvalues are therefore
\begin{equation}
E_\pm(\lambda,\mathbf{k})
=
\pm
\sqrt{
(1-2\lambda)^2
+
4\lambda(1-\lambda)|S|^2
}.
\end{equation}
For $\lambda\in[0,1]$, both terms under the square root are
non-negative. The gap therefore closes if and only if
\begin{equation}
\lambda=\frac12,
\qquad
S(\mathbf{k})=0.
\end{equation}
This establishes Eq.~\eqref{eq:gap_closing_overlap_zero}: in the single-band case, the zeros of the comparison section coincide exactly
with the gap closings of the projector interpolation.

\subsection{Local Weyl Hamiltonian and Chern-number transfer}

Let $\mathbf{k}_*$ be a simple zero of the scalar overlap
$S(\mathbf{k})$. At the zero,
Eq.~\eqref{eq:app_st_normalization} gives
$|T(\mathbf{k}_*)|=1$. We can therefore choose the local phase of
$\ket{\chi^\alpha}$ such that
\begin{equation}
T(\mathbf{k}_*)=1.
\end{equation}
Introducing
\begin{equation}
\mathbf{q}
=
\mathbf{k}-\mathbf{k}_*,
\qquad
\eta
=
\lambda-\frac12,
\end{equation}
and keeping only terms linear in $\mathbf q$ and $\eta$,
Eq.~\eqref{eq:app_interpolating_matrix} reduces to
\begin{equation}
H(\lambda,\mathbf{k})
\simeq
\begin{pmatrix}
-2\eta & S(\mathbf{k})\\
[S(\mathbf{k})]^* & 2\eta
\end{pmatrix}.
\label{eq:app_linearized_matrix}
\end{equation}
Writing
\begin{equation}
S(\mathbf{k})
=
S_1(\mathbf{k})
+
iS_2(\mathbf{k}),
\end{equation}
we obtain
\begin{equation}
H_{\mathrm{eff}}
\simeq
S_1(\mathbf{k})\,\sigma_1
-
S_2(\mathbf{k})\,\sigma_2
-
2\eta\,\sigma_3.
\label{eq:app_weyl_hamiltonian}
\end{equation}
For a simple zero, $S_1$ and $S_2$ vary independently to linear order
in $(q_x,q_y)$, and $\sigma_1,\sigma_2,\sigma_3$ denote the Pauli matrices. The crossing is therefore linear in all three local
coordinates $(q_x,q_y,\eta)$ and defines an auxiliary Weyl point in
the parameter space $(k_x,k_y,\lambda)$. This is the local form of the
bulk crossing shown schematically in
Fig.~\ref{fig:bulk_spectral_flow}.

The same local derivatives determine both the winding of the
zero of the comparison section and the chirality of the Weyl point. For a simple zero,
define
\begin{equation}
J_S(\mathbf{k}_*)
=
\left.
\left(
\frac{\partial S_1}{\partial k_x}
\frac{\partial S_2}{\partial k_y}
-
\frac{\partial S_1}{\partial k_y}
\frac{\partial S_2}{\partial k_x}
\right)
\right|_{\mathbf{k}=\mathbf{k}_*}.
\end{equation}
Its winding, defined generally in
Eq.~\eqref{eq:zero_winding_general}, is then
\begin{equation}
\nu_{\mathbf{k}_*}
=
\operatorname{sgn}J_S(\mathbf{k}_*).
\label{eq:app_overlap_winding}
\end{equation}

From Eq.~\eqref{eq:app_weyl_hamiltonian}, the three components of the
effective Hamiltonian are
\begin{equation}
(d_1,d_2,d_3)
=
(S_1,-S_2,-2\eta).
\end{equation}
Defining the Weyl chirality by the orientation of the map
$(k_x,k_y,\lambda)\mapsto(d_1,d_2,d_3)$ gives
\begin{align}
\chi_{\mathbf{k}_*}
&=
\operatorname{sgn}
\det
\left[
\frac{\partial(d_1,d_2,d_3)}
{\partial(k_x,k_y,\lambda)}
\right]_{\mathbf{k}=\mathbf{k}_*,\,\lambda=1/2}
\nonumber\\
&=
\operatorname{sgn}
\left[
2
\left(
\frac{\partial S_1}{\partial k_x}
\frac{\partial S_2}{\partial k_y}
-
\frac{\partial S_1}{\partial k_y}
\frac{\partial S_2}{\partial k_x}
\right)
\right]_{\mathbf{k}=\mathbf{k}_*}
\nonumber\\
&=
\nu_{\mathbf{k}_*}.
\label{eq:app_weyl_winding_relation}
\end{align}
The local charge of the zero of the comparison section is therefore precisely the
chirality of the corresponding auxiliary Weyl point.

It remains to relate this local charge to the change in Chern number
along the interpolation. With our flattening convention
$Q^\mu=2P^\mu-\mathbb 1$, the occupied state lies on the
positive-energy branch. Using the Berry-connection convention adopted
throughout the paper,
$\mathcal A=-i\braket{\psi|\mathrm d\psi}$, we define
\begin{equation}
C
=
-\frac{1}{2\pi}
\int_{\mathrm{BZ}^2}
\mathrm d\mathcal A.
\label{eq:app_chern_convention}
\end{equation}
With these conventions, each Weyl point of chirality
$\chi_{\mathbf{k}_*}$ contributes
$-\chi_{\mathbf{k}_*}$ to the change in Chern number
of the positive-energy branch as $\lambda$ increases
through the crossing. Since the interpolation runs
from $\alpha$ at $\lambda=0$ to $\sigma$ at $\lambda=1$,
summing over all crossings gives
\begin{equation}
C_\sigma-C_\alpha
=
-
\sum_{\mathbf{k}_*}
\chi_{\mathbf{k}_*}.
\end{equation}
Using Eq.~\eqref{eq:app_weyl_winding_relation}, we recover
\begin{equation}
C_\alpha-C_\sigma
=
\sum_{\mathbf{k}_*}
\nu_{\mathbf{k}_*},
\end{equation}
in agreement with Eq.~\eqref{eq:classA_main_result}.

The Weyl picture therefore gives a spectral interpretation of the
local charges introduced through the singular connection. Away from
the crossings, each fixed-$\lambda$ slice is a gapped two-dimensional
system whose Chern number cannot change. The transfer of Chern number
between the two endpoint phases occurs only at the auxiliary Weyl
points, with each crossing contributing the charge carried by the
corresponding zero of the comparison section.

\section{Haldane-model conventions}
\label{app:haldane_model}

For the calculation in
Sec.~\ref{subsec:chern_insulator_application}, we use the Hamiltonian
of Eq.~\eqref{eq:Haldane_Hamiltonian} with
\begin{equation}
\begin{aligned}
d_1(\mathbf{k})
&=
t_1\sum_{j=1}^{3}
\cos(\mathbf{k}\cdot\mathbf{a}_j),
\\
d_2(\mathbf{k})
&=
-t_1\sum_{j=1}^{3}
\sin(\mathbf{k}\cdot\mathbf{a}_j),
\\
d_3(\mathbf{k})
&=
M
+
2t_2\sum_{j=1}^{3}
\sin(\mathbf{k}\cdot\mathbf{b}_j).
\end{aligned}
\label{eq:haldane_d_components}
\end{equation}
With these conventions, $M$ denotes the staggered sublattice
potential and $t_2$ the time-reversal-breaking
next-nearest-neighbor hopping. We omit the term proportional to the
identity, since it changes the spectrum but not the eigenstates and
therefore does not affect the Berry curvature, Chern number, or
zeros of the comparison section. The parameters used in
Fig.~\ref{fig:overlap_haldane} are given in
Sec.~\ref{subsec:chern_insulator_application}.

Our nearest-neighbor vectors are
\begin{equation}
\begin{aligned}
\mathbf{a}_1&=(0,1),
&
\mathbf{a}_2&=
\left(
\frac{\sqrt{3}}{2},-\frac12
\right),
&
\mathbf{a}_3&=
\left(
-\frac{\sqrt{3}}{2},-\frac12
\right),
\end{aligned}
\label{eq:haldane_nn_vectors}
\end{equation}
while the next-nearest-neighbor vectors are
\begin{align}
\mathbf{b}_1
&=
\mathbf{a}_2-\mathbf{a}_3
=
\left(
\sqrt{3},0
\right),
\\
\mathbf{b}_2
&=
\mathbf{a}_3-\mathbf{a}_1
=
\left(
-\frac{\sqrt{3}}{2},-\frac{3}{2}
\right),
\\
\mathbf{b}_3
&=
\mathbf{a}_1-\mathbf{a}_2
=
\left(
-\frac{\sqrt{3}}{2},\frac{3}{2}
\right).
\end{align}

We choose the primitive Bravais-lattice vectors
\begin{equation}
\mathbf A_1
=
\mathbf b_1
=
(\sqrt3,0),
\qquad
\mathbf A_2
=
\mathbf b_3
=
\left(
-\frac{\sqrt3}{2},
\frac32
\right).
\end{equation}
The corresponding reciprocal-lattice vectors,
$\mathbf A_i\cdot\mathbf G_j=2\pi\delta_{ij}$, are
\begin{equation}
\mathbf G_1
=
\left(
\frac{2\pi}{\sqrt3},
\frac{2\pi}{3}
\right),
\qquad
\mathbf G_2
=
\left(
0,
\frac{4\pi}{3}
\right).
\label{eq:haldane_reciprocal_vectors}
\end{equation}
With these conventions, we take
\begin{equation}
\Gamma=(0,0),
\qquad
K=
\left(
\frac{4\pi}{3\sqrt3},
0
\right),
\qquad
K'=-K.
\label{eq:haldane_high_symmetry_points}
\end{equation}

\section{Time-reversal-compatible sections and the relative
$\mathbb{Z}_2$ index}
\label{app:AII_details}

In this appendix we give the technical details underlying the
class AII construction of
Sec.~\ref{sec:time_reversal_singular_connection}. We first derive the
time-reversal structure of the determinant line and the corresponding
constraint on compatible local bases. We then apply this structure to
the reference-overlap, Pfaffian, and relative-overlap sections, and
clarify how their zero winding reduces to a parity on a time-reversal
fundamental domain.

\subsection{Quaternionic structure and the determinant line bundle}

Let $\mathcal E\rightarrow\mathrm{BZ}^2$ be a rank-two occupied bundle
with an antiunitary time-reversal operator satisfying
\begin{equation}
\Theta^2=-1,
\qquad
\Theta:
\mathcal E_{\mathbf{k}}
\longrightarrow
\mathcal E_{-\mathbf{k}}.
\end{equation}
The pair $(\mathcal E,\Theta)$ therefore defines a Quaternionic vector
bundle over the Brillouin zone with involution
$\mathbf{k}\mapsto-\mathbf{k}$.
For a local occupied basis
\begin{equation}
U(\mathbf{k})
=
\bigl(
\ket{\psi_1(\mathbf{k})},
\ket{\psi_2(\mathbf{k})}
\bigr),
\end{equation}
the determinant line
$\det\mathcal E=\wedge^2\mathcal E$ has local basis vector
\begin{equation}
e_{\mathcal E}(\mathbf{k})
=
\ket{\psi_1(\mathbf{k})}
\wedge
\ket{\psi_2(\mathbf{k})}.
\end{equation}

Time reversal induces an anti-linear map on the determinant line,
\begin{equation}
\Theta_{\det}
\left(
\ket{\psi_1}\wedge\ket{\psi_2}
\right)
=
\Theta\ket{\psi_1}
\wedge
\Theta\ket{\psi_2}.
\end{equation}
Applying this map twice gives
\begin{align}
\Theta_{\det}^2
\left(
\ket{\psi_1}\wedge\ket{\psi_2}
\right)
&=
\Theta^2\ket{\psi_1}
\wedge
\Theta^2\ket{\psi_2}
\nonumber\\
&=
(-\ket{\psi_1})
\wedge
(-\ket{\psi_2})
\nonumber\\
&=
\ket{\psi_1}\wedge\ket{\psi_2},
\end{align}
and therefore
\begin{equation}
\Theta_{\det}^2=+1.
\end{equation}
Thus, although time reversal squares to $-1$ on the occupied states,
its induced action on the determinant line squares to the identity.

At a TRIM $\Gamma_i$, where
$-\Gamma_i\equiv\Gamma_i$ modulo a reciprocal lattice vector, the
Quaternionic structure provides additional information. Since
$\Theta$ acts within the same fiber,
\begin{equation}
\Theta:
\mathcal E_{\Gamma_i}
\longrightarrow
\mathcal E_{\Gamma_i},
\qquad
\Theta^2=-1,
\end{equation}
any normalized state
$\ket{u}\in\mathcal E_{\Gamma_i}$ and its Kramers partner
$\Theta\ket{u}$ define the nonzero determinant vector
\begin{equation}
e_{K}(\Gamma_i)
=
\ket{u}\wedge\Theta\ket{u}.
\label{eq:app_AII_canonical_vector}
\end{equation}
This vector is independent of the normalized state $\ket{u}$ used in
its construction. Indeed, for another normalized state
\begin{equation}
\ket{u'}
=
a\ket{u}
+
b\,\Theta\ket{u},
\qquad
|a|^2+|b|^2=1,
\end{equation}
anti-linearity and $\Theta^2=-1$ give
\begin{equation}
\Theta\ket{u'}
=
-b^{*}\,\ket{u}
+
a^{*}\,\Theta\ket{u},
\end{equation}
and hence
\begin{equation}
\ket{u'}\wedge\Theta\ket{u'}
=
\left(
|a|^2+|b|^2
\right)
\ket{u}\wedge\Theta\ket{u}
=
\ket{u}\wedge\Theta\ket{u}.
\end{equation}
Moreover,
\begin{equation}
\Theta_{\det}e_{K}(\Gamma_i)
=
e_{K}(\Gamma_i).
\end{equation}
The Kramers pair therefore defines a nonzero,
time-reversal-invariant vector in the determinant
line at each TRIM, independent of the choice of
normalized state used to construct it.

We now determine how this structure constrains changes of local basis.
Consider local determinant-line bases on two time-reversal-related
regions chosen such that
\begin{equation}
e_{\mathcal E}(-\mathbf{k})
=
\Theta_{\det}e_{\mathcal E}(\mathbf{k}).
\label{eq:app_AII_compatible_frame}
\end{equation}
Let
\begin{equation}
e'_{\mathcal E}(\mathbf{k})
=
h(\mathbf{k})e_{\mathcal E}(\mathbf{k}),
\qquad
h(\mathbf{k})\in U(1),
\label{eq:app_eprime}
\end{equation}
be another choice. Requiring it to satisfy the same
time-reversal compatibility condition gives
\begin{equation}
e'_{\mathcal E}(-\mathbf{k})
=
\Theta_{\det}e'_{\mathcal E}(\mathbf{k}).
\end{equation}
Using the anti-linearity of $\Theta_{\det}$ on Eq.~\eqref{eq:app_eprime}
\begin{equation}
\Theta_{\det}e'_{\mathcal E}(\mathbf{k})
=
(h(\mathbf{k}))^{*}
\Theta_{\det}e_{\mathcal E}(\mathbf{k}).
\end{equation}
At the same time evaluating Eq.~\eqref{eq:app_eprime} at $-\mathbf{k}$ gives:
\begin{equation}
e'_{\mathcal E}(-\mathbf{k})
=
h(-\mathbf{k})
e_{\mathcal E}(-\mathbf{k}),
\end{equation}
These two equations together with
Eq.~\eqref{eq:app_AII_compatible_frame} imply
\begin{equation}
h(-\mathbf{k})
=
(h(\mathbf{k}))^{*}.
\label{eq:app_AII_real_gauge}
\end{equation}

At a time-reversal-invariant momentum, Eq.~\eqref{eq:app_AII_real_gauge} reduces to
\begin{equation}
h(\mathbf{k})
=
(h(\mathbf{k}))^{*}.
\end{equation}
Thus, $h(\Gamma_i)$ must be real. Since
$h(\Gamma_i)\in U(1)$, it can only take the values
\begin{equation}
h(\Gamma_i)=\pm1.
\end{equation}
Therefore, time-reversal symmetry alone fixes the
determinant-line basis at a TRIM only up to a sign.
Requiring this basis to coincide with the vector
$e_K(\Gamma_i)$ defined in
Eq.~\eqref{eq:app_AII_canonical_vector}
removes this remaining ambiguity.

Equation~\eqref{eq:app_AII_real_gauge} relates the gauge choices at any
pair of time-reversed momenta. When the Brillouin zone is represented
by a time-reversal fundamental domain $\mathrm{BZ}^2_+$ and its image,
the same relation determines how the local bases on the two halves are
matched along their boundary. Generic boundary points are paired as
$\mathbf{k}\leftrightarrow-\mathbf{k}$, while only the
time-reversal-invariant momenta are fixed individually. Time-reversal symmetry therefore constrains how the determinant-line bases at $\mathbf{k}$ and $-\mathbf{k}$ are related. At each TRIM, the Kramers pairs further provide a unique nonzero determinant vector, independent of the choice of orthonormal Kramers basis.

\subsection{Reference-overlap section}

We now verify the gauge, time-reversal, and fixed-point properties of
the reference-overlap section introduced in
Sec.~\ref{subsec:AII_absolute}. The fixed reference Kramers pair of
Eq.~\eqref{eq:AII_reference_pair} satisfies
\begin{equation}
\ket{R_2}=\Theta\ket{R_1},
\qquad
\Theta\ket{R_2}=-\ket{R_1},
\end{equation}
and its determinant-line basis
\begin{equation}
e_R
=
\ket{R_1}\wedge\ket{R_2}
\end{equation}
is invariant under the induced time-reversal action:
\begin{align}
\Theta_{\det}e_R
&=
\Theta\ket{R_1}
\wedge
\Theta\ket{R_2}
\nonumber\\
&=
\ket{R_2}\wedge(-\ket{R_1})
\nonumber\\
&=
e_R.
\end{align}
Thus $e_R$ gives a global time-reversal-compatible basis of
$\det R$. Because this is a fixed reference bundle, at a time-reversal-invariant momentum it trivially coincides with
the determinant vector of the reference bundle.

For a local occupied basis
\begin{equation}
U(\mathbf{k})
=
\bigl(
\ket{\psi_1(\mathbf{k})},
\ket{\psi_2(\mathbf{k})}
\bigr),
\end{equation}
the reference-matrix of overlaps is
\begin{equation}
O_{ij}(\mathbf{k})
=
\braket{\psi_i(\mathbf{k})|R_j}.
\end{equation}
Under a gauge transformation of the occupied basis,
\begin{equation}
U(\mathbf{k})
\longrightarrow
U(\mathbf{k})W(\mathbf{k}),
\qquad
W(\mathbf{k})\in U(2),
\end{equation}
the matrix of overlaps transforms as
\begin{equation}
O(\mathbf{k})
\longrightarrow
W^\dagger(\mathbf{k})O(\mathbf{k}),
\end{equation}
and hence
\begin{equation}
\det O(\mathbf{k})
\longrightarrow
\det W(\mathbf{k})^{-1}
\det O(\mathbf{k}).
\label{eq:app_AII_O_gauge}
\end{equation}
At the same time,
\begin{equation}
e_{\mathcal E}(\mathbf{k})
\longrightarrow
\det W(\mathbf{k})
e_{\mathcal E}(\mathbf{k}).
\label{app:eq:basis_change_basis}
\end{equation}
The two factors therefore cancel in
\begin{equation}
a_O(\mathbf{k})
=
\det O(\mathbf{k})\,
e_{\mathcal E}(\mathbf{k})\otimes e_R^*,
\end{equation}
showing explicitly that $a_O$ is a globally defined section of
$\mathcal L_{\mathcal E,R}$.

We next derive its time-reversal constraint. It is convenient to use
the projected reference states
\begin{equation}
\ket{v_j(\mathbf{k})}
=
P_{\mathcal E}(\mathbf{k})\ket{R_j},
\end{equation}
where
\begin{equation}
P_{\mathcal E}(\mathbf{k})
=
\sum_{i=1}^{2}
\ket{\psi_i(\mathbf{k})}
\bra{\psi_i(\mathbf{k})}
\end{equation}
is the projector onto the occupied subspace
$\mathcal E_{\mathbf{k}}$.
Expanding them in the occupied basis gives:
\begin{equation}
\ket{v_j(\mathbf{k})} = \sum_{i=1}^{2} O_{ij}(\mathbf{k})\, \ket{\psi_i(\mathbf{k})},
\end{equation}
the antisymmetry of the wedge product yields
\begin{align}
\ket{v_1(\mathbf{k})}\wedge \ket{v_2(\mathbf{k})} &= (O_{11}(\mathbf{k})\ket{\psi_1(\mathbf{k})}+O_{21}(\mathbf{k})\ket{\psi_2(\mathbf{k})}) \wedge (O_{12}(\mathbf{k})\ket{\psi_1(\mathbf{k})}+O_{22}(\mathbf{k})\ket{\psi_2(\mathbf{k})})\nonumber\\
&= (O_{11}(\mathbf{k})O_{22}(\mathbf{k})-O_{12}(\mathbf{k})O_{21}(\mathbf{k}))\, e_{\mathcal{E}}(\mathbf{k})\nonumber\\
&= \det{(O(\mathbf{k}))}\, e_{\mathcal{E}}(\mathbf{k}),
\label{eq:app_AII_projected_wedge}
\end{align}
which is Eq.~\eqref{eq:AII_projected_area} of the main text.

Because the occupied projector respects time-reversal
symmetry, it satisfies
\begin{equation}
P_{\mathcal E}(-\mathbf{k})
=
\Theta P_{\mathcal E}(\mathbf{k})\Theta^{-1}.
\end{equation}
Using $\Theta^{-1}=-\Theta$ and the Kramers relations
of the reference states, we obtain
\begin{align*}
\ket{v_1(-\mathbf{k})}
&=
P_{\mathcal E}(-\mathbf{k})\ket{R_1}=\Theta P_{\mathcal E}(\mathbf{k})
\Theta^{-1}\ket{R_1}=
-\Theta P_{\mathcal E}(\mathbf{k})\ket{R_2}=
-\Theta\ket{v_2(\mathbf{k})}.
\end{align*}
Similarly,
\begin{align*}
\ket{v_2(-\mathbf{k})}
&=
P_{\mathcal E}(-\mathbf{k})\ket{R_2}=\Theta P_{\mathcal E}(\mathbf{k})
\Theta^{-1}\ket{R_2}=
\Theta P_{\mathcal E}(\mathbf{k})\ket{R_1}=
\Theta\ket{v_1(\mathbf{k})}.
\end{align*}
Therefore, the projected reference states satisfy
\begin{align}
\ket{v_1(-\mathbf{k})}
&=
-\Theta\ket{v_2(\mathbf{k})},
\\
\ket{v_2(-\mathbf{k})}
&=
\Theta\ket{v_1(\mathbf{k})}.
\end{align}
Their wedge product therefore transforms as
\begin{align}
\ket{v_1(-\mathbf{k})}
\wedge
\ket{v_2(-\mathbf{k})}
&=
-\Theta\ket{v_2(\mathbf{k})}
\wedge
\Theta\ket{v_1(\mathbf{k})}=
\Theta\ket{v_1(\mathbf{k})}
\wedge
\Theta\ket{v_2(\mathbf{k})}=
\Theta_{\det}
\left[
\ket{v_1(\mathbf{k})}
\wedge
\ket{v_2(\mathbf{k})}
\right].
\end{align}
Using Eq.~\eqref{eq:app_AII_projected_wedge} together with a determinant-line basis satisfying
Eq.~\eqref{eq:app_AII_compatible_frame}, the anti-linearity of
$\Theta_{\det}$ gives
\begin{equation}
\det O(-\mathbf{k})
=
(\det O(\mathbf{k}))^{*}.
\label{eq:app_AII_O_TR}
\end{equation}
A zero at $\mathbf{k}_*$ is therefore accompanied
by one at $-\mathbf{k}_*$. Since complex conjugation
reverses the phase winding, while
$\mathbf{k}\mapsto-\mathbf{k}$ preserves orientation
in two dimensions, the two zeros have opposite
winding numbers,
\begin{equation}
\nu^{\mathcal E}_{-\mathbf{k}_*}
=
-\nu^{\mathcal E}_{\mathbf{k}_*}.
\label{eq:app_AII_winding_pair}
\end{equation}

We now show that the time-reversal properties of the occupied and reference states ensure that the overlap determinant is nonnegative at every TRIM. At a TRIM $\Gamma_i$, choose the occupied basis to be a Kramers pair,
\begin{equation}
\ket{\psi_2}
=
\Theta\ket{\psi_1}.
\end{equation}
Writing
\begin{equation}
a
=
\braket{\psi_1|R_1},
\qquad
b
=
\braket{\psi_1|R_2},
\end{equation}
the overlap matrix takes the form
\begin{equation}
O(\Gamma_i)
=
\begin{pmatrix}
a & b\\
-b^{*} & a^{*}
\end{pmatrix}.
\label{eq:app_AII_O_TRIM_matrix}
\end{equation}
Its determinant is therefore
\begin{equation}
\det O(\Gamma_i)
=
|a|^2+|b|^2
\geq0.
\label{eq:app_AII_O_TRIM_positive}
\end{equation}
Since the Kramers basis satisfies
\begin{equation}
e_{\mathcal E}(\Gamma_i)
=
e_{K}(\Gamma_i),
\end{equation}
the comparison section at the fixed point is
\begin{equation}
a_O(\Gamma_i)
=
\det O(\Gamma_i)\,
e_{K}(\Gamma_i)\otimes e_R^*.
\label{eq:app_AII_O_fixed_point}
\end{equation}
Thus, whenever it is nonzero, $a_O(\Gamma_i)$
is a positive multiple of
$e_K(\Gamma_i)\otimes e_R^*$.

\subsection{Generic nonvanishing at the TRIMs}
\label{app:AII_generic_reference}

We now show that the reference $R$ can be chosen so that
$\det O(\Gamma_i)>0$ at all four TRIMs. Let the ambient Hilbert space be $\mathcal H=\mathbb C^{n}\cong\mathbb H^{n/2}$, with $n$ even, equipped with the standard Quaternionic structure. A fixed reference containing $m$ Kramers pairs corresponds to an $m$-dimensional Quaternionic subspace $R\subset\mathcal H$. The possible choices of $R$ form the Quaternionic Grassmannian $\mathrm{Gr}_m(\mathbb H^{n/2})$. For the rank-two construction considered here, $m=1$.

At a TRIM $\Gamma_i$, the overlap matrix $O(\Gamma_i)$ represents the orthogonal projection
\begin{equation}
P_{\mathcal E}(\Gamma_i)|_R:
R\longrightarrow\mathcal E_{\Gamma_i}
\end{equation}
in orthonormal Kramers bases. Its determinant is nonnegative, as established in Eq.~\eqref{eq:app_AII_O_TRIM_positive} for $m=1$. This property also holds for general $m$. Indeed, in Kramers bases, $O(\Gamma_i)$ commutes with the standard Quaternionic structure. If $O$ is invertible, its polar decomposition
\begin{equation}
O=U(O^\dagger O)^{1/2}
\end{equation}
has both factors commuting with this structure. The eigenvalues of $U$ occur in complex-conjugate pairs, while its real eigenvalues, necessarily $\pm1$, have even multiplicity. Consequently, $\det U=1$ and
\begin{equation}
\det O=\det(O^\dagger O)^{1/2}>0.
\end{equation}

We now show that invertibility is a generic property. For a fixed $\Gamma_i$, define
\begin{equation}
\mathcal U_i=
\left\{
R\in\mathrm{Gr}_m(\mathbb H^n):
\det O(\Gamma_i)>0
\right\}.
\end{equation}
Continuity of the overlap determinant implies that $\mathcal U_i$ is open.

To establish density, consider any reference $R$ with an orthonormal Kramers basis $\{\ket{R_j},\Theta\ket{R_j}\}_{j=1}^m$, and an orthonormal Kramers basis $\{\ket{\psi_j},\Theta\ket{\psi_j}\}_{j=1}^m$ of $\mathcal E_{\Gamma_i}$. For a small real parameter $\varepsilon$, replace
\begin{equation}
\ket{R_j}\longmapsto
\ket{R_j}+\varepsilon\ket{\psi_j},
\end{equation}
together with the corresponding Kramers partners. For sufficiently small $\varepsilon$, these vectors remain linearly independent and span a Quaternionic subspace $R_\varepsilon$ arbitrarily close to $R$.

In these generally nonorthonormal spanning bases, the projection onto $\mathcal E_{\Gamma_i}$ has overlap matrix
\begin{equation}
O(\Gamma_i)+\varepsilon\mathbb 1_{2m}.
\end{equation}
Its determinant is a polynomial of degree $2m$ in $\varepsilon$, with leading coefficient one. It therefore has only finitely many roots. Hence, for arbitrarily small $\varepsilon$, the projection from $R_\varepsilon$ onto $\mathcal E_{\Gamma_i}$ is invertible. Passing to an orthonormal Kramers basis does not change invertibility, and its determinant is then strictly positive. Thus $\mathcal U_i$ is dense.

Since the Brillouin zone has only four TRIMs, the intersection
\begin{equation}
\mathcal U=\bigcap_{i=0}^{3}\mathcal U_i
\end{equation}
is also open and dense in $\mathrm{Gr}_m(\mathbb H^{n/2})$.
In particular, a single momentum-independent reference $R$ can be chosen such that $\det O(\Gamma_i)>0$ at every TRIM.

\subsection{Pfaffian section}

We next establish the corresponding properties of the Pfaffian section
introduced in Sec.~\ref{subsec:AII_pfaffian}. For a local occupied
basis
\begin{equation}
U(\mathbf{k})
=
\bigl(
\ket{\psi_1(\mathbf{k})},
\ket{\psi_2(\mathbf{k})}
\bigr),
\end{equation}
define
\begin{equation}
M_{ij}(\mathbf{k})
=
\braket{
\psi_i(\mathbf{k})
|
\Theta\psi_j(\mathbf{k})
}.
\end{equation}
Because $\Theta^2=-1$,
\begin{equation}
\braket{\phi|\Theta\psi}
=
-\braket{\psi|\Theta\phi},
\end{equation}
so $M(\mathbf{k})$ is antisymmetric. For a rank-two occupied bundle it
therefore has the form
\begin{equation}
M(\mathbf{k})
=
\begin{pmatrix}
0 & p(\mathbf{k})\\
-p(\mathbf{k}) & 0
\end{pmatrix},
\qquad
p(\mathbf{k})
=
\operatorname{Pf}M(\mathbf{k})=
\braket{\psi_1(\mathbf{k})|
\Theta\psi_2(\mathbf{k})}..
\end{equation}

Under a change of occupied basis,
\begin{equation}
U(\mathbf{k})
\longrightarrow
U(\mathbf{k})W(\mathbf{k}),
\qquad
W(\mathbf{k})\in U(2),
\end{equation}
antiunitarity of $\Theta$ gives
\begin{equation}
M(\mathbf{k})
\longrightarrow
W^\dagger(\mathbf{k})
M(\mathbf{k})
(W^\dagger)^T.
\end{equation}
The Pfaffian identity
\begin{equation}
\operatorname{Pf}(AMA^T)
=
\det A\,\operatorname{Pf}M
\end{equation}
implies that
\begin{equation}
p(\mathbf{k})
\longrightarrow
\det W(\mathbf{k})^{-1}
p(\mathbf{k}).
\label{eq:app_AII_pf_gauge}
\end{equation}
This is precisely the inverse transformation of the determinant-line
basis given in Eq.~\eqref{app:eq:basis_change_basis}. Hence
\begin{equation}
a_{\mathrm{Pf}}(\mathbf{k})
=
p(\mathbf{k})\,
e_{\mathcal E}(\mathbf{k})
\otimes e_R^*
\end{equation}
is a globally defined section of the same line bundle
$\mathcal L_{\mathcal E,R}$ as the reference-overlap section $a_O$.

To determine its time-reversal transformation, choose the occupied
bases at a pair of momenta away from TRIMs such that the basis at $-\mathbf{k}$ is the time-reversed image of the one at $\mathbf{k}$,
\begin{equation}
\ket{\psi_i(-\mathbf{k})}
=
\Theta\ket{\psi_i(\mathbf{k})}.
\end{equation}
The corresponding determinant bases satisfy
Eq.~\eqref{eq:app_AII_compatible_frame}. In this basis,
\begin{align}
M_{12}(-\mathbf{k})
&=
\braket{
\Theta\psi_1(\mathbf{k})
|
\Theta^2\psi_2(\mathbf{k})
}
\nonumber\\
&=
-\braket{
\Theta\psi_1(\mathbf{k})
|
\psi_2(\mathbf{k})
}
\nonumber\\
&=
(
\braket{
\psi_1(\mathbf{k})
|
\Theta\psi_2(\mathbf{k})
}
)^{*},
\end{align}
and therefore
\begin{equation}
p(-\mathbf{k})
=
(p(\mathbf{k}))^{*}.
\label{eq:app_AII_pf_TR}
\end{equation}
Thus the zeros of $a_{\mathrm{Pf}}$ also occur in time-reversed pairs
with opposite winding. At a time-reversal-invariant momentum choose the canonical Kramers
basis
\begin{equation}
\ket{\psi_2}
=
\Theta\ket{\psi_1}.
\end{equation}
With the Pfaffian convention above,
\begin{equation}
p(\Gamma_i)
=
\braket{\psi_1|\Theta\psi_2}
=
-1.
\label{eq:app_AII_pf_TRIM}
\end{equation}
In particular, the Pfaffian section never vanishes at a
time-reversal-invariant momentum, and
\begin{equation}
a_{\mathrm{Pf}}(\Gamma_i)
=
-
e_{K}(\Gamma_i)\otimes e_R^*.
\end{equation}
Multiplication of the entire section by $-1$ changes neither its zeros
nor their winding numbers. We may therefore define
\begin{equation}
\widetilde a_{\mathrm{Pf}}
=
-a_{\mathrm{Pf}},
\end{equation}
for which
\begin{equation}
\widetilde a_{\mathrm{Pf}}(\Gamma_i)
=
e_{K}(\Gamma_i)\otimes e_R^*.
\label{eq:app_AII_pf_fixed_point}
\end{equation}
As shown in Eqs.~\eqref{eq:app_AII_O_fixed_point}
and \eqref{eq:app_AII_pf_fixed_point}, the
reference-overlap and the normalized Pfaffian
sections are both time-reversal compatible and
have positive coefficients at every TRIM.

\subsection{Relative determinant-overlap section}

We finally establish the gauge, time-reversal, and fixed-point
properties of the comparison section used in
Sec.~\ref{subsec:AII_relative}. Let
$\mathcal E^\alpha$ and $\mathcal E^\sigma$ be two rank-two occupied
bundles with local occupied bases
\begin{equation}
U^\alpha
=
\bigl(
\ket{\psi_1^\alpha},
\ket{\psi_2^\alpha}
\bigr),
\qquad
U^\sigma
=
\bigl(
\ket{\psi_1^\sigma},
\ket{\psi_2^\sigma}
\bigr).
\end{equation}
The matrix of overlaps is the one introduced in
Eq.~\eqref{eq:overlap_matrix},
\begin{equation}
S(\mathbf{k})
=
[U^\alpha(\mathbf{k})]^\dagger
U^\sigma(\mathbf{k}).
\end{equation}
Under independent changes of occupied basis,
\begin{equation}
U^\alpha
\longrightarrow
U^\alpha W^\alpha,
\qquad
U^\sigma
\longrightarrow
U^\sigma W^\sigma,
\end{equation}
the matrix of overlaps transforms as
\begin{equation}
S
\longrightarrow
(W^\alpha)^\dagger
S
W^\sigma.
\end{equation}
Consequently,
\begin{equation}
\det S
\longrightarrow
\det(W^\alpha)^{-1}
\det S
\det(W^\sigma).
\label{eq:app_AII_relative_det_gauge}
\end{equation}
The local basis of the relative determinant line
\begin{equation}
\mathcal L_{\alpha\sigma}
=
\det\mathcal E^\alpha
\otimes
(\det\mathcal E^\sigma)^*
\end{equation}
transforms oppositely,
\begin{equation}
e_\alpha\otimes e_\sigma^*
\longrightarrow
\det(W^\alpha)\det(W^\sigma)^{-1}
\,e_\alpha\otimes e_\sigma^*.
\end{equation}
Thus
\begin{equation}
a_S
=
\det S\,
e_\alpha\otimes e_\sigma^*
\end{equation}
is independent of the occupied-basis choices and therefore defines a
global section of $\mathcal L_{\alpha\sigma}$.

Choose occupied bases at time-reversed momenta
related by
\begin{equation}
U^\mu(-\mathbf{k})
=
(\Theta U^\mu(\mathbf{k}))B_\Theta^\mu(\mathbf{k}),
\qquad
\mu\in\{\alpha,\sigma\},
\label{eq:app_AII_frame_sewing}
\end{equation}
where $B_\Theta^\mu(\mathbf{k})\in U(2)$.
Writing
$b_{ij}=[B_\Theta^\mu(\mathbf{k})]_{ij}$
for simplicity, this relation gives
\begin{align}
\ket{\psi_1^\mu(-\mathbf{k})}
&=
b_{11}\Theta\ket{\psi_1^\mu(\mathbf{k})}
+
b_{21}\Theta\ket{\psi_2^\mu(\mathbf{k})},
\\
\ket{\psi_2^\mu(-\mathbf{k})}
&=
b_{12}\Theta\ket{\psi_1^\mu(\mathbf{k})}
+
b_{22}\Theta\ket{\psi_2^\mu(\mathbf{k})}.
\end{align}
Taking the wedge product of these two states,
we obtain
\begin{align}
e_\mu(-\mathbf{k})
&=
\ket{\psi_1^\mu(-\mathbf{k})}
\wedge
\ket{\psi_2^\mu(-\mathbf{k})}
\nonumber\\
&=
\left(
b_{11}\Theta\ket{\psi_1^\mu}
+
b_{21}\Theta\ket{\psi_2^\mu}
\right)
\wedge
\left(
b_{12}\Theta\ket{\psi_1^\mu}
+
b_{22}\Theta\ket{\psi_2^\mu}
\right)
\nonumber\\
&=
(b_{11}b_{22}-b_{12}b_{21})
\left(
\Theta\ket{\psi_1^\mu}
\wedge
\Theta\ket{\psi_2^\mu}
\right)
\nonumber\\
&=
\det B_\Theta^\mu(\mathbf{k})\,
\Theta_{\det}e_\mu(\mathbf{k}),
\end{align}
where, in the intermediate steps, the states
without explicit momentum arguments are
evaluated at $\mathbf{k}$. We used the
antisymmetry of the wedge product and the
definition of $\Theta_{\det}$.
For the time-reversal-compatible determinant bases
of Eq.~\eqref{eq:app_AII_compatible_frame},
\begin{equation}
\det B_\Theta^\mu(\mathbf{k})=1.
\label{eq:app_AII_det_sewing}
\end{equation}

Using Eq.~\eqref{eq:app_AII_frame_sewing},
the overlap matrix at $-\mathbf{k}$ becomes
\begin{align}
S(-\mathbf{k})
&=
[U^\alpha(-\mathbf{k})]^\dagger
U^\sigma(-\mathbf{k})
=
[B_\Theta^\alpha(\mathbf{k})]^\dagger
[\Theta U^\alpha(\mathbf{k})]^\dagger
[\Theta U^\sigma(\mathbf{k})]
B_\Theta^\sigma(\mathbf{k}).
\end{align}
Since $\Theta$ is antiunitary,
\begin{align}
\braket{\Theta\psi_i^\alpha|
\Theta\psi_j^\sigma}
&=
\braket{\psi_j^\sigma|\psi_i^\alpha}
=
\braket{\psi_i^\alpha|\psi_j^\sigma}^{*}
=
S_{ij}(\mathbf{k})^*.
\end{align}
We therefore obtain
\begin{equation}
S(-\mathbf{k})
=
[B_\Theta^\alpha(\mathbf{k})]^\dagger
S(\mathbf{k})^*
B_\Theta^\sigma(\mathbf{k}).
\end{equation}
Taking the determinant and using
Eq.~\eqref{eq:app_AII_det_sewing}, we obtain
\begin{equation}
\det S(-\mathbf{k})
=
(\det S(\mathbf{k}))^{*}.
\label{eq:app_AII_relative_TR}
\end{equation}
The zeros of the comparison section therefore occur in time-reversed pairs
with opposite winding,
\begin{equation}
\nu_{-\mathbf{k}_*}
=
-\nu_{\mathbf{k}_*}.
\end{equation}
We now show that time-reversal symmetry ensures that the relative overlap determinant is nonnegative at every TRIM. At a TRIM $\Gamma_i$, choose Kramers bases
in both occupied spaces,
\begin{equation}
\ket{\psi_2^\mu}
=
\Theta\ket{\psi_1^\mu},
\qquad
\mu\in\{\alpha,\sigma\}.
\end{equation}
Writing
\begin{equation}
a
=
\braket{\psi_1^\alpha|\psi_1^\sigma},
\qquad
b
=
\braket{\psi_1^\alpha|\psi_2^\sigma},
\end{equation}
the overlap matrix takes the form
\begin{equation}
S(\Gamma_i)
=
\begin{pmatrix}
a & b\\
- b^{*} & a^{*}
\end{pmatrix},
\label{eq:app_AII_relative_TRIM_matrix}
\end{equation}
and therefore
\begin{equation}
\det S(\Gamma_i)
=
|a|^2+|b|^2
\geq0.
\label{eq:app_AII_relative_TRIM_positive}
\end{equation}

At each TRIM, the Kramers pairs of the two
occupied subspaces define the determinant vectors
$e_\alpha^K(\Gamma_i)$ and $e_\sigma^K(\Gamma_i)$,
as introduced above. These give the following
vector in the relative determinant line:
\begin{equation}
e_{\alpha\sigma}^{K}(\Gamma_i)
=
e_\alpha^{K}(\Gamma_i)
\otimes
\left(
e_\sigma^{K}(\Gamma_i)
\right)^*.
\end{equation}
Hence
\begin{equation}
a_S(\Gamma_i)
=
\det S(\Gamma_i)\,
e_{\alpha\sigma}^{K}(\Gamma_i).
\label{eq:app_AII_relative_fixed_point}
\end{equation}
Since $\det S(\Gamma_i)\geq0$, whenever the
overlap is nonzero, $a_S(\Gamma_i)$ is a
positive multiple of $e_{\alpha\sigma}^{K}(\Gamma_i)$.
However, the overlap determinant may vanish at a TRIM. Such a zero is nongeneric, since
Eq.~\eqref{eq:app_AII_relative_TRIM_positive}
shows that $\det S(\Gamma_i)=0$ requires
$a=b=0$. In this case, the two occupied Kramers pairs are orthogonal. We now show that a small time-reversal-preserving deformation can make the overlap nonzero at that momentum without changing the class AII $\mathbb Z_2$ invariant of either system.

Let
$\{\ket{\psi_1^\mu},
\ket{\psi_2^\mu}=\Theta\ket{\psi_1^\mu}\}$
be orthonormal Kramers bases of the occupied
subspaces $\mathcal E^\mu_{\Gamma_i}$,
with $\mu\in\{\alpha,\sigma\}$.
We wish to rotate the occupied plane of
$\sigma$ slightly toward that of $\alpha$.
To this end, define
\begin{equation}
X
=
\sum_{j=1}^{2}
\left(
\ket{\psi_j^\alpha}\bra{\psi_j^\sigma}
-
\ket{\psi_j^\sigma}\bra{\psi_j^\alpha}
\right).
\end{equation}
Because the two planes are orthogonal,
\begin{equation}
X\ket{\psi_j^\sigma}
=
\ket{\psi_j^\alpha},
\qquad
X\ket{\psi_j^\alpha}
=
-\ket{\psi_j^\sigma}.
\end{equation}
Thus, $X$ generates rotations between the
corresponding states of the two Kramers pairs.
It is anti-Hermitian and satisfies
$\Theta X\Theta^{-1}=X$.

To perform this rotation only near $\Gamma_i$,
we choose a smooth real function
$\chi(\mathbf{k})=\chi(-\mathbf{k})$,
supported in a small neighborhood of $\Gamma_i$,
with $\chi(\Gamma_i)=1$.
We define the unitary deformation
\begin{equation}
V_t(\mathbf{k})
=
\exp[t\epsilon\chi(\mathbf{k})X],
\qquad t\in[0,1],
\end{equation}
where $\epsilon$ is a small rotation angle.
At $\Gamma_i$, its action is
\begin{equation}
V_t(\Gamma_i)\ket{\psi_j^\sigma}
=
\cos(t\epsilon)\ket{\psi_j^\sigma}
+
\sin(t\epsilon)\ket{\psi_j^\alpha}.
\end{equation}
At $t=1$, the overlap matrix between the
occupied states of $\alpha$ and the deformed
occupied states of $\sigma$ becomes
\begin{equation}
\begin{aligned}
S_{ab}(\Gamma_i)
&=
\bra{\psi_a^\alpha}
V_1(\Gamma_i)
\ket{\psi_b^\sigma}\\
&=
\cos\epsilon\,
\braket{\psi_a^\alpha|\psi_b^\sigma}
+
\sin\epsilon\,
\braket{\psi_a^\alpha|\psi_b^\alpha}\\
&=
\sin\epsilon\,\delta_{ab},
\end{aligned}
\end{equation}
where we used the orthogonality of the
original occupied subspaces and the
orthonormality of the Kramers bases.
Consequently,
$\det S(\Gamma_i)=\sin^2\epsilon>0$.

The deformed occupied projector
\begin{equation}
P_t^\sigma(\mathbf{k})
=
V_t(\mathbf{k})P^\sigma(\mathbf{k})
V_t(\mathbf{k})^\dagger
\end{equation}
remains time-reversal symmetric because
$\Theta V_t(\mathbf{k})\Theta^{-1}=V_t(-\mathbf{k})$.
Let $H^\sigma(\mathbf{k})$ be the original
Bloch Hamiltonian of system $\sigma$, whose
occupied states define $P^\sigma(\mathbf{k})$. We can construct the deformed Hamiltonian
\begin{equation}
H_t^\sigma(\mathbf{k})
=
V_t(\mathbf{k})H^\sigma(\mathbf{k})
V_t(\mathbf{k})^\dagger.
\end{equation}
Since $V_t(\mathbf{k})$ is unitary,
$H_t^\sigma(\mathbf{k})$ has exactly the same
spectrum as $H^\sigma(\mathbf{k})$ at every
momentum. The energy gap therefore remains
open throughout the deformation. Since
time-reversal symmetry is also preserved,
the class AII $\mathbb Z_2$ invariant
remains unchanged.

This deformation may break additional
crystalline symmetries. In particular,
a TRIM zero enforced by a crystalline
representation mismatch can be removed
in this way only when those additional
symmetries are relaxed. The resulting
configuration can nevertheless be used
to evaluate the underlying class AII parity.

\subsection{Parity on a time-reversal fundamental domain}

We now show that the parity of the total winding
in a time-reversal fundamental domain does not
depend on the particular global section used
to compute it, provided the sections respect
time reversal and satisfy the same positivity
condition at the TRIMs.

Let $s$ be a global section that respects
time-reversal symmetry. We require that, at
every TRIM,
\begin{equation}
s(\Gamma_i)
=
r_{\Gamma_i}e_K(\Gamma_i),
\qquad
r_{\Gamma_i}>0.
\label{eq:app_AII_admissible}
\end{equation}
Here $e_K(\Gamma_i)$ denotes the nonzero
determinant-line vector defined using the Kramers
pairs, including the appropriate dual factor
when necessary. For the reference-overlap and
relative-overlap sections, $r_{\Gamma_i}$ is
respectively $\det O(\Gamma_i)$ or
$\det S(\Gamma_i)$ in Kramers bases. For the
normalized Pfaffian section
$\widetilde a_{\mathrm{Pf}}$, we have
$r_{\Gamma_i}=1$ at every TRIM.

Choose a time-reversal fundamental domain
$\mathrm{BZ}^2_+$ and assume first that the
section has no zeros on its boundary.
Since every zero away from the TRIMs
occurs in a pair
$\{\mathbf{k}_*,-\mathbf{k}_*\}$
with opposite winding numbers, the domain
contains one representative of each pair.
We define
\begin{equation}
N_s
=
\sum_{\mathbf{k}_*\in Z_s^+}
\nu_{\mathbf{k}_*},
\label{eq:app_AII_half_BZ_sum}
\end{equation}
where $Z_s^+$ denotes the zeros of $s$
inside $\mathrm{BZ}^2_+$.

We now consider how this sum changes under
a continuous deformation of the section
that preserves time-reversal symmetry and
the positivity condition of
Eq.~\eqref{eq:app_AII_admissible}.
Zeros may move within the domain, or be
created and annihilated with opposite winding
numbers, without changing their total winding.
However, the winding sum may change when a
zero crosses the boundary of the domain. Suppose a zero with winding $\nu$ leaves
$\mathrm{BZ}^2_+$. Time reversal requires
its partner, with winding $-\nu$, to enter
through the time-reversal-related boundary
point. Thus, the contribution $\nu$ is
replaced by $-\nu$, giving
\begin{equation}
\Delta N_s=-2\nu.
\end{equation}
Although the total winding in the domain
may change, its parity remains unchanged.

The same argument shows that the parity
does not depend on the choice of
time-reversal fundamental domain. Different
domains may contain opposite members of
some time-reversed pairs, but replacing
$\nu$ by $-\nu$ changes the winding sum
only by an even integer. If a zero lies on the boundary away from
a TRIM, a small deformation of the boundary
or of the section allows us to select one
of the two time-reversed representatives.
The two choices give the same winding
parity. At a TRIM, the positivity condition
of Eq.~\eqref{eq:app_AII_admissible}
ensures that the section cannot vanish.
Consequently, a single time-reversed pair
of zeros cannot annihilate by meeting at
a TRIM while this condition is preserved. We have therefore shown that the winding
parity is independent of the chosen
fundamental domain and remains unchanged
under continuous time-reversal-preserving
deformations that satisfy the positivity
condition at every TRIM.

We have shown that the winding parity is
independent of the chosen fundamental domain
and remains unchanged under continuous deformations
that preserve time-reversal symmetry and the
positivity condition at every TRIM. We now show
that any two sections of the same determinant
line bundle satisfying these requirements can
be connected by such a deformation.

Consider two global sections $s_0$ and $s_1$
of the same determinant line bundle that
respect time-reversal symmetry and satisfy
Eq.~\eqref{eq:app_AII_admissible}.
We can connect them through
\begin{equation}
s_t=(1-t)s_0+t s_1,
\qquad t\in[0,1].
\end{equation}

Since both sections respect time reversal,
so does $s_t$. At every TRIM, we have
\begin{equation}
s_t(\Gamma_i)
=
\left[
(1-t)r_{0,\Gamma_i}
+
t r_{1,\Gamma_i}
\right]e_K(\Gamma_i).
\end{equation}
Because $r_{0,\Gamma_i}>0$ and
$r_{1,\Gamma_i}>0$, the coefficient
remains strictly positive throughout
the interpolation. Thus, $s_t$ never
vanishes at a TRIM. Since the interpolation
preserves time-reversal symmetry and the
positivity condition at every TRIM, the
preceding argument shows that the winding
parity is the same for both sections. Notice that the positivity condition
is essential, if the coefficients of
two sections had opposite signs at a
TRIM, their interpolation would
necessarily vanish at that momentum.

We can now apply this result to the
reference-overlap and normalized Pfaffian
sections. Both sections respect time reversal
and belong to the same determinant line bundle.
Moreover, Eqs.~\eqref{eq:app_AII_O_fixed_point}
and \eqref{eq:app_AII_pf_fixed_point}
show that both have positive coefficients
at every TRIM. Their winding parities therefore
agree,
\begin{equation}
\sum_{\mathbf{k}_*\in Z_O^+}
\nu^{\mathcal E}_{\mathbf{k}_*}
=
\sum_{\mathbf{k}_*\in Z_{\mathrm{Pf}}^+}
\nu^{\mathrm{Pf}}_{\mathbf{k}_*}
\quad\mathrm{mod}\;2.
\label{eq:app_AII_pf_parity}
\end{equation}
Since the Pfaffian winding parity gives
the Kane-Mele $\mathbb Z_2$ invariant,
the reference-overlap section provides
the same invariant.

\subsection{Relative $\mathbb{Z}_2$ index}

We have shown that the winding parity in a
time-reversal fundamental domain is independent
of the particular global section, provided the
sections belong to the same determinant line
bundle, respect time reversal, and have positive
coefficients at every TRIM. We now use this result
to show that the winding parity of the relative
overlap gives the difference between the class AII
$\mathbb Z_2$ indices of the two systems. Consider two occupied bundles
$\mathcal E^\alpha$ and $\mathcal E^\sigma$.
Using the same trivial Kramers reference $R$
introduced in the absolute construction, define
\begin{equation}
\mathcal L_{\alpha R}
=
\det\mathcal E^\alpha\otimes(\det R)^*,
\qquad
\mathcal L_{\sigma R}
=
\det\mathcal E^\sigma\otimes(\det R)^*.
\end{equation}
As shown in  Sec.~\ref{app:AII_generic_reference}, the reference can be chosen so that its overlap with both occupied bundles is nonzero at every TRIM. Because the reference bundle is trivial, the relative determinant line can be written as
\begin{equation}
\mathcal L_{\alpha\sigma}
\simeq
\mathcal L_{\alpha R}
\otimes
\mathcal L_{\sigma R}^{*}.
\label{eq:app_AII_relative_factorization}
\end{equation}
We can therefore construct a global section of the relative determinant line by combining the two reference-overlap sections,
\begin{equation}
a_{\alpha\sigma}^{(R)}
=
a_O^\alpha\otimes(a_O^\sigma)^*.
\end{equation}
In local occupied bases, its coefficient is
\begin{equation}
\det O^\alpha(\mathbf{k})\,
[\det O^\sigma(\mathbf{k})]^*.
\end{equation}
Since the winding of a product is the sum of
the individual windings, while complex
conjugation reverses their signs, we obtain
\begin{align}
\sum_{\mathbf{k}_*\in Z_R^+}
\nu_{\mathbf{k}_*}
&=
\sum_{\mathbf{k}_*\in Z_{O^\alpha}^+}
\nu_{\mathbf{k}_*}^{\mathcal E^\alpha}
-
\sum_{\mathbf{k}_*\in Z_{O^\sigma}^+}
\nu_{\mathbf{k}_*}^{\mathcal E^\sigma}
\quad\mathrm{mod}\;2
\nonumber\\
&=
\nu_{\mathbb Z_2}(\mathcal E^\alpha)
-
\nu_{\mathbb Z_2}(\mathcal E^\sigma)
\quad\mathrm{mod}\;2,
\label{eq:app_AII_relative_additivity}
\end{align}
where $Z_R^+$ denotes the zeros of
$a_{\alpha\sigma}^{(R)}$ in the time-reversal
fundamental domain. In the second equality,
we used the result of the absolute construction,
which identifies the winding parity of each
reference-overlap section with the corresponding
class AII $\mathbb Z_2$ index.

Finally, consider the direct overlap section
$a_S$ between $\mathcal E^\alpha$ and
$\mathcal E^\sigma$. Both $a_S$ and
$a_{\alpha\sigma}^{(R)}$ are global sections
of $\mathcal L_{\alpha\sigma}$ that respect
time reversal. With $R$ chosen so that both reference overlaps are nonzero at every TRIM, and for a generic direct comparison, both sections have positive coefficients at these momenta, as established above.
The result of the preceding subsection
therefore guarantees that their winding
parities agree,
\begin{equation}
\sum_{\mathbf{k}_*\in Z_{\alpha\sigma}^+}
\nu_{\mathbf{k}_*}
=
\sum_{\mathbf{k}_*\in Z_R^+}
\nu_{\mathbf{k}_*}
\quad\mathrm{mod}\;2.
\end{equation}
Combining this with
Eq.~\eqref{eq:app_AII_relative_additivity},
we obtain
\begin{equation}
\nu_{\mathbb Z_2}(\mathcal E^\alpha)
-
\nu_{\mathbb Z_2}(\mathcal E^\sigma)
=
\sum_{\mathbf{k}_*\in Z_{\alpha\sigma}^+}
\nu_{\mathbf{k}_*}
\quad\mathrm{mod}\;2,
\end{equation}
which is Eq.~\eqref{eq:relative_Z2_det_overlap}
of the main text.

\subsection{Extension to higher even ranks}
\label{app:AII_higher_rank}

We now show that the preceding construction extends to
occupied bundles of arbitrary even rank $2m$. The main
difference from the rank-two case is that the occupied
subspace contains $m$ Kramers pairs rather than one.

Time reversal acts on the determinant line
$\det\mathcal E=\wedge^{2m}\mathcal E$ through
\begin{equation}
\Theta_{\det}
(u_1\wedge\cdots\wedge u_{2m})
=
\Theta u_1\wedge\cdots\wedge\Theta u_{2m}.
\end{equation}
Since $\Theta^2=-1$ and the rank is even,
$\Theta_{\det}^2=+1$, as in the rank-two case.
At a TRIM, an orthonormal Kramers basis defines
the determinant vector
\begin{equation}
e_K=
\bigwedge_{j=1}^{m}(u_j\wedge\Theta u_j),
\end{equation}
which is independent of the orthonormal Kramers
frame, as established in
Sec.~\ref{app:AII_generic_reference}.
The same subsection also shows that the overlap
determinant between two occupied subspaces of rank
$2m$ is nonnegative in these bases and strictly
positive whenever the overlap is invertible.

The reference-overlap construction extends directly.
Choose a fixed reference $R$ containing $m$ Kramers
pairs and define
\begin{equation}
a_O(\mathbf k)
=
\det O(\mathbf k)\,
e_{\mathcal E}(\mathbf k)\otimes e_R^*,
\end{equation}
where $O$ is now a $2m\times2m$ overlap matrix.
Since the occupied projector respects time-reversal
symmetry and $R$ consists of Kramers pairs, the
reference-overlap section $a_O$ also respects
time reversal. The argument of
Sec.~\ref{app:AII_generic_reference}
guarantees that $R$ can be chosen so that
$a_O$ is nonzero and has a positive coefficient
at every TRIM when the occupied states are
arranged in Kramers pairs.

To identify its winding parity with the conventional
class AII invariant, consider the Pfaffian construction
for the same occupied bundle. For an orthonormal
occupied basis $\{\ket{\psi_i(\mathbf k)}\}_{i=1}^{2m}$,
define
\begin{equation}
M_{ij}(\mathbf k)
=
\braket{\psi_i(\mathbf k)|
\Theta\psi_j(\mathbf k)}.
\end{equation}
This is an antisymmetric $2m\times2m$ matrix.
Under an occupied-basis transformation
$U\mapsto UW$, its Pfaffian transforms as
\begin{equation}
\operatorname{Pf}M
\longmapsto
(\det W)^{-1}\operatorname{Pf}M.
\end{equation}
Therefore,
\begin{equation}
a_{\mathrm{Pf}}(\mathbf k)
=
\operatorname{Pf}M(\mathbf k)\,
e_{\mathcal E}(\mathbf k)\otimes e_R^*
\end{equation}
defines a global section of the same determinant
line as $a_O$.

In time-reversal-related occupied bases,
$M(-\mathbf k)=M(\mathbf k)^*$,
so $a_{\mathrm{Pf}}$ also respects time reversal.
At a TRIM, choosing an orthonormal Kramers
basis gives
\begin{equation}
M(\Gamma_i)
=
\bigoplus_{j=1}^{m}
\begin{pmatrix}
0&-1\\
1&0
\end{pmatrix}.
\end{equation}
Since the Pfaffian of a block-diagonal antisymmetric matrix
is the product of the Pfaffians of its blocks,
\begin{equation}
\operatorname{Pf}(A\oplus B)
=
\operatorname{Pf}(A)\operatorname{Pf}(B),
\end{equation}
and each block has Pfaffian $-1$, we obtain
\begin{equation}
\operatorname{Pf}M(\Gamma_i)=(-1)^m.
\end{equation}
Thus, the section
\begin{equation}
\widetilde a_{\mathrm{Pf}}
=
(-1)^m a_{\mathrm{Pf}}
\end{equation}
has a positive coefficient at every TRIM
in the Kramers determinant basis.

Both $a_O$ and $\widetilde a_{\mathrm{Pf}}$
are therefore global sections of the same
determinant line bundle. As established above, their winding parities
must agree,
\begin{equation}
\sum_{\mathbf k_*\in Z_O^+}
\nu_{\mathbf k_*}^{\mathcal{E}}
=
\sum_{\mathbf k_*\in Z_{\mathrm{Pf}}^+}
\nu_{\mathbf k_*}^{\mathrm{Pf}}
\quad\mathrm{mod}\;2.
\end{equation}
The Pfaffian winding parity is the conventional
Kane-Mele $\mathbb Z_2$ invariant, whose
Pfaffian formulation applies to arbitrary
even occupied rank~\cite{Kane_Mele_2005,Fu_Kane_2006}.
Hence the reference-overlap construction gives
the same invariant for rank $2m$.

Finally, consider two occupied bundles
$\mathcal E^\alpha$ and $\mathcal E^\sigma$
of equal rank $2m$. Their relative overlap
defines a time-reversal-compatible section
$a_S$ of
\begin{equation}
\mathcal L_{\alpha\sigma}
=
\det\mathcal E^\alpha\otimes
(\det\mathcal E^\sigma)^*.
\end{equation}
For a generic comparison section, the overlap determinant
is nonzero at every TRIM. However, for two particular systems, it may vanish at one of these momenta. We now show that the deformation used for a single occupied Kramers pair extends to arbitrary even rank. The difference is that we must identify the Kramers pairs responsible for the vanishing determinant and apply a small rotation to each of them. As before, this deformation preserves time-reversal symmetry without changing the topological class of either system.

To see this, consider the projection map
\begin{equation}
T=P^\alpha(\Gamma_i)|_{\mathcal E^\sigma_{\Gamma_i}}:
\mathcal E^\sigma_{\Gamma_i}
\longrightarrow
\mathcal E^\alpha_{\Gamma_i}.
\end{equation}
The overlap matrix $S(\Gamma_i)$ represents $T$ in orthonormal occupied bases. At a TRIM, both occupied subspaces are
invariant under time reversal. Consequently,
the projection map satisfies
\begin{equation}
T\Theta\ket{v}
=
\Theta T\ket{v},
\qquad
\ket{v}\in\mathcal E^\sigma_{\Gamma_i}.
\end{equation}
It follows that $T^\dagger T$ also commutes
with time reversal, so its eigenvectors
can be chosen in orthonormal Kramers pairs
$\{\ket{v_{j,1}},\ket{v_{j,2}}\}$, where
$\ket{v_{j,2}}=\Theta\ket{v_{j,1}}$ and
\begin{equation}
T^\dagger T\ket{v_{j,a}}
=s_j^2\ket{v_{j,a}},
\qquad
0\leq s_j\leq1,
\quad a=1,2.
\end{equation}
Here $s_j$ are the singular values of the overlap matrix, each occurring twice because of Kramers degeneracy.

For $s_j>0$, we define
\begin{equation}
\ket{u_{j,a}}=\frac{1}{s_j}T\ket{v_{j,a}}
=\frac{1}{s_j}P^\alpha(\Gamma_i)\ket{v_{j,a}}.
\end{equation}
These states form orthonormal Kramers pairs in $\mathcal E^\alpha_{\Gamma_i}$. They are the normalized projections of the $\ket{v_{j,a}}$ states onto the occupied subspace of $\alpha$.

For $s_j=0$, this definition cannot be used. Indeed,
\begin{equation}
T\ket{v_{j,a}}=0,
\end{equation}
so $\ket{v_{j,a}}$ is orthogonal to every occupied state of $\alpha$. Because the two occupied subspaces have the same dimension, $\ker T$ and $\ker T^\dagger$
have the same dimension. We choose the remaining orthonormal Kramers
pairs $\ket{u_{j,a}}$ in $\ker T^\dagger$. Since
\begin{equation}
T^\dagger=P^\sigma(\Gamma_i)|_{\mathcal E^\alpha_{\Gamma_i}},
\end{equation}
these states satisfy
\begin{equation}
P^\sigma(\Gamma_i)\ket{u_{j,a}}=0,
\end{equation}
and are therefore orthogonal to every occupied state of $\sigma$. This, for $s_j=0$, the $\ket{u_{j,a}}$ states are chosen independently of the $\ket{v_{j,a}}$ states, rather than obtained by projecting them.

With these choices, the two sets of states satisfy
\begin{equation}
\braket{u_{i,a}|v_{j,b}}
=s_j\,\delta_{ij}\delta_{ab},
\end{equation}
so the overlap matrix takes the diagonal form
\begin{equation}
S(\Gamma_i)
=\operatorname{diag}
\bigl(s_1\mathbb{1}_2,\ldots,s_m\mathbb{1}_2\bigr),
\qquad
\mathbb{1}_2=\begin{pmatrix}1&0\\0&1\end{pmatrix}.
\end{equation}
This is the singular-value decomposition in Kramers-paired bases. In particular,
\begin{equation}
\det S(\Gamma_i)=\prod_{j=1}^m s_j^2.
\end{equation}

Let $Z=\{j:s_j=0\}$ and denote the number of Kramers pairs with vanishing singular value by $r=|Z|$.
For each $j\in Z$, the pair $\ket{v_{j,a}}$ is orthogonal to all of $\mathcal E^\alpha_{\Gamma_i}$, while $\ket{u_{j,a}}$ is orthogonal to all of $\mathcal E^\sigma_{\Gamma_i}$. We can therefore rotate these mutually orthogonal Kramers pairs toward one another. Define
\begin{equation}
X=\sum_{j\in Z}\sum_{a=1}^{2}
\bigl(
\ket{u_{j,a}}\bra{v_{j,a}}
-\ket{v_{j,a}}\bra{u_{j,a}}
\bigr).
\end{equation}
The operator is anti-Hermitian and time-reversal invariant:
\begin{equation}
X^\dagger=-X,
\qquad \Theta X\Theta^{-1}=X.
\end{equation}
Choose a smooth, real function $\chi(\mathbf k)$ supported near
$\Gamma_i$, satisfying $\chi(-\mathbf k)=\chi(\mathbf k)$ and $\chi(\Gamma_i)=1$. Working in a local time-reversal-compatible representation, define
\begin{equation}
V_t(\mathbf k)
=\exp\!\bigl[t\varepsilon\chi(\mathbf k)X\bigr],
\qquad t\in[0,1].
\end{equation}
The unitary operator $V_t$ is the identity outside the chosen neighborhood and satisfies $\Theta V_t(\mathbf k)\Theta^{-1}=V_t(-\mathbf k)$. At the TRIM, for each $j\in Z$,
\begin{equation}
V_t(\Gamma_i)\ket{v_{j,a}}
=\cos(t\varepsilon)\ket{v_{j,a}}
+\sin(t\varepsilon)\ket{u_{j,a}}.
\end{equation}
Since $X$ acts only on the Kramers pairs corresponding to vanishing singular values, the remaining occupied states are unchanged.

The deformed overlap at $t=1$ is therefore
\begin{equation}
S_{\varepsilon}(\Gamma_i)
=\operatorname{diag}
\bigl(\widetilde s_1\mathbb{1}_2,\ldots,
\widetilde s_m\mathbb{1}_2\bigr),
\qquad
\widetilde s_j=
\begin{cases}
s_j,&j\notin Z,\\
\sin\varepsilon,&j\in Z.
\end{cases}
\end{equation}
Consequently,
\begin{equation}
\det S_{\varepsilon}(\Gamma_i)
=\left(\prod_{j\notin Z}s_j^2\right)
(\sin\varepsilon)^{2r}>0
\end{equation}
for sufficiently small $\varepsilon>0$.

Finally, the deformed Hamiltonian
\begin{equation}
H_t^\sigma(\mathbf k) =V_t(\mathbf k)H^\sigma(\mathbf k) V_t^\dagger(\mathbf k)
\end{equation}
preserves time reversal and has exactly the same spectrum as $H^\sigma$ at every momentum. The gap therefore remains open throughout the deformation, and the class AII $\mathbb Z_2$ invariant of $\mathcal E^\sigma$ is unchanged. In this case $\mathcal E^\alpha$ is not deformed. Thus, also at arbitrary even rank, a vanishing overlap determinant at a TRIM can be made strictly positive by an arbitrarily small time-reversal-preserving deformation without changing the class AII invariant of either system. 

Using the factorization of Eq.~\eqref{eq:app_AII_relative_factorization} and the winding-parity argument established in the preceding subsection, we obtain
\begin{equation}
\sum_{\mathbf k_*\in Z_{\alpha\sigma}^+}
\nu_{\mathbf k_*}
=
\nu_{\mathbb Z_2}(\mathcal E^\alpha)
-
\nu_{\mathbb Z_2}(\mathcal E^\sigma)
\quad\mathrm{mod}\;2.
\end{equation}
Thus, both the absolute and relative overlap constructions extend to arbitrary even occupied rank.

\section{Projector interpolation and gap-closing conditions}
\label{app:z2_spectral_flow_details}

In this appendix we provide the details underlying the spectral flow
interpretation of Sec.~\ref{subsec:spectral_flow_Z2}. We first derive the spectrum of the projector interpolation from the principal overlaps between the two occupied subspaces. We then analyze the interpolation near a simple zero of the determinant overlap and derive the corresponding two-level Hamiltonian. Finally, we discuss how time reversal relates the resulting crossings and how the same structure appears in the boundary spectral flow.

\subsection{Spectrum of the projector interpolation}

Although we consider rank-two occupied subspaces, the following spectral calculation does not rely on time-reversal symmetry. It follows from the singular-value decomposition of the overlap matrix and extends to an arbitrary number of occupied bands. Time-reversal symmetry enters only when we discuss how the resulting crossings are related.

Let $\mathcal E^\alpha_{\mathbf{k}}$ and
$\mathcal E^\sigma_{\mathbf{k}}$ be two rank-two occupied subspaces of
the same microscopic Bloch Hilbert space, with orthonormal occupied bases
\begin{equation}
U^\alpha(\mathbf{k})
=
\bigl(
\ket{\psi_1^\alpha(\mathbf{k})},
\ket{\psi_2^\alpha(\mathbf{k})}
\bigr),
\qquad
U^\sigma(\mathbf{k})
=
\bigl(
\ket{\psi_1^\sigma(\mathbf{k})},
\ket{\psi_2^\sigma(\mathbf{k})}
\bigr).
\end{equation}
Their occupied projectors are
\begin{equation}
P^\mu(\mathbf{k})
=
U^\mu(\mathbf{k})
[U^\mu(\mathbf{k})]^\dagger,
\qquad
\mu\in\{\alpha,\sigma\},
\end{equation}
and their matrix of overlaps is
\begin{equation}
S(\mathbf{k})
=
[U^\alpha(\mathbf{k})]^\dagger
U^\sigma(\mathbf{k}).
\end{equation}
We consider the interpolation
\begin{equation}
H(\lambda,\mathbf{k})
=
(1-\lambda)P^\alpha(\mathbf{k})
+
\lambda P^\sigma(\mathbf{k}),
\qquad
\lambda\in[0,1],
\label{eq:app_z2_projector_interpolation}
\end{equation}
introduced in Eq.~\eqref{eq:z2_projector_interpolation}. Since
$P^\alpha$ and $P^\sigma$ are positive semidefinite projectors, the
spectrum of $H$ lies in $[0,1]$.

To determine this spectrum, we perform the singular-value decomposition
\begin{equation}
S = V\Sigma W^\dagger,
\qquad
\Sigma = \operatorname{diag} (\rho_1,\rho_2), \qquad 0\leq\rho_i\leq1.
\label{eq:app_z2_svd}
\end{equation}
The singular values $\rho_i$ are the principal overlaps between the
two occupied subspaces. Rotating the occupied bases according to
\begin{equation}
\widetilde U^\alpha
=
U^\alpha V,
\qquad
\widetilde U^\sigma
=
U^\sigma W,
\end{equation}
gives
\begin{equation}
\braket{
\widetilde\psi_i^\alpha
|
\widetilde\psi_j^\sigma
}
=
\rho_i\delta_{ij}.
\label{eq:app_z2_principal_overlap}
\end{equation}
For a principal direction with $\rho_i<1$, define
\begin{equation}
\ket{\chi_i^\alpha}
=
\frac{
\ket{\widetilde\psi_i^\sigma}
-
\rho_i\ket{\widetilde\psi_i^\alpha}
}{
\sqrt{1-\rho_i^2}
}.
\end{equation}
By construction,
$\ket{\chi_i^\alpha}$ is normalized and orthogonal to the
$\alpha$ occupied subspace, and
\begin{equation}
\ket{\widetilde\psi_i^\sigma}
=
\rho_i
\ket{\widetilde\psi_i^\alpha}
+
\sqrt{1-\rho_i^2}\,
\ket{\chi_i^\alpha}.
\label{eq:app_z2_principal_decomposition}
\end{equation}
The principal-overlap relation
Eq.~\eqref{eq:app_z2_principal_overlap} ensures that the two-dimensional
space spanned by
$\ket{\widetilde\psi_i^\alpha}$ and
$\ket{\chi_i^\alpha}$ is invariant under both projectors. In the basis
$\{\ket{\widetilde\psi_i^\alpha},\ket{\chi_i^\alpha}\}$, their
restrictions are
\begin{equation}
P_i^\alpha
=
\begin{pmatrix}
1&0\\
0&0
\end{pmatrix},
\qquad
P_i^\sigma
=
\begin{pmatrix}
\rho_i^2
&
\rho_i\sqrt{1-\rho_i^2}
\\
\rho_i\sqrt{1-\rho_i^2}
&
1-\rho_i^2
\end{pmatrix}.
\end{equation}

The interpolation in this principal sector is therefore
\begin{equation}
H_i(\lambda)
=
\begin{pmatrix}
1-\lambda+\lambda\rho_i^2
&
\lambda\rho_i\sqrt{1-\rho_i^2}
\\
\lambda\rho_i\sqrt{1-\rho_i^2}
&
\lambda(1-\rho_i^2)
\end{pmatrix}.
\label{eq:app_z2_principal_interpolation}
\end{equation}
Its trace and determinant are
\begin{equation}
\operatorname{Tr}H_i=1,
\qquad
\det H_i
=
\lambda(1-\lambda)(1-\rho_i^2),
\end{equation}
so the two eigenvalues are
\begin{equation}
\Omega_{i,\pm}
=
\frac12
\left[
1
\pm
\sqrt{
1-
4\lambda(1-\lambda)
(1-\rho_i^2)
}
\right].
\label{eq:app_z2_spectrum}
\end{equation}

If $\rho_i=1$, the corresponding principal states coincide,
$\ket{\widetilde\psi_i^\alpha}
=
\ket{\widetilde\psi_i^\sigma}$,
and this common occupied direction remains an eigenstate of
$H(\lambda,\mathbf{k})$ with eigenvalue $1$ throughout the
interpolation. Such a sector cannot participate in a closing of the
gap at $1/2$. Indeed, in the limit $\rho_i\to1$,
Eq.~\eqref{eq:app_z2_spectrum} gives $\Omega_{i,+}=1$ and $\Omega_{i,-}=0$, independently of $\lambda$.

States orthogonal to both occupied subspaces lie in the common kernel
of $P^\alpha$ and $P^\sigma$ and remain at zero energy throughout the
interpolation. They likewise do not affect the spectral gap across
$1/2$. Away from a gap closing, the two branches
$\Omega_{i,+}$ form the rank-two spectral subspace above this gap,
which connects continuously to $\mathcal E^\alpha$ at $\lambda=0$ and
to $\mathcal E^\sigma$ at $\lambda=1$.

Equation~\eqref{eq:app_z2_spectrum} shows that a principal sector can
reach the middle of the gap only if
\begin{equation}
1-
4\lambda(1-\lambda)(1-\rho_i^2)
=
0.
\end{equation}
Since
\begin{equation}
4\lambda(1-\lambda)\leq1,
\qquad
1-\rho_i^2\leq1,
\end{equation}
equality requires both factors to attain their maximal value. Hence
\begin{equation}
\lambda=\frac12,
\qquad
\rho_i=0
\end{equation}
for at least one $i$. Finally, because
\begin{equation}
|\det S(\mathbf{k})|
=
\rho_1(\mathbf{k})\rho_2(\mathbf{k}),
\end{equation}
this condition is equivalent to
\begin{equation}
\lambda=\frac12,
\qquad
\det S(\mathbf{k})=0.
\label{eq:app_z2_gap_closing}
\end{equation}
The zeros of the determinant-overlap section are therefore precisely
the momenta at which the projector interpolation closes its gap across
$1/2$.

\subsection{Local two-level crossing}

We next determine the local form of the interpolation near an isolated
simple zero. Let $\mathbf{k}_*$ satisfy
\begin{equation}
\det S(\mathbf{k}_*)=0.
\end{equation}
For a simple zero, $S(\mathbf{k}_*)$ has rank one: one principal overlap vanishes, while the other remains nonzero. From Eq.~\eqref{eq:app_z2_spectrum}, the sector associated with the nonzero principal overlap remains gapped across $1/2$ in a neighborhood of $(\mathbf{k}_*,1/2)$. The gap closing is therefore controlled by the remaining two-dimensional sector.

Because the nonzero principal overlap remains separated from zero, its
corresponding states can be chosen smoothly in a sufficiently small
neighborhood of $\mathbf{k}_*$. Their orthogonal complements can then
also be chosen smoothly. In these local occupied bases, the overlap in
the closing sector is a complex scalar
\begin{equation}
z(\mathbf{k})
=
z_1(\mathbf{k})
+
i z_2(\mathbf{k}),
\qquad
z(\mathbf{k}_*)=0.
\label{eq:app_z2_local_complex_overlap}
\end{equation}
Here $z(\mathbf{k})$ is the complex overlap coefficient in the
gap-closing sector, not its singular value. Its magnitude is the
vanishing principal overlap,
\begin{equation}
|z(\mathbf{k})|
=
\rho_1(\mathbf{k}),
\end{equation}
while its phase carries the local winding information.
Since the second principal sector remains nonsingular, the determinant
overlap factorizes locally as
\begin{equation}
\det S(\mathbf{k})
=
g(\mathbf{k})\,z(\mathbf{k}),
\label{eq:app_z2_local_factorization}
\end{equation}
where $g(\mathbf{k})$ is smooth and nonzero in a neighborhood of
$\mathbf{k}_*$. Choosing a sufficiently small positively oriented loop
$C_{\mathbf{k}_*}$ around the zero, we therefore have
\begin{align}
\frac{1}{2\pi i}
\oint_{C_{\mathbf{k}_*}}
\mathrm d\log\det S
&=
\frac{1}{2\pi i}
\oint_{C_{\mathbf{k}_*}}
\mathrm d\log z
+
\frac{1}{2\pi i}
\oint_{C_{\mathbf{k}_*}}
\mathrm d\log g
\nonumber\\
&=
\frac{1}{2\pi i}
\oint_{C_{\mathbf{k}_*}}
\mathrm d\log z.
\label{eq:app_z2_same_winding}
\end{align}
The last term vanishes because $g$ is smooth and nonzero throughout
the region enclosed by $C_{\mathbf{k}_*}$. Thus the complex overlap
$z$ carries exactly the winding of the determinant-overlap zero.

Let $\ket{\widetilde\psi^\alpha}$ and
$\ket{\widetilde\psi^\sigma}$ denote the two states spanning the closing sector. Choosing a normalized state $\ket{\chi^\alpha}$ orthogonal to $\ket{\widetilde\psi^\alpha}$ within this two-dimensional plane, we may write
\begin{equation}
\ket{\widetilde\psi^\sigma}
=
z(\mathbf{k})
\ket{\widetilde\psi^\alpha}
+
\sqrt{1-|z(\mathbf{k})|^2}\,
\ket{\chi^\alpha}.
\label{eq:app_z2_local_state_decomposition}
\end{equation}
In the basis $\{\ket{\widetilde\psi^\alpha},\ket{\chi^\alpha}\}$, the restriction of Eq.~\eqref{eq:app_z2_projector_interpolation} to the gap-closing sector is
\begin{equation}
H_{\mathrm{eff}}(\lambda,\mathbf{k})
=
\begin{pmatrix}
1-\lambda+\lambda |z|^2
&
\lambda z\sqrt{1-|z|^2}
\\
\lambda z^*\sqrt{1-|z|^2}
&
\lambda(1-|z|^2)
\end{pmatrix}.
\label{eq:app_z2_exact_two_level}
\end{equation}
To expand around the closing, define
\begin{equation}
\eta
=
\lambda-\frac12.
\end{equation}
Near $(\mathbf{k}_*,1/2)$, both $z$ and $\eta$ are small. Keeping only terms linear in these quantities gives
\begin{equation}
H_{\mathrm{eff}}
-
\frac12\mathbb 1
\simeq
\begin{pmatrix}
-\eta & z/2\\
z^*/2 & \eta
\end{pmatrix}.
\end{equation}
Writing $z=z_1+iz_2$, this becomes
\begin{equation}
H_{\mathrm{eff}}
-
\frac12\mathbb 1
\simeq
\frac12 z_1\,\sigma_1
-
\frac12 z_2\,\sigma_2
-
\eta\,\sigma_3.
\label{eq:app_z2_local_weyl}
\end{equation}
Thus a simple determinant-overlap zero produces an isolated two-level crossing in the three-dimensional parameter space $(k_x,k_y,\lambda)$ at $(\mathbf{k}_*,1/2)$.

To relate the orientation of this crossing to the winding of the
zero of the comparison section, define
\begin{equation}
J_z(\mathbf{k}_*)
=
\left.
\left(
\frac{\partial z_1}{\partial k_x}
\frac{\partial z_2}{\partial k_y}
-
\frac{\partial z_1}{\partial k_y}
\frac{\partial z_2}{\partial k_x}
\right)
\right|_{\mathbf{k}=\mathbf{k}_*}.
\label{eq:app_z2_Jz}
\end{equation}
For a simple zero, $J_z(\mathbf{k}_*)\neq0$, and the local winding is
\begin{equation}
\nu_{\mathbf{k}_*}
=
\operatorname{sgn}J_z(\mathbf{k}_*).
\label{eq:app_z2_local_winding}
\end{equation}
The effective Hamiltonian of Eq.~\eqref{eq:app_z2_local_weyl} has the form $\mathbf d\cdot\boldsymbol{\sigma}$ with
\begin{equation}
(d_1,d_2,d_3)
=
\left(
\frac{z_1}{2},
-\frac{z_2}{2},
-\eta
\right).
\end{equation}
Its local orientation in the coordinates
$(k_x,k_y,\lambda)$ is determined by
\begin{align}
\det
\left[
\frac{\partial(d_1,d_2,d_3)}
{\partial(k_x,k_y,\lambda)}
\right]_{\mathbf{k}=\mathbf{k}_*,\,\lambda=1/2}
&=
\frac14
\left(
\frac{\partial z_1}{\partial k_x}
\frac{\partial z_2}{\partial k_y}
-
\frac{\partial z_1}{\partial k_y}
\frac{\partial z_2}{\partial k_x}
\right)_{\mathbf{k}=\mathbf{k}_*}
\nonumber\\
&=
\frac14 J_z(\mathbf{k}_*).
\end{align}
Hence, with the orientation $(k_x,k_y,\lambda)$ used here, the orientation of the auxiliary Weyl crossing agrees with the winding of the determinant-overlap zero. This is the local bulk structure shown schematically in Fig.~\ref{fig:z2_spectral_flow}(a).

\subsection{Time-reversal pairing and the $\mathbb Z_2$ change}

The projector interpolation preserves time reversal for every value of $\lambda$,
\begin{equation}
\Theta
H(\lambda,\mathbf{k})
\Theta^{-1}
=
H(\lambda,-\mathbf{k}).
\label{eq:app_z2_interpolation_TR}
\end{equation}
As shown in Appendix~\ref{app:AII_details}, when the determinant-line bases are chosen compatibly with time reversal, the local coefficient of the comparison section satisfies
\begin{equation}
\det S(-\mathbf{k})
=
(\det S(\mathbf{k}))^{*}.
\label{eq:app_z2_detS_TR}
\end{equation}
For a generic zero $\mathbf{k}_*$ away from a
time-reversal-invariant momentum, a zero at $\mathbf{k}_*$ is
therefore accompanied by one at $-\mathbf{k}_*$. Their windings are
opposite,
\begin{equation}
\nu_{-\mathbf{k}_*}
=
-\nu_{\mathbf{k}_*}.
\label{eq:app_z2_TR_winding_pair}
\end{equation}
By Eq.~\eqref{eq:app_z2_gap_closing}, the corresponding gap closings of
the projector interpolation occur at
\begin{equation}
(\mathbf{k}_*,1/2)
\qquad\text{and}\qquad
(-\mathbf{k}_*,1/2).
\end{equation}
The signed charges of the two crossings therefore cancel when both members of the time-reversal pair are included.

We now relate these crossings to the change of the class AII invariant along the interpolation. Whenever the gap across $1/2$ is open, let $\mathcal E_\lambda$ denote the rank-two spectral bundle formed by the two eigenstates above this gap. At the endpoints,
\begin{equation}
\mathcal E_0
=
\mathcal E^\alpha,
\qquad
\mathcal E_1
=
\mathcal E^\sigma.
\end{equation}
Equation~\eqref{eq:app_z2_interpolation_TR} implies that $\mathcal E_\lambda$ is itself time-reversal symmetric whenever the gap is open, and therefore carries a class AII $\mathbb Z_2$ invariant. This invariant cannot change under a continuous variation of $\lambda$ unless the gap closes.

The spectrum derived in
Eq.~\eqref{eq:app_z2_spectrum} shows that such a closing can occur only
at
\begin{equation}
\lambda=\frac12,
\qquad
\det S(\mathbf{k})=0.
\end{equation}
Consequently, $\mathcal E_\lambda$ is continuously connected to
$\mathcal E^\alpha$ for $\lambda<1/2$ and to
$\mathcal E^\sigma$ for $\lambda>1/2$, provided no other gap closes.

The relative zero-counting result of
Eq.~\eqref{eq:relative_Z2_det_overlap} gives
\begin{equation}
\nu_{\mathbb Z_2}(\mathcal E^\alpha)
-
\nu_{\mathbb Z_2}(\mathcal E^\sigma)
=
\sum_{\mathbf{k}_*\in Z_{\alpha\sigma}^+}
\nu_{\mathbf{k}_*}
\quad
\mathrm{mod}\;2.
\label{eq:app_z2_spectral_parity}
\end{equation}
The preceding subsection showed that the local winding
$\nu_{\mathbf{k}_*}$ is also the orientation of the
auxiliary two-level crossing at
$(\mathbf{k}_*,1/2)$. Equation~\eqref{eq:app_z2_spectral_parity} therefore states that the change of the class AII invariant between the two sides of the interpolation is the parity of the oriented gap-closing charges contained in a time-reversal fundamental domain.

For generic simple zeros,
\begin{equation}
\nu_{\mathbf{k}_*}=\pm1.
\end{equation}
Since $+1$ and $-1$ differ by an even integer, each zero contributes equally to the winding parity, regardless of its sign. Therefore, Eq.~\eqref{eq:app_z2_spectral_parity} reduces to
\begin{equation}
\nu_{\mathbb Z_2}(\mathcal E^\alpha)
-
\nu_{\mathbb Z_2}(\mathcal E^\sigma)
=
\# Z_{\alpha\sigma}^+
\quad
\mathrm{mod}\;2.
\label{eq:app_z2_simple_zero_count}
\end{equation}
Thus, for simple zeros, a nontrivial relative
$\mathbb Z_2$ invariant corresponds to an odd
number of auxiliary gap closings in the chosen
time-reversal fundamental domain. For higher-order zeros, the topological quantity is instead the parity of their winding charges, as in Eq.~\eqref{eq:app_z2_spectral_parity}.

Zeros on the boundary of the chosen fundamental
domain can be handled by a small time-reversal-preserving perturbation, as discussed in Appendix~\ref{app:AII_details}. Zeros at TRIMs are excluded by requiring the overlap determinant to be nonzero at these momenta, as established previously.

\subsection{Boundary projection of the auxiliary crossings}

We finally describe how the auxiliary bulk crossings appear in the boundary parameter space used in Fig.~\ref{fig:z2_spectral_flow}(c). Consider a system that is open along $y$ and remains periodic along $x$. The bulk interpolation is parametrized by $(k_x,k_y,\lambda)$, whereas the boundary spectrum depends only on $(k_x,\lambda)$.

A bulk closing at
\begin{equation}
\left(
(k_{x})_{*},(k_{y})_{*},\frac12
\right)
\end{equation}
therefore projects to
\begin{equation}
\left(
(k_{x})_{*},\frac12
\right)
\end{equation}
in the boundary parameter space. For a generic closing away from a
time-reversal-invariant momentum, time reversal produces the partner
\begin{equation}
\left(
-(k_{x})_{*},-(k_{y})_{*},\frac12
\right),
\end{equation}
which projects to
\begin{equation}
\left(
-(k_{x})_{*},\frac12
\right).
\end{equation}
Thus the projected bulk closings occur in time-reversal-related pairs
in $(k_x,\lambda)$.

For each fixed $\lambda$ away from a bulk closing, the rank-two
spectral bundle above the gap at $1/2$ defines a two-dimensional
class AII insulator. Its $\mathbb Z_2$ index determines the parity of the corresponding helical boundary spectral flow. Consequently, when the two sides of the interpolation differ by a nontrivial relative $\mathbb Z_2$ invariant, the boundary spectrum must undergo an odd change in this parity as $\lambda$ passes through the bulk closing plane.

Time reversal acts on the auxiliary bulk parameter space as
\begin{equation}
(k_x,k_y,\lambda)
\longmapsto
(-k_x,-k_y,\lambda),
\end{equation}
and therefore on the boundary parameter space as
\begin{equation}
(k_x,\lambda)
\longmapsto
(-k_x,\lambda).
\end{equation}
The boundary branches shown schematically in
Fig.~\ref{fig:z2_spectral_flow}(c) should therefore be understood as
the time-reversal-related spectral flow associated with the change of
the two-dimensional class AII invariant.

The analogy with Fermi arcs is only geometric. The coordinate
$\lambda$ is an auxiliary interpolation parameter rather than a
physical momentum and is not periodic. The boundary branches are
therefore not Fermi arcs of a three-dimensional crystal, but a
visualization of the boundary spectral flow accompanying the relative
$\mathbb Z_2$ transition.

\section{Kane-Mele model conventions}
\label{app:kane_mele_model}

We use the Kane-Mele model on the same honeycomb lattice as the Haldane model of Appendix~\ref{app:haldane_model}. The Bloch Hamiltonian used in the numerical calculations is written in a different momentum and Bloch-basis convention. In this convention, the high-symmetry momenta relevant to the discussion are
\begin{equation}
\Gamma=(0,0),
\qquad
K=\left(\frac{4\pi}{3},0\right),
\qquad
K'=-K.
\label{eq:kane_mele_high_symmetry_points}
\end{equation}
These correspond to the same center and valleys of the honeycomb Brillouin zone as in Appendix~\ref{app:haldane_model}.
We work in the basis
\begin{equation}
\{
\ket{A,\uparrow},
\ket{A,\downarrow},
\ket{B,\uparrow},
\ket{B,\downarrow}
\},
\label{eq:kane_mele_basis}
\end{equation}
where $A,B$ label the two honeycomb sublattices and $\uparrow,\downarrow$ the spin degree of freedom. We denote by $\tau_i$ and $s_i$ the Pauli matrices acting on sublattice and spin, respectively, with $\tau_0$ and $s_0$ the corresponding identity matrices.
It is convenient to introduce the matrices
\begin{align}
\gamma_1&=\tau_x\otimes s_0,
&
\gamma_2&=\tau_z\otimes s_0,
&
\gamma_3&=\tau_y\otimes s_x,
\\
\gamma_4&=\tau_y\otimes s_y,
&
\gamma_5&=\tau_y\otimes s_z,
\end{align}
together with
\begin{equation}
\gamma_{ab}
=
\frac{1}{2i}
[\gamma_a,\gamma_b].
\end{equation}
The Bloch Hamiltonian~\cite{Kane_Mele_2005} is then written as
\begin{equation}
H_{\mathrm{KM}}(\mathbf{k})
=
\sum_{a\in\mathcal I}
d_a(\mathbf{k})\gamma_a,
\qquad
\mathcal I=\{1,2,3,4,12,15,23,24\}.
\label{eq:kane_mele_hamiltonian}
\end{equation}
Introducing the shorthand
\begin{align}
C_x&=\cos\frac{k_x}{2},
&
S_x&=\sin\frac{k_x}{2},
&
C_y&=\cos\frac{\sqrt3 k_y}{2},
&
S_y&=\sin\frac{\sqrt3 k_y}{2},
\end{align}
the nonzero coefficients in
Eq.~\eqref{eq:kane_mele_hamiltonian} are
\begin{align}
d_1(\mathbf{k})
&=
t(1+2C_xC_y),
&
d_2(\mathbf{k})
&=
\lambda_v,
\\
d_3(\mathbf{k})
&=
\lambda_R(1-C_xC_y),
&
d_4(\mathbf{k})
&=
-\sqrt3\lambda_R S_xS_y,
\\
d_{12}(\mathbf{k})
&=
-2tC_xS_y,
&
d_{15}(\mathbf{k})
&=
2\lambda_{\mathrm{SO}}
(\sin k_x-2S_xC_y),
\\
d_{23}(\mathbf{k})
&=
-\lambda_R C_xS_y,
&
d_{24}(\mathbf{k})
&=
\sqrt3\lambda_R S_xC_y.
\label{eq:kane_mele_coefficients}
\end{align}
Here $t$ is the nearest-neighbor hopping,
$\lambda_v$ is the staggered sublattice potential,
$\lambda_R$ is the Rashba spin-orbit coupling, and
$\lambda_{\mathrm{SO}}$ is the intrinsic spin-orbit coupling.

With the momenta of
Eq.~\eqref{eq:kane_mele_high_symmetry_points}, the trigonometric
factors at the two valleys are
\begin{equation}
C_x(K)=C_x(K')=-\frac12,
\qquad
S_x(K)=-S_x(K')=\frac{\sqrt3}{2},
\qquad
C_y(K)=C_y(K')=1,
\qquad
S_y(K)=S_y(K')=0.
\label{eq:kane_mele_valley_trig}
\end{equation}
In particular,
\begin{equation}
d_1(K)=d_{12}(K)=0,
\qquad
d_1(K')=d_{12}(K')=0,
\end{equation}
as expected at the two honeycomb valleys. These relations will also
be useful when analyzing the symmetry content of the occupied states.

Time reversal is represented in the basis of
Eq.~\eqref{eq:kane_mele_basis} by
\begin{equation}
\Theta
=
\tau_0\otimes i s_y\,\mathcal K,
\label{eq:kane_mele_time_reversal}
\end{equation}
where $\mathcal K$ denotes complex conjugation. It satisfies
\begin{equation}
\Theta^2=-1,
\qquad
\Theta
H_{\mathrm{KM}}(\mathbf{k})
\Theta^{-1}
=
H_{\mathrm{KM}}(-\mathbf{k}).
\end{equation}
At half filling, the two lowest-energy bands form the rank-two
occupied bundle used in Sec.~\ref{subsec:kane_mele_application}.

\subsection{Anisotropy and winding measurement}
\label{app:km_winding_numerics}

Here, we show that while time-reversal symmetry constrains the winding parity of the zeros of the comparison section, it does not fix their positions. In the Kane-Mele model, the zeros at the valleys are instead pinned by threefold rotation symmetry. Breaking this symmetry therefore allows the zeros to move away from the valleys while preserving their winding parity.

To illustrate this, we introduce an anisotropy in the nearest-neighbor hoppings of the Kane-Mele model of Eq.~\eqref{eq:kane_mele_hamiltonian}. The modified Hamiltonian is
\begin{equation}
H^\mu_\eta(\mathbf k)
=H^\mu_{\mathrm{KM}}(\mathbf k)
+t\eta_\mu\Bigg[
\left(1+0.55\cos\frac{k_x+\sqrt3 k_y}{2}\right)\gamma_1-0.55\sin\frac{k_x+\sqrt3 k_y}{2}\,\gamma_{12}
\Bigg],
\label{eq:km_anisotropic_hamiltonian}
\end{equation}
where $H^\mu_{\mathrm{KM}}$ is given by
Eq.~\eqref{eq:kane_mele_hamiltonian},
and $\eta_\mu$ controls the anisotropy in system $\mu\in\{\alpha,\sigma\}$.
The three nearest-neighbor hopping amplitudes become $t$, $t(1+\eta_\mu)$, and $t(1+0.55\eta_\mu)$.
Since these hoppings remain real, time-reversal
symmetry is preserved,
\begin{equation}
\Theta H^\mu_\eta(\mathbf k)\Theta^{-1}
=H^\mu_\eta(-\mathbf k).
\end{equation}
However, for $\eta_\mu\neq0$, the three bond amplitudes
are different, breaking threefold rotation symmetry.

For Fig.~\ref{fig:km_anisotropic_test}, we introduce the anisotropy only in system $\sigma$, taking $\eta_\alpha=0$ and $\eta_\sigma=\eta$. We keep $\lambda_{\mathrm{SO}}/t=0.06$ and $\lambda_R/t=0.05$ in both systems, with $\lambda_v^\sigma/t=0.6$, corresponding to a topologically trivial insulator. We consider two choices for system $\alpha$: $\lambda_v^\alpha/t=-0.1$, corresponding to a topological insulator ($\nu_{\mathbb Z_2}^\alpha=1$), and $\lambda_v^\alpha/t=-0.6$, corresponding to a trivial insulator ($\nu_{\mathbb Z_2}^\alpha=0$). These give relative $\mathbb Z_2$ indices of one and zero, respectively. The occupied and unoccupied bands remain separated by a finite gap throughout $0\leq\eta\leq1.55$.
We also verify numerically that the comparison section remains nonzero at all four TRIMs throughout the deformation.

As the anisotropy increases, the zeros of the
comparison section move away from the valleys
without changing their winding charges, as illustrated in Fig.~\ref{fig:km_anisotropic_test}.
We locate the zeros numerically from the
overlap determinant and determine their winding charges using a smooth local gauge.
The charges are verified by computing the
phase winding around small loops enclosing
each zero. 
At $\eta=1.55$, for the case with a nontrivial relative $\mathbb Z_2$ invariant, the comparison section has one zero at $\mathbf k_*=(4.775,0.198)$, with winding $-1$. For the case with a trivial relative invariant, it has two zeros at $(4.496,0.140)$ and $(4.957,0.209)$, each with winding $-1$. Time reversal relates each zero to another at the opposite momentum with winding $+1$, so the total winding over the full Brillouin zone vanishes in both cases.

We also investigate the annihilation of zeros
in Fig.~\ref{fig:km_pair_annihilation}.
Here, the same anisotropy is applied to both
systems, $\eta_\alpha=\eta_\sigma=\eta$, with
$\lambda_v^\alpha/t=-0.6$,
$\lambda_v^\sigma/t=0.6$,
$\lambda_{\mathrm{SO}}/t=0.06$, and
$\lambda_R/t=0.1$.
Both systems remain gapped and topologically
trivial along this deformation.
Furthermore, the comparison section remains
nonzero at all four TRIMs.
As $\eta$ increases, the two zeros of winding
$-1$ move away from $K$, together with their
time-reversed partners of winding $+1$.
At $\eta_c\simeq2.2228$, zeros of opposite
winding meet and annihilate at a generic
momentum, close to the
TRIM $M$. For $\eta>\eta_c$, the comparison
section has no zeros.
The winding in a time-reversal fundamental
domain therefore changes from an even,
nonzero value to zero, while its parity
remains unchanged.

\section{Crystalline covariance and Haldane-model details}
\label{app:crystalline_details}

In this appendix we provide the technical details underlying
Sec.~\ref{sec:crystalline_symmetries}. We first derive the gauge
transformation of the crystalline sewing matrices and the resulting
transformation of the comparison section and its local winding. We
then derive the two-band overlap identity used in the Haldane-model
application. Finally, using the lattice conventions of
Appendix~\ref{app:haldane_model}, we determine the action of $C_3$ on
the sublattice Bloch states at the rotation-invariant momenta and
derive the corresponding rotation eigenvalues.

\subsection{Crystalline covariance of the comparison section}

We start by deriving the crystalline transformation of the matrix of overlaps.
For each occupied bundle, the sewing matrix is defined by
\begin{equation}
\hat g(\mathbf{k})U^\mu(\mathbf{k})
=
U^\mu(g\mathbf{k})B_g^\mu(\mathbf{k}),
\qquad
\mu\in\{\alpha,\sigma\}.
\label{eq:app_crystalline_sewing}
\end{equation}
Since $B_g^\mu(\mathbf{k})$ is unitary,
\begin{equation}
U^\mu(g\mathbf{k})
=
\hat g(\mathbf{k})
U^\mu(\mathbf{k})
[B_g^\mu(\mathbf{k})]^\dagger.
\end{equation}
Using the matrix of overlaps
\begin{equation}
S(\mathbf{k})
=
[U^\alpha(\mathbf{k})]^\dagger
U^\sigma(\mathbf{k}),
\end{equation}
we obtain
\begin{align}
S(g\mathbf{k})
&=
[U^\alpha(g\mathbf{k})]^\dagger
U^\sigma(g\mathbf{k})
\nonumber\\
&=
B_g^\alpha(\mathbf{k})
[U^\alpha(\mathbf{k})]^\dagger
\hat g(\mathbf{k})^\dagger
\hat g(\mathbf{k})
U^\sigma(\mathbf{k})
[B_g^\sigma(\mathbf{k})]^\dagger
\nonumber\\
&=
B_g^\alpha(\mathbf{k})
S(\mathbf{k})
[B_g^\sigma(\mathbf{k})]^\dagger,
\label{eq:app_crystalline_overlap_covariance}
\end{align}
where the last line follows from the unitarity of
$\hat g(\mathbf{k})$.

Taking the determinant gives
\begin{equation}
\det S(g\mathbf{k})
=
\det B_g^\alpha(\mathbf{k})
\left[
\det B_g^\sigma(\mathbf{k})
\right]^*
\det S(\mathbf{k}),
\label{eq:app_crystalline_det_covariance}
\end{equation}
which is Eq.~\eqref{eq:crystalline_det_overlap_covariance} of the main
text.

We next verify that Eq.~\eqref{eq:app_crystalline_det_covariance} is preserved under independent changes of the occupied bases. Consider an independent change of occupied basis in each system,
\begin{equation}
U^\mu(\mathbf{k})
\longrightarrow
U'^\mu(\mathbf{k})
=
U^\mu(\mathbf{k})W^\mu(\mathbf{k}),
\qquad
W^\mu(\mathbf{k})\in U(N_{\mathrm{occ}}).
\end{equation}
We first determine how the sewing matrix changes. In the new basis it
is defined, as before, by
\begin{equation}
\hat g(\mathbf{k})U'^\mu(\mathbf{k})
=
U'^\mu(g\mathbf{k})
B_g^{\mu\,\prime}(\mathbf{k}).
\end{equation}
Substituting
$U'^\mu(\mathbf{k})=U^\mu(\mathbf{k})W^\mu(\mathbf{k})$
and
$U'^\mu(g\mathbf{k})
=
U^\mu(g\mathbf{k})W^\mu(g\mathbf{k})$
gives
\begin{equation}
\hat g(\mathbf{k})
U^\mu(\mathbf{k})W^\mu(\mathbf{k})
=
U^\mu(g\mathbf{k})
W^\mu(g\mathbf{k})
B_g^{\mu\,\prime}(\mathbf{k}).
\end{equation}
Using Eq.~\eqref{eq:app_crystalline_sewing} we obtain
\begin{equation}
U^\mu(g\mathbf{k})
B_g^\mu(\mathbf{k})W^\mu(\mathbf{k})
=
U^\mu(g\mathbf{k})
W^\mu(g\mathbf{k})
B_g^{\mu\,\prime}(\mathbf{k}).
\end{equation}
Multiplying on the left by
$[U^\mu(g\mathbf{k})]^\dagger$ and using the orthonormality of the
occupied basis yields
\begin{equation}
B_g^\mu(\mathbf{k})W^\mu(\mathbf{k})
=
W^\mu(g\mathbf{k})
B_g^{\mu\,\prime}(\mathbf{k}),
\end{equation}
and therefore
\begin{equation}
B_g^{\mu\,\prime}(\mathbf{k})
=
W^\mu(g\mathbf{k})^\dagger
B_g^\mu(\mathbf{k})
W^\mu(\mathbf{k}).
\label{eq:app_crystalline_sewing_gauge}
\end{equation}
The two basis transformations appear at different momenta because the
sewing matrix maps the occupied space at $\mathbf{k}$ to that at
$g\mathbf{k}$.

The transformation of the matrix of overlaps follows directly from its
definition. In the new bases,
\begin{align}
S'(\mathbf{k})
&=
[U'^\alpha(\mathbf{k})]^\dagger
U'^\sigma(\mathbf{k})
\nonumber\\
&=
[U^\alpha(\mathbf{k})W^\alpha(\mathbf{k})]^\dagger
U^\sigma(\mathbf{k})W^\sigma(\mathbf{k})
\nonumber\\
&=
W^\alpha(\mathbf{k})^\dagger
S(\mathbf{k})
W^\sigma(\mathbf{k}).
\label{eq:app_crystalline_overlap_gauge}
\end{align}
Taking the determinant gives
\begin{align}
\det S'(\mathbf{k})
&=
\det[W^\alpha(\mathbf{k})^\dagger]\,
\det S(\mathbf{k})\,
\det W^\sigma(\mathbf{k})
\nonumber\\
&=
\det W^\alpha(\mathbf{k})^{-1}
\det S(\mathbf{k})
\det W^\sigma(\mathbf{k}),
\label{eq:app_crystalline_det_gauge}
\end{align}
where we used the unitarity of $W^\mu$ in the last line.
Thus $\det S(\mathbf{k})$ is not itself invariant under a change of occupied basis, it is multiplied by a nonzero phase. Its zeros, however, are unchanged. This is the same transformation law discussed in Sec.~\ref{sec:singular_connections}, where $\det S$ appears as the local coefficient of the gauge-independent comparison section.

We now verify that the crystalline covariance relation,
Eq.~\eqref{eq:app_crystalline_det_covariance},
remains valid under changes of occupied bases.
From Eq.~\eqref{eq:app_crystalline_sewing_gauge},
\begin{align}
\det B_g^{\alpha\,\prime}(\mathbf{k})
&=
\det W^\alpha(g\mathbf{k})^{-1}
\det B_g^\alpha(\mathbf{k})
\det W^\alpha(\mathbf{k}),
\\
\left[
\det B_g^{\sigma\,\prime}(\mathbf{k})
\right]^*
&=
\det W^\sigma(g\mathbf{k})
\left[
\det B_g^\sigma(\mathbf{k})
\right]^*
\det W^\sigma(\mathbf{k})^{-1}.
\end{align}
Combining these expressions with
Eq.~\eqref{eq:app_crystalline_det_gauge},
we obtain
\begin{align}
&\det B_g^{\alpha\,\prime}(\mathbf{k})
\left[
\det B_g^{\sigma\,\prime}(\mathbf{k})
\right]^*
\det S'(\mathbf{k})
\nonumber\\
&=
\left[
\det W^\alpha(g\mathbf{k})^{-1}
\det B_g^\alpha(\mathbf{k})
\det W^\alpha(\mathbf{k})
\right]\times
\left[
\det W^\sigma(g\mathbf{k})
\left[\det B_g^\sigma(\mathbf{k})\right]^*
\det W^\sigma(\mathbf{k})^{-1}
\right]
\nonumber\\
&\quad\times
\left[
\det W^\alpha(\mathbf{k})^{-1}
\det S(\mathbf{k})
\det W^\sigma(\mathbf{k})
\right]
\nonumber\\
&=
\det W^\alpha(g\mathbf{k})^{-1}
\det W^\sigma(g\mathbf{k})
\det B_g^\alpha(\mathbf{k})
\left[\det B_g^\sigma(\mathbf{k})\right]^*
\det S(\mathbf{k})
\nonumber\\
&=
\det W^\alpha(g\mathbf{k})^{-1}
\det W^\sigma(g\mathbf{k})
\det S(g\mathbf{k})
\nonumber\\
&=
\det S'(g\mathbf{k}).
\end{align}
In the second equality, the gauge factors at
$\mathbf{k}$ cancel. The third equality follows
from the original crystalline covariance relation,
Eq.~\eqref{eq:app_crystalline_det_covariance},
while the last follows from
Eq.~\eqref{eq:app_crystalline_det_gauge}.
Thus, both sides transform identically, and
the crystalline covariance relation holds
independently of the choice of occupied bases.

\subsection{Transformation of the local winding}

We now derive how the winding of an isolated zero of the comparison section transforms
under the crystalline symmetry. Let $\mathbf{k}_*$ be an isolated zero
of $\det S(\mathbf{k})$, and let $C_{\mathbf{k}_*}$ be a sufficiently
small positively oriented loop around it. The symmetry maps
$\mathbf{k}_*$ to $g\mathbf{k}_*$ and maps
$C_{\mathbf{k}_*}$ to a loop surrounding the symmetry-related zero.

Let $R_g$ denote the action of $g$ on momentum space. If
$\det R_g>0$, the orientation of the loop is preserved, whereas if
$\det R_g<0$, it is reversed. Pulling the winding integral around
$g\mathbf{k}_*$ back to the original loop therefore gives
\begin{equation}
\nu_{g\mathbf{k}_*}
=
\operatorname{sgn}(\det R_g)
\frac{1}{2\pi i}
\oint_{C_{\mathbf{k}_*}}
\mathrm d
\log
\det S(g\mathbf{k}).
\label{eq:app_crystalline_winding_pullback}
\end{equation}
The factor $\operatorname{sgn}(\det R_g)$ accounts for the possible
reversal of the loop orientation under the momentum-space action of
$g$.

Using the crystalline transformation derived in
Eq.~\eqref{eq:app_crystalline_det_covariance}, we obtain
\begin{equation}
\mathrm d\log\det S(g\mathbf{k})
=
\mathrm d\log\det S(\mathbf{k})
+
\mathrm d\log
\left\{
\det B_g^\alpha(\mathbf{k})
\left[
\det B_g^\sigma(\mathbf{k})
\right]^*
\right\}.
\label{eq:app_crystalline_winding_decomposition}
\end{equation}
The second term does not contribute to the local winding. Indeed, in a
sufficiently small neighborhood of $\mathbf{k}_*$ the occupied bases
can be chosen smoothly, so the corresponding sewing matrices are
smooth functions of momentum. Since the sewing matrices are unitary,
their determinants never vanish. Therefore
\begin{equation}
f_g(\mathbf{k})
=
\det B_g^\alpha(\mathbf{k})
\left[
\det B_g^\sigma(\mathbf{k})
\right]^*
\end{equation}
is smooth and nonzero throughout the region enclosed by
$C_{\mathbf{k}_*}$. Its phase consequently has zero winding around the
loop,
\begin{equation}
\frac{1}{2\pi i}
\oint_{C_{\mathbf{k}_*}}
\mathrm d\log f_g(\mathbf{k})
=
0.
\end{equation}
Substituting this result into
Eq.~\eqref{eq:app_crystalline_winding_pullback} leaves only the winding
of the original zero of the comparison section,
\begin{equation}
\nu_{g\mathbf{k}_*}
=
\operatorname{sgn}(\det R_g)
\nu_{\mathbf{k}_*},
\end{equation}
which is Eq.~\eqref{eq:crystalline_winding_transformation} of the main
text. Thus orientation-preserving crystalline operations relate zeros
with the same winding, whereas orientation-reversing operations relate
zeros with opposite winding.

\subsection{Rotation eigenvalues and winding at invariant momenta}

We now derive Eq.~\eqref{eq:crystalline_winding_eigenvalues}.
Let $\mathbf{k}_\star$ be invariant under a proper rotation
$g$ of order $n$, through a signed angle $\phi_g$,
taken positive for counterclockwise rotations.
Near $\mathbf{k}_\star$, define
\begin{equation}
f(\mathbf q)
=
\det S(\mathbf{k}_\star+\mathbf q),
\qquad
\Lambda_g(\mathbf{k})
=
\det B_g^\alpha(\mathbf{k})
\left[\det B_g^\sigma(\mathbf{k})\right]^*.
\end{equation}
With the reciprocal-lattice identification understood,
the covariance law
Eq.~\eqref{eq:app_crystalline_det_covariance} gives
\begin{equation}
f(R_g\mathbf q)
=
\Lambda_g(\mathbf{k}_\star+\mathbf q)f(\mathbf q).
\label{eq:app_rotation_local_covariance}
\end{equation}

Suppose $f$ has an isolated zero at $\mathbf q=0$.
On a small circle around the zero, we can write
$f=|f|e^{i\vartheta_r(\theta)}$, where $\theta$
is the angular coordinate. By definition of the winding,
\begin{equation}
\vartheta_r(\theta+2\pi)-\vartheta_r(\theta)
=
2\pi\nu_{\mathbf{k}_\star}.
\end{equation}
A rotation through $\phi_g$ takes $\theta$ to
$\theta+\phi_g$. Since the sewing matrices are
unitary, $|\Lambda_g(\mathbf{k})|=1$, and
Eq.~\eqref{eq:app_rotation_local_covariance}
implies $|f(R_g\mathbf q)|=|f(\mathbf q)|$.
Comparing the phases on both sides of the
same equation, we obtain
\begin{equation}
e^{i[\vartheta_r(\theta+\phi_g)-\vartheta_r(\theta)]}
=
\Lambda_g(\mathbf{k}_\star+\mathbf q).
\end{equation}

Since $\Lambda_g$ is smooth and nonvanishing,
its phase approaches a constant as the circle shrinks
toward $\mathbf{k}_\star$. Thus, in the limit
$r\to0$, each application of the rotation produces
the same continuous phase increment, which we denote by $\Delta$.
Since $g$ is a rotation of order $n$ (for example,
$n=3$ for $C_3$), applying it $n$ times completes
an integer number of full revolutions around the zero.
The total phase change is then $n\Delta$. By the
definition of the winding number, this must equal
$n\phi_g\nu_{\mathbf{k}_\star}$, giving
\begin{equation}
n\Delta=n\phi_g\nu_{\mathbf{k}_\star}.
\end{equation}
Since $e^{i\Delta}=\Lambda_g(\mathbf{k}_\star)$,
we obtain
\begin{equation}
e^{i\phi_g\nu_{\mathbf{k}_\star}}
=
\det B_g^\alpha(\mathbf{k}_\star)
\left[\det B_g^\sigma(\mathbf{k}_\star)\right]^*,
\label{app:eq:crystalline_winding_eigenvalues}
\end{equation}
which is Eq.~\eqref{eq:crystalline_winding_eigenvalues}.

At the rotation-invariant momentum $\mathbf{k}_\star$,
the sewing matrix represents the action of the
rotation operator within the occupied subspace,
\begin{equation}
\hat g\ket{\psi_j^\mu(\mathbf{k}_\star)}
=
\sum_i
\ket{\psi_i^\mu(\mathbf{k}_\star)}
[B_g^\mu(\mathbf{k}_\star)]_{ij},
\end{equation}
where the reciprocal-lattice identification is
understood. We can therefore choose the occupied
basis so that the sewing matrix is diagonal,
\begin{equation}
B_g^\mu(\mathbf{k}_\star)
=
\operatorname{diag}
\bigl(
\xi_{g,1}^\mu,\ldots,
\xi_{g,N_{\mathrm{occ}}}^\mu
\bigr),
\end{equation}
where $\xi_{g,j}^\mu$ are the rotation eigenvalues
of the occupied states. Consequently,
\begin{equation}
\det B_g^\mu(\mathbf{k}_\star)
=
\prod_{j=1}^{N_{\mathrm{occ}}}\xi_{g,j}^\mu.
\end{equation}
Equation~\eqref{app:eq:crystalline_winding_eigenvalues}
can thus be written as
\begin{equation}
e^{i\phi_g\nu_{\mathbf{k}_\star}}
=
\prod_{j=1}^{N_{\mathrm{occ}}}
\xi_{g,j}^\alpha
\left[\xi_{g,j}^\sigma\right]^*.
\end{equation}
Since $g$ is a rotation of order $n$,
the left-hand side is unchanged under
$\nu_{\mathbf{k}_\star}\to\nu_{\mathbf{k}_\star}+n$.
The rotation eigenvalues therefore constrain
the winding modulo $n$, rather than determining
its full integer value.

\subsection{Valence-band overlap in the Haldane model}

We now derive the overlap identity used in Sec.~\ref{subsec:c3_haldane}. For the Haldane Hamiltonian of Eq.~\eqref{eq:Haldane_Hamiltonian}, define
\begin{equation}
\hat{\mathbf d}^\mu(\mathbf{k})
=
\frac{\mathbf d^\mu(\mathbf{k})}
{|\mathbf d^\mu(\mathbf{k})|}.
\end{equation}
The projector onto the valence band is
\begin{equation}
P_-^\mu(\mathbf{k})
=
\frac12
\left[
\mathbb 1
-
\hat{\mathbf d}^\mu(\mathbf{k})
\cdot\boldsymbol{\sigma}
\right].
\end{equation}
For a normalized valence eigenstate
$\ket{\psi^\mu(\mathbf{k})}$,
\begin{equation}
P_-^\mu(\mathbf{k})
=
\ket{\psi^\mu(\mathbf{k})}
\bra{\psi^\mu(\mathbf{k})}.
\end{equation}
Since there is a single occupied band, the matrix of overlaps reduces to the scalar
\begin{equation}
S(\mathbf{k})
=
\braket{
\psi^\alpha(\mathbf{k})
|
\psi^\sigma(\mathbf{k})
}.
\end{equation}
Its magnitude can be written in terms of the two projectors as
\begin{align}
|S(\mathbf{k})|^2
&=
\braket{\psi^\alpha|\psi^\sigma}
\braket{\psi^\sigma|\psi^\alpha}
\nonumber\\
&=
\operatorname{Tr}
\left[
P_-^\alpha(\mathbf{k})
P_-^\sigma(\mathbf{k})
\right].
\end{align}
Substituting the projector expressions gives
\begin{equation}
|S(\mathbf{k})|^2
=
\frac14
\operatorname{Tr}
\Big[ \mathbb 1 - \hat{\mathbf d}^\alpha\cdot\boldsymbol{\sigma} - \hat{\mathbf d}^\sigma\cdot\boldsymbol{\sigma}+(\hat{\mathbf d}^\alpha\cdot\boldsymbol{\sigma})(\hat{\mathbf d}^\sigma\cdot\boldsymbol{\sigma})\Big].
\end{equation}
Using
\begin{equation}
\operatorname{Tr}\sigma_i=0,
\qquad
\operatorname{Tr}(\sigma_i\sigma_j) = 2\delta_{ij},
\end{equation}
we obtain
\begin{equation}
|S(\mathbf{k})|^2 = \frac{1+ \hat{\mathbf d}^\alpha(\mathbf{k}) \cdot \hat{\mathbf d}^\sigma(\mathbf{k})}{2},
\label{eq:app_haldane_overlap_bloch_vectors}
\end{equation}
which gives Eq.~\eqref{eq:haldane_overlap_bloch_vectors} of the main text. The overlap therefore vanishes if and only if the two normalized Bloch vectors are antiparallel,
\begin{equation}
\hat{\mathbf d}^\alpha(\mathbf{k})
=
-
\hat{\mathbf d}^\sigma(\mathbf{k}).
\label{eq:app_haldane_antiparallel}
\end{equation}
We now apply this condition to the two Hamiltonians considered in Fig.~\ref{fig:overlap_haldane}. From Eq.~\eqref{eq:haldane_d_components}, $d_1$ and $d_2$ depend only on $t_1$. Since the two Hamiltonians have the same $t_1$, 
\begin{equation}
d_1^\alpha(\mathbf{k})
=
d_1^\sigma(\mathbf{k}),
\qquad
d_2^\alpha(\mathbf{k})
=
d_2^\sigma(\mathbf{k}).
\end{equation}
Equation~\eqref{eq:app_haldane_antiparallel} is equivalent to
\begin{equation}
\mathbf d^\alpha(\mathbf{k})
=
-c(\mathbf{k})
\mathbf d^\sigma(\mathbf{k}),
\qquad
c(\mathbf{k})>0.
\end{equation}
Comparing the first two components gives
\begin{equation}
[1+c(\mathbf{k})]d_1(\mathbf{k})=0,
\qquad
[1+c(\mathbf{k})]d_2(\mathbf{k})=0.
\end{equation}
Since $c(\mathbf{k})>0$, a zero of the overlap therefore requires
\begin{equation}
d_1(\mathbf{k})=d_2(\mathbf{k})=0.
\end{equation}
For the honeycomb lattice, these two conditions are satisfied only at
the two inequivalent valleys $K$ and $K'$, modulo reciprocal lattice
vectors.

Using Eq.~\eqref{eq:haldane_d_components} at the two valleys gives
\begin{equation}
d_3^\mu(K)
=
M-3\sqrt3\,t_2^\mu,
\qquad
d_3^\mu(K')
=
M+3\sqrt3\,t_2^\mu.
\end{equation}
For the topological Hamiltonian,
$M=0.8$ and $t_2^\alpha=0.2$, giving
\begin{align}
d_3^\alpha(K)
&=
0.8-3\sqrt3(0.2)
\simeq -0.239,
\\
d_3^\alpha(K')
&=
0.8+3\sqrt3(0.2)
\simeq 1.839.
\end{align}
For the trivial Hamiltonian,
$t_2^\sigma=0.02$ at the same $M$, giving
\begin{align}
d_3^\sigma(K)
&=
0.8-3\sqrt3(0.02)
\simeq 0.696,
\\
d_3^\sigma(K')
&=
0.8+3\sqrt3(0.02)
\simeq 0.904.
\end{align}
Thus $d_3^\alpha(K)$ and $d_3^\sigma(K)$ have opposite signs, while
$d_3^\alpha(K')$ and $d_3^\sigma(K')$ have the same sign.

Since $d_1=d_2=0$ at both valleys, the Hamiltonian there is
proportional to $\sigma_3$. With
$\sigma_3\ket{A}=+\ket{A}$ and
$\sigma_3\ket{B}=-\ket{B}$, opposite signs of $d_3$ correspond to
opposite occupied sublattice states, whereas equal signs correspond to
the same occupied sublattice state. Consequently,
\begin{equation}
S(K)=0,
\qquad
|S(K')|=1.
\end{equation}

Finally, Eq.~\eqref{eq:haldane_d_components} shows that the
$t_2$-dependent contribution to $d_3$ vanishes at $\Gamma$.
Since the two Hamiltonians also have the same $t_1$ and $M$, their
Bloch vectors coincide there, and hence
\begin{equation}
|S(\Gamma)|=1.
\end{equation}

\subsection{Threefold-rotation action on the Haldane Bloch states}

We now derive the $C_3$ representation used in
Sec.~\ref{subsec:c3_haldane}. We use the lattice conventions of
Appendix~\ref{app:haldane_model} and choose an $A$ site as the center
of rotation. The two sites within each unit cell are taken at
\begin{equation}
\boldsymbol{\tau}_A=0,
\qquad
\boldsymbol{\tau}_B=\mathbf a_1,
\end{equation}
where $\mathbf a_1$ is one of the nearest-neighbor vectors defined in
Eq.~\eqref{eq:haldane_nn_vectors}.

For each sublattice $\ell\in\{A,B\}$, we define the Bloch basis state
\begin{equation}
\ket{\ell,\mathbf k}
=
\frac{1}{\sqrt{N}}
\sum_{\mathbf R}
e^{i\mathbf k\cdot(\mathbf R+\boldsymbol{\tau}_\ell)}
\ket{\mathbf R+\boldsymbol{\tau}_\ell},
\label{eq:app_haldane_bloch_basis}
\end{equation}
where $\mathbf R$ runs over the Bravais lattice. The phase therefore
contains the position of the site within the unit cell, in addition to
the Bravais-lattice vector.

We represent $C_3$ by a clockwise rotation through $2\pi/3$,
\begin{equation}
R_3
=
\begin{pmatrix}
-\frac12 & \frac{\sqrt3}{2}\\[1mm]
-\frac{\sqrt3}{2} & -\frac12
\end{pmatrix}.
\label{eq:app_haldane_R3}
\end{equation}
With the nearest-neighbor vectors of
Eq.~\eqref{eq:haldane_nn_vectors}, this rotation acts as
\begin{equation}
\mathbf a_1
\longrightarrow
\mathbf a_2
\longrightarrow
\mathbf a_3
\longrightarrow
\mathbf a_1.
\label{eq:app_haldane_C3_nn_action}
\end{equation}
It therefore preserves the honeycomb lattice and maps each sublattice
onto itself.

For the spinless orbital considered here, the on-site orbital carries
no additional $C_3$ phase. Acting on
Eq.~\eqref{eq:app_haldane_bloch_basis} gives
\begin{align}
\hat C_3\ket{\ell,\mathbf k}
&=
\frac{1}{\sqrt{N}}
\sum_{\mathbf R}
e^{i\mathbf k\cdot(\mathbf R+\boldsymbol{\tau}_\ell)}
\ket{
R_3\mathbf R+R_3\boldsymbol{\tau}_\ell
}
\nonumber\\
&=
\ket{\ell,R_3\mathbf k},
\label{eq:app_haldane_C3_bloch_action}
\end{align}
where in the second line the Bravais-lattice sum has been relabeled.
Thus, between the fibers at $\mathbf k$ and $R_3\mathbf k$, the matrix
representing $C_3$ in the $(A,B)$ Bloch basis is simply the identity. This does not imply that the $C_3$ eigenvalues at a rotation-invariant momentum are all equal to one. Such a momentum need
only satisfy
\begin{equation}
R_3\mathbf k_\star
=
\mathbf k_\star+\mathbf G
\end{equation}
for some reciprocal lattice vector $\mathbf G$. In the Bloch basis of
Eq.~\eqref{eq:app_haldane_bloch_basis}, momenta differing by a
reciprocal lattice vector obey
\begin{equation}
\ket{\ell,\mathbf k+\mathbf G}
=
e^{i\mathbf G\cdot\boldsymbol{\tau}_\ell}
\ket{\ell,\mathbf k}.
\label{eq:app_haldane_bloch_reciprocal_shift}
\end{equation}
The identification of the two Bloch fibers therefore introduces a
sublattice-dependent phase.

For the high-symmetry momenta defined in
Eq.~\eqref{eq:haldane_high_symmetry_points}, the rotation acts as
\begin{equation}
R_3K
=
K-\mathbf G_1,
\qquad
R_3K'
=
K'+\mathbf G_1,
\qquad
R_3\Gamma
=
\Gamma,
\label{eq:app_haldane_C3_high_symmetry}
\end{equation}
where $\mathbf G_1$ is the reciprocal vector defined in
Eq.~\eqref{eq:haldane_reciprocal_vectors}. Since
\begin{equation}
\mathbf G_1\cdot\boldsymbol{\tau}_A=0,
\qquad
\mathbf G_1\cdot\boldsymbol{\tau}_B
=
\mathbf G_1\cdot\mathbf a_1
=
\frac{2\pi}{3},
\end{equation}
it is convenient to define
\begin{equation}
\omega=e^{2\pi i/3}.
\end{equation}
Equations~\eqref{eq:app_haldane_C3_bloch_action} and
\eqref{eq:app_haldane_bloch_reciprocal_shift} then give
\begin{align}
\hat C_3(K)\ket{A,K}
&=
\ket{A,K},
&
\hat C_3(K)\ket{B,K}
&=
\omega^{-1}\ket{B,K},
\label{eq:app_haldane_C3_K}
\\
\hat C_3(K')\ket{A,K'}
&=
\ket{A,K'},
&
\hat C_3(K')\ket{B,K'}
&=
\omega\ket{B,K'},
\label{eq:app_haldane_C3_Kp}
\\
\hat C_3(\Gamma)\ket{A,\Gamma}
&=
\ket{A,\Gamma},
&
\hat C_3(\Gamma)\ket{B,\Gamma}
&=
\ket{B,\Gamma}.
\label{eq:app_haldane_C3_Gamma}
\end{align}

Thus, after identifying the symmetry-related Bloch fibers, the
rotation operator is represented in the $(A,B)$ basis by
\begin{equation}
\hat C_3(K)
=
\begin{pmatrix}
1&0\\
0&\omega^{-1}
\end{pmatrix},
\qquad
\hat C_3(K')
=
\begin{pmatrix}
1&0\\
0&\omega
\end{pmatrix},
\qquad
\hat C_3(\Gamma)
=
\mathbb 1.
\label{eq:app_haldane_C3_matrices}
\end{equation}
We now determine the $C_3$ eigenvalues of the valence states. At $K$
and $K'$, the two sublattice states carry different rotation
eigenvalues. The symmetry condition on the Hamiltonian therefore
forbids a matrix element between them at these momenta. Equivalently,
$d_1=d_2=0$, as found directly from the Haldane Hamiltonian in the
preceding subsection, and
\begin{equation}
H^\mu(K)
=
d_3^\mu(K)\sigma_3,
\qquad
H^\mu(K')
=
d_3^\mu(K')\sigma_3.
\label{eq:app_haldane_valley_hamiltonian}
\end{equation}
With
\begin{equation}
\sigma_3\ket{A}=+\ket{A},
\qquad
\sigma_3\ket{B}=-\ket{B},
\end{equation}
the valence state at either valley is therefore
\begin{equation}
\ket{\psi^\mu_-(\mathbf k_\star)}
=
\begin{cases}
\ket{A,\mathbf k_\star},
&
d_3^\mu(\mathbf k_\star)<0,
\\[1mm]
\ket{B,\mathbf k_\star},
&
d_3^\mu(\mathbf k_\star)>0,
\end{cases}
\qquad
\mathbf k_\star\in\{K,K'\}.
\label{eq:app_haldane_valence_sublattice}
\end{equation}
The sign of $d_3$ thus selects the occupied sublattice, while the
$C_3$ representation of that sublattice fixes the rotation eigenvalue
of the valence state.
At $K$, the values obtained in the preceding subsection satisfy
\begin{equation}
d_3^\alpha(K)<0,
\qquad
d_3^\sigma(K)>0.
\end{equation}
Hence
\begin{equation}
\ket{\psi_-^\alpha(K)}
=
\ket{A,K},
\qquad
\ket{\psi_-^\sigma(K)}
=
\ket{B,K},
\end{equation}
and their rotation eigenvalues are
\begin{equation}
\xi_{C_3}^\alpha(K)=1,
\qquad
\xi_{C_3}^\sigma(K)=\omega^{-1}.
\label{eq:app_haldane_valence_C3_K}
\end{equation}
The two occupied states therefore carry different $C_3$
representations. Equation~\eqref{eq:crystalline_eigenvalue_zero} then
requires
\begin{equation}
S(K)=0.
\end{equation}
These rotation eigenvalues also constrain the winding
of the overlap determinant around this zero.
For the clockwise $C_3$ rotation,
$\phi_{C_3}=-2\pi/3$. Using
Eq.~\eqref{app:eq:crystalline_winding_eigenvalues}
and the rotation eigenvalues obtained above, we find
\begin{equation}
e^{-2\pi i\nu_K/3}
=
\xi_{C_3}^\alpha(K)
\left[\xi_{C_3}^\sigma(K)\right]^*
=
\omega.
\end{equation}
Hence $\nu_K\equiv-1\pmod 3$. For the single
simple zero found in the Haldane example, this
gives $\nu_K=-1$, consistent with the total
winding $C_\alpha-C_\sigma=-1$.
At $K'$, by contrast,
\begin{equation}
d_3^\alpha(K')>0,
\qquad
d_3^\sigma(K')>0,
\end{equation}
so both valence states belong to the $B$ sublattice,
\begin{equation}
\ket{\psi_-^\alpha(K')}
=
\ket{\psi_-^\sigma(K')}
=
\ket{B,K'}.
\end{equation}
They consequently have the same rotation eigenvalue,
\begin{equation}
\xi_{C_3}^\alpha(K')
=
\xi_{C_3}^\sigma(K')
=
\omega.
\end{equation}
Crystalline symmetry therefore does not force the overlap to vanish at
$K'$. For these two parameter regimes, the states in fact coincide
there, giving $|S(K')|=1$.

Finally, at $\Gamma$ the rotation operator is the identity in the
sublattice basis. Both valence states therefore carry $C_3$ eigenvalue
one. Moreover, as shown in the preceding subsection,
$H^\alpha(\Gamma)=H^\sigma(\Gamma)$, so the two valence states
coincide and
\begin{equation}
|S(\Gamma)|=1.
\end{equation}

The results are summarized in Table~\ref{tab:haldane_C3_summary}.
\begin{table}[t]
\centering
\begin{tabular}{c|c|c|c|c}
\hline
$\mathbf k_\star$
&
$\hat C_3(\mathbf k_\star)$
&
occupied states
&
$(\xi^\alpha,\xi^\sigma)$
&
$|S|$
\\
\hline
$K$
&
$\mathrm{diag}(1,\omega^{-1})$
&
$\alpha:A,\ \sigma:B$
&
$(1,\omega^{-1})$
&
$0$
\\
$K'$
&
$\mathrm{diag}(1,\omega)$
&
$\alpha:B,\ \sigma:B$
&
$(\omega,\omega)$
&
$1$
\\
$\Gamma$
&
$\mathbb 1$
&
$\psi_-^\alpha=\psi_-^\sigma$
&
$(1,1)$
&
$1$
\\
\hline
\end{tabular}
\caption{
Threefold-rotation representations of the Haldane valence states for
the two parameter regimes shown in Fig.~\ref{fig:overlap_haldane}. At $K$, the two
occupied states belong to different sublattices and carry different
$C_3$ eigenvalues, enforcing the zero of the comparison section. At $K'$ and
$\Gamma$, they carry the same $C_3$ eigenvalue and
the overlap remains nonzero.
}
\label{tab:haldane_C3_summary}
\end{table}

The explicit phases in
Eq.~\eqref{eq:app_haldane_C3_matrices} depend on the choice of rotation
center and Bloch convention. What is relevant here is the relative
symmetry content: at $K$ the two occupied states carry different
$C_3$ eigenvalues, whereas at $K'$ and $\Gamma$ they carry the same one.

\subsection{Threefold-rotation constraints on the Kane-Mele zeros}
\label{app:kane_mele_c3}

In the Kane-Mele application of
Sec.~\ref{subsec:kane_mele_application}, the reference determinant,
the Pfaffian, and the relative determinant overlap all have a zero at
the valley $K$ in the chosen time-reversal fundamental domain. The
class AII invariant fixes the parity of the total winding of these
zeros, but does not by itself determine their positions. We now show
that, for the parameters used there, the valley zeros are enforced by
the threefold rotation symmetry of the model.

We use the same $A$-centered clockwise rotation as in
Eq.~\eqref{eq:app_haldane_R3}. In the Kane-Mele
lattice convention, the sublattice positions within
the unit cell can be chosen as
$\boldsymbol\tau_A=0$ and
$\boldsymbol\tau_B=(0,1/\sqrt3)$.
Under this rotation,
\begin{equation}
R_3\boldsymbol\tau_A=\boldsymbol\tau_A,
\qquad
R_3\boldsymbol\tau_B
=
\boldsymbol\tau_B+\mathbf t_B,
\end{equation}
where $\mathbf t_B=(1/2,-\sqrt3/2)$ is a Bravais-lattice vector.

In the periodic Bloch convention, the sublattice basis states are
\begin{equation}
\ket{\ell,\mathbf{k}}
=
\frac{1}{\sqrt{N}}
\sum_{\mathbf R}
e^{i\mathbf{k}\cdot\mathbf R}
\ket{\mathbf R+\boldsymbol\tau_\ell},
\qquad \ell=A,B.
\end{equation}
The translation of the $B$ site under rotation
therefore introduces a momentum-dependent phase,
\begin{equation}
\begin{aligned}
\hat C_3\ket{A,\mathbf{k}}
&=
\ket{A,R_3\mathbf{k}},\\
\hat C_3\ket{B,\mathbf{k}}
&=
e^{-i(R_3\mathbf{k})\cdot\mathbf t_B}
\ket{B,R_3\mathbf{k}}.
\end{aligned}
\end{equation}
Thus, the spinless sublattice rotation
representation is
\begin{equation}
\hat C_3^{\mathrm{orb}}(\mathbf{k})
=
\operatorname{diag}
\left(
1,e^{-i(R_3\mathbf{k})\cdot\mathbf t_B}
\right).
\end{equation}
At $K$, $K'$, and $\Gamma$, this reproduces
the rotation eigenvalues of
Eq.~\eqref{eq:app_haldane_C3_matrices}.

The Kane-Mele model also contains spin.
For a clockwise rotation through $2\pi/3$,
the spin part is
\begin{equation}
U_{\mathrm{spin}}(C_3)
=
e^{i\pi s_z/3}.
\end{equation}
Defining
\begin{equation}
\eta_+
=
e^{i\pi/3},
\qquad
\eta_-
=
e^{-i\pi/3},
\end{equation}

The full spinful rotation operator is the tensor
product of the orbital and spin rotations.
In the basis of Eq.~\eqref{eq:kane_mele_basis},
$\{A\uparrow,A\downarrow,B\uparrow,B\downarrow\}$,
we therefore obtain
\begin{align}
\hat C_3(K)
&=
\hat C_3^{\mathrm{orb}}(K)
\otimes U_{\mathrm{spin}}(C_3)
\nonumber\\
&=
\operatorname{diag}
\left(
\eta_+,\eta_-,
\omega^{-1}\eta_+,\omega^{-1}\eta_-
\right)
\nonumber\\
&=
\operatorname{diag}
\left(
\eta_+,\eta_-,\eta_-,-1
\right),
\label{eq:app_km_C3_K}
\end{align}

\begin{align}
\hat C_3(K')
&=
\hat C_3^{\mathrm{orb}}(K')
\otimes U_{\mathrm{spin}}(C_3)
\nonumber\\
&=
\operatorname{diag}
\left(
\eta_+,\eta_-,
\omega\eta_+,\omega\eta_-
\right)
\nonumber\\
&=
\operatorname{diag}
\left(
\eta_+,\eta_-,-1,\eta_+
\right),
\label{eq:app_km_C3_Kp}
\end{align}

\begin{align}
\hat C_3(\Gamma)
&=
\hat C_3^{\mathrm{orb}}(\Gamma)
\otimes U_{\mathrm{spin}}(C_3)
\nonumber\\
&=
\operatorname{diag}
\left(
\eta_+,\eta_-,\eta_+,\eta_-
\right).
\label{eq:app_km_C3_Gamma}
\end{align}
In particular, at $K$ the states
$\ket{A,\downarrow}$ and $\ket{B,\uparrow}$ carry the same rotation
eigenvalue $\eta_-$, while at $K'$ the states
$\ket{A,\uparrow}$ and $\ket{B,\downarrow}$ carry the same eigenvalue
$\eta_+$. The spinful matrices satisfy
$\hat C_3^3=-\mathbb 1$, as expected for spin-$1/2$ states.

We next determine the representation matrices obtained by restricting
this action to the occupied subspace. 
Using Eqs.~\eqref{eq:kane_mele_hamiltonian} and
\eqref{eq:kane_mele_coefficients}, the Hamiltonian at $K$ becomes
\begin{equation}
H_{\mathrm{KM}}(K)
=
\begin{pmatrix}
\lambda_v-\Delta & 0 & 0 & 0\\
0 & \lambda_v+\Delta & -3i\lambda_R & 0\\
0 & 3i\lambda_R & -\lambda_v+\Delta & 0\\
0 & 0 & 0 & -\lambda_v-\Delta
\end{pmatrix},
\label{eq:app_km_H_K}
\end{equation}
with $\Delta
=
3\sqrt3\,\lambda_{\mathrm{SO}}$.
The block structure agrees with
Eq.~\eqref{eq:app_km_C3_K}: the Rashba coupling mixes
$\ket{A,\downarrow}$ and $\ket{B,\uparrow}$, which belong to the same
$\eta_-$ sector, while
$\ket{A,\uparrow}$ and $\ket{B,\downarrow}$ form one-dimensional
sectors with eigenvalues $\eta_+$ and $-1$, respectively.

The corresponding energies are
\begin{equation}
E_{A\uparrow}
=
\lambda_v-\Delta,
\qquad
E_{B\downarrow}
=
-\lambda_v-\Delta,
\end{equation}
together with
\begin{equation}
E_{\eta_-,\pm}
=
\Delta
\pm
\sqrt{\lambda_v^2+9\lambda_R^2}
\label{eq:app_km_K_energies}
\end{equation}
for the two states in the $\eta_-$ sector.

For
$\lambda_R/t=0.05$ and
$\lambda_{\mathrm{SO}}/t=0.06$,
\begin{equation}
\frac{\Delta}{t}
=
3\sqrt3(0.06)
\simeq 0.312.
\end{equation}
For either of the topological parameter values
$\lambda_v/t=\pm0.1$ used in
Sec.~\ref{subsec:kane_mele_application}, the two one-dimensional
levels in Eq.~\eqref{eq:app_km_H_K} are occupied, while both states of
the $\eta_-$ block lie above them. Hence
\begin{equation}
\mathcal E^{\mathrm{top}}_K
=
\operatorname{span}
\left\{
\ket{A,\uparrow},
\ket{B,\downarrow}
\right\}.
\end{equation}
In this occupied basis, the $C_3$ sewing matrix is
\begin{equation}
B_{C_3}^{\mathrm{top}}(K)
=
\begin{pmatrix}
\eta_+&0\\
0&-1
\end{pmatrix}.
\label{eq:app_km_topological_B_K}
\end{equation}

At $K'$, time reversal exchanges the spin sectors. Explicitly,
\begin{equation}
H_{\mathrm{KM}}(K')
=
\begin{pmatrix}
\lambda_v+\Delta & 0 & 0 & -3i\lambda_R\\
0 & \lambda_v-\Delta & 0 & 0\\
0 & 0 & -\lambda_v-\Delta & 0\\
3i\lambda_R & 0 & 0 & -\lambda_v+\Delta
\end{pmatrix}.
\label{eq:app_km_H_Kp}
\end{equation}
The occupied subspace in the topological phase is therefore
\begin{equation}
\mathcal E^{\mathrm{top}}_{K'}
=
\operatorname{span}
\left\{
\ket{A,\downarrow},
\ket{B,\uparrow}
\right\},
\end{equation}
and the corresponding sewing matrix is
\begin{equation}
B_{C_3}^{\mathrm{top}}(K')
=
\begin{pmatrix}
\eta_-&0\\
0&-1
\end{pmatrix}.
\label{eq:app_km_topological_B_Kp}
\end{equation}
At $\Gamma$, all spin-orbit terms in
Eq.~\eqref{eq:kane_mele_coefficients} vanish and
\begin{equation}
H_{\mathrm{KM}}(\Gamma)
=
\left(
3t\,\tau_x+\lambda_v\tau_z
\right)\otimes s_0.
\label{eq:app_km_H_Gamma}
\end{equation}
If $\ket{v_{\Gamma}}$ denotes the negative-energy sublattice
eigenstate of $3t\,\tau_x+\lambda_v\tau_z$, the occupied subspace is
\begin{equation}
\mathcal E_{\Gamma}
=
\operatorname{span}
\left\{
\ket{v_{\Gamma},\uparrow},
\ket{v_{\Gamma},\downarrow}
\right\}.
\end{equation}
Since the $A$ and $B$ states of a fixed spin carry the same rotation
eigenvalue at $\Gamma$ (see Eq.~\eqref{eq:app_km_C3_Gamma}), the occupied sewing matrix in this basis is
\begin{equation}
B_{C_3}(\Gamma)
=
\begin{pmatrix}
\eta_+&0\\
0&\eta_-
\end{pmatrix},
\label{eq:app_km_B_Gamma}
\end{equation}
independently of $\lambda_v$.

We now apply these representation matrices to the three sections used
in Fig.~\ref{fig:kane_mele_z2_overlap}.

\paragraph{Reference determinant.}

For the absolute construction, the trivial reference of
Sec.~\ref{subsec:kane_mele_application} is
\begin{equation}
R
=
\operatorname{span}
\left\{
\ket{B,\downarrow},
\ket{B,\uparrow}
\right\}.
\end{equation}
In the reference basis
$\{\ket{B,\downarrow},\ket{B,\uparrow}\}$,
Eqs.~\eqref{eq:app_km_C3_K}--\eqref{eq:app_km_C3_Gamma} give
\begin{align}
B_{C_3}^{R}(K)
&=
\begin{pmatrix}
-1&0\\
0&\eta_-
\end{pmatrix},
\label{eq:app_km_reference_B_K}
\\
B_{C_3}^{R}(K')
&=
\begin{pmatrix}
\eta_+&0\\
0&-1
\end{pmatrix},
\label{eq:app_km_reference_B_Kp}
\\
B_{C_3}^{R}(\Gamma)
&=
\begin{pmatrix}
\eta_-&0\\
0&\eta_+
\end{pmatrix}.
\label{eq:app_km_reference_B_Gamma}
\end{align}

At $K$, the occupied sewing matrix
$B_{C_3}^{\mathcal E}(K)$ is
Eq.~\eqref{eq:app_km_topological_B_K},
whereas the reference sewing matrix is
Eq.~\eqref{eq:app_km_reference_B_K}. Their eigenvalue spectra are
\begin{equation}
\{\eta_+,-1\},
\qquad
\{\eta_-,-1\},
\end{equation}
respectively, so the two representations are inequivalent. The
fixed-point argument of
Eq.~\eqref{eq:crystalline_representation_zero}, with the reference
space in place of the second occupied bundle, therefore gives
\begin{equation}
\det O(K)=0.
\label{eq:app_km_reference_zero_K}
\end{equation}
The rank loss can also be seen directly. Reordering the occupied basis
as
$\{\ket{B,\downarrow},\ket{A,\uparrow}\}$ while keeping the reference
basis
$\{\ket{B,\downarrow},\ket{B,\uparrow}\}$ gives
\begin{equation}
O(K)
=
\begin{pmatrix}
1&0\\
0&0
\end{pmatrix}.
\end{equation}
At $K'$,
Eqs.~\eqref{eq:app_km_topological_B_Kp} and
\eqref{eq:app_km_reference_B_Kp} have eigenvalue spectra
\begin{equation}
\{\eta_-,-1\},
\qquad
\{\eta_+,-1\},
\end{equation}
and are again inequivalent. Hence
\begin{equation}
\det O(K')=0.
\label{eq:app_km_reference_zero_Kp}
\end{equation}
Time reversal relates these two valley zeros.

At $\Gamma$,
Eqs.~\eqref{eq:app_km_B_Gamma} and
\eqref{eq:app_km_reference_B_Gamma} differ only by an exchange of the
two basis states and therefore describe equivalent representations.
Threefold rotation consequently does not force the reference overlap
to vanish. In fact, Eq.~\eqref{eq:app_km_H_Gamma} gives
\begin{equation}
\left|\det O(\Gamma)\right|
=
\frac12
\left(
1+
\frac{\lambda_v}
{\sqrt{9t^2+\lambda_v^2}}
\right),
\end{equation}
which is nonzero for the parameters of
Eq.~\eqref{eq:KM_absolute_parameters}. Thus the reference determinant
zeros at the two valleys are symmetry enforced, whereas no such
constraint occurs at $\Gamma$.

\paragraph{Pfaffian section.}

We now determine how $C_3$ constrains the Pfaffian coefficient. Recall
from Eq.~\eqref{eq:AII_pfaffian_matrix} that
\begin{equation}
M_{ij}(\mathbf{k})
=
\braket{
\psi_i(\mathbf{k})
|
\Theta\psi_j(\mathbf{k})
},
\qquad
p(\mathbf{k})
=
\operatorname{Pf}M(\mathbf{k}).
\end{equation}
Let $B_{C_3}(\mathbf{k})$ denote the sewing matrix of the occupied
subspace defined by Eq.~\eqref{eq:crystalline_sewing_matrix}. Unlike $S$ and
$O$, the matrix $M$ pairs the fiber at $\mathbf{k}$ with the time-reversed fiber
at $-\mathbf{k}$, compared through their common embedding in $\mathbb C^4$. Using
$\Theta\hat C_3(\mathbf{k})=\hat C_3(-\mathbf{k})\Theta$ and repeating the
derivation of Eq.~\eqref{eq:app_crystalline_overlap_covariance}, we obtain
\begin{equation}
M(R_3\mathbf{k})
=
B_{C_3}(\mathbf{k})
M_\Phi(\mathbf{k})
B_{C_3}(\mathbf{k})^T,
\qquad
(M_\Phi)_{ij}
=
\braket{\psi_i(\mathbf{k})|\Phi(\mathbf{k})\,\Theta\psi_j(\mathbf{k})},
\label{eq:app_km_pfaffian_matrix_covariance}
\end{equation}
with $\Phi(\mathbf{k})=\hat C_3(\mathbf{k})^\dagger\hat C_3(-\mathbf{k})$. The
transpose follows from the antiunitarity of $\Theta$. Only when $\Phi$ acts
trivially does this reduce to $M(R_3\mathbf{k})=B_{C_3}M B_{C_3}^T$, and then
$\operatorname{Pf}(AMA^T)=\det(A)\operatorname{Pf}(M)$ gives
\begin{equation}
p(R_3\mathbf{k})
=
\det B_{C_3}(\mathbf{k})\,
p(\mathbf{k}).
\label{eq:app_km_pfaffian_covariance}
\end{equation}
This holds at $\Gamma$, where $\Phi(\Gamma)=\mathbb 1$, but not at the valleys:
Eqs.~\eqref{eq:app_km_C3_K} and~\eqref{eq:app_km_C3_Kp} give
$\Phi(K)=\operatorname{diag}(1,1,\omega^{-1},\omega^{-1})$, a sublattice-dependent
phase. Consequently, $\det B_{C_3}(K)\neq1$ does not by itself force the Pfaffian
to vanish. The trivial system with $\lambda_v/t=0.6$ is an explicit
counterexample: its occupied sewing matrix,
Eq.~\eqref{eq:app_km_relative_B_sigma_K}, has $\det B_{C_3}(K)=-\eta_-=\omega$,
while $p(K)=\braket{\chi_-(K)|\Theta|B,\downarrow}
=\braket{\chi_-(K)|B,\uparrow}\neq0$; numerically $|p(K)|\simeq0.99$.

In the topological phase, the valley zero nevertheless has a crystalline origin.
The $\eta_+$ and $-1$ eigenspaces of $\hat C_3(K)$ in Eq.~\eqref{eq:app_km_C3_K}
are one-dimensional, spanned by $\ket{A,\uparrow}$ and $\ket{B,\downarrow}$. The
occupied representation of Eq.~\eqref{eq:app_km_topological_B_K}, with
eigenvalues $\{\eta_+,-1\}$ and
$\det B_{C_3}(K)=-\eta_+=\omega^{-1}$, therefore fixes
$\mathcal E^{\mathrm{top}}_K=\operatorname{span}\{\ket{A,\uparrow},\ket{B,\downarrow}\}$
independently of the remaining details of the Hamiltonian. Time reversal maps this
subspace onto $\operatorname{span}\{\ket{A,\downarrow},\ket{B,\uparrow}\}$,
which is orthogonal to it in $\mathbb C^4$. Hence $M(K)=0$ and
\begin{equation}
p(K)=0.
\label{eq:app_km_pfaffian_zero_K}
\end{equation}
Explicitly,
\begin{align}
p(K)
&=
\braket{
A,\uparrow
|
\Theta
|
B,\downarrow
}
\nonumber\\
&=
\braket{
A,\uparrow
|
B,\uparrow
}
=
0,
\end{align}
where we used Eq.~\eqref{eq:kane_mele_time_reversal}. Since $p$ transforms
by a nonzero factor under a change of occupied basis, the zero does not depend on
this choice.

At $K'$, the occupied representation of Eq.~\eqref{eq:app_km_topological_B_Kp},
with eigenvalues $\{\eta_-,-1\}$, likewise fixes
$\mathcal E^{\mathrm{top}}_{K'}=\operatorname{span}\{\ket{A,\downarrow},\ket{B,\uparrow}\}$,
whose time-reversed image $\operatorname{span}\{\ket{A,\uparrow},\ket{B,\downarrow}\}$
is orthogonal to it. Hence
\begin{equation}
p(K')=0.
\label{eq:app_km_pfaffian_zero_Kp}
\end{equation}
The two valley zeros are related by time reversal.

At $\Gamma$, Eq.~\eqref{eq:app_km_B_Gamma} gives
\begin{equation}
\det B_{C_3}(\Gamma)
=
\eta_+\eta_-
=
1.
\end{equation}
Since $\Phi(\Gamma)=\mathbb 1$, Eq.~\eqref{eq:app_km_pfaffian_covariance}
applies at $\Gamma$, and threefold rotation does not force the Pfaffian to vanish there.
This agrees with the explicit occupied states. Using
Eq.~\eqref{eq:app_km_H_Gamma}, we may choose
\begin{equation}
\ket{\psi_1(\Gamma)}
=
\ket{v_{\Gamma},\uparrow},
\qquad
\ket{\psi_2(\Gamma)}
=
\ket{v_{\Gamma},\downarrow},
\end{equation}
for which
\begin{equation}
|p(\Gamma)|=1.
\end{equation}
Thus the Pfaffian zeros at $K$ and $K'$ are enforced by the occupied $C_3$
representation through its one-dimensional sectors, whereas no such constraint is
present at $\Gamma$.

For the parameters used in Fig.~\ref{fig:kane_mele_z2_overlap}, the
numerical calculations find one zero of each of the three sections in the
chosen half of the Brillouin zone, at $K$, with winding $-1$ in the
orientation $(k_x,k_y)$, and no additional zeros on the sampled meshes.
For the trivial system ($\lambda_v/t=0.6$), the reference-overlap and
Pfaffian sections have no zeros on those meshes.

\paragraph{Relative determinant overlap.}

Finally, consider the two parameter regimes of
Eq.~\eqref{eq:KM_relative_parameters}. For the topological system
$\alpha$, with $\lambda_v^\alpha/t=-0.1$, the occupied sewing matrices
at the two valleys are again
\begin{equation}
B_{C_3}^{\alpha}(K)
=
\begin{pmatrix}
\eta_+&0\\
0&-1
\end{pmatrix},
\qquad
B_{C_3}^{\alpha}(K')
=
\begin{pmatrix}
\eta_-&0\\
0&-1
\end{pmatrix}.
\label{eq:app_km_relative_B_alpha}
\end{equation}

For the trivial system $\sigma$, with
$\lambda_v^\sigma/t=0.6$, the energy
$E_{A\uparrow}=\lambda_v^\sigma-\Delta$ is positive, while
$E_{B\downarrow}=-\lambda_v^\sigma-\Delta$ and the lower eigenvalue
of the $\eta_-$ block,
\begin{equation}
E_{\eta_-,-}
=
\Delta
-
\sqrt{
(\lambda_v^\sigma)^2+9\lambda_R^2
},
\end{equation}
are negative. The occupied subspace at $K$ is therefore
\begin{equation}
\mathcal E^\sigma_K
=
\operatorname{span}
\left\{
\ket{\chi_-(K)},
\ket{B,\downarrow}
\right\},
\end{equation}
where
\begin{equation}
\ket{\chi_-(K)}
\in
\operatorname{span}
\left\{
\ket{A,\downarrow},
\ket{B,\uparrow}
\right\}.
\end{equation}
Both components of $\ket{\chi_-(K)}$ carry the rotation eigenvalue
$\eta_-$. In the occupied basis
$\{\ket{\chi_-(K)},\ket{B,\downarrow}\}$,
\begin{equation}
B_{C_3}^{\sigma}(K)
=
\begin{pmatrix}
\eta_-&0\\
0&-1
\end{pmatrix}.
\label{eq:app_km_relative_B_sigma_K}
\end{equation}
Comparing this with
$B_{C_3}^{\alpha}(K)$ in
Eq.~\eqref{eq:app_km_relative_B_alpha}, the two occupied
representations have different eigenvalue spectra and are therefore
inequivalent. Equation~\eqref{eq:crystalline_representation_zero}
then requires
\begin{equation}
\det S(K)=0.
\label{eq:app_km_relative_zero_K}
\end{equation}

At $K'$, the occupied subspace of the trivial system is
\begin{equation}
\mathcal E^\sigma_{K'}
=
\operatorname{span}
\left\{
\ket{\chi_+(K')},
\ket{B,\uparrow}
\right\},
\end{equation}
where
\begin{equation}
\ket{\chi_+(K')}
\in
\operatorname{span}
\left\{
\ket{A,\uparrow},
\ket{B,\downarrow}
\right\}.
\end{equation}
Both components of $\ket{\chi_+(K')}$ carry rotation eigenvalue
$\eta_+$, while $\ket{B,\uparrow}$ carries eigenvalue $-1$. Hence
\begin{equation}
B_{C_3}^{\sigma}(K')
=
\begin{pmatrix}
\eta_+&0\\
0&-1
\end{pmatrix}.
\label{eq:app_km_relative_B_sigma_Kp}
\end{equation}
This is inequivalent to
$B_{C_3}^{\alpha}(K')$ in
Eq.~\eqref{eq:app_km_relative_B_alpha}, and therefore
\begin{equation}
\det S(K')=0.
\label{eq:app_km_relative_zero_Kp}
\end{equation}
At $\Gamma$, the occupied subspace of either system has the form
\begin{equation}
\mathcal E^\mu_{\Gamma}
=
\operatorname{span}
\left\{
\ket{v_{\Gamma}^\mu,\uparrow},
\ket{v_{\Gamma}^\mu,\downarrow}
\right\},
\qquad
\mu\in\{\alpha,\sigma\}.
\end{equation}
Since the rotation action is independent of $\lambda_v$,
\begin{equation}
B_{C_3}^{\alpha}(\Gamma)
=
B_{C_3}^{\sigma}(\Gamma)
=
\begin{pmatrix}
\eta_+&0\\
0&\eta_-
\end{pmatrix}.
\label{eq:app_km_relative_B_Gamma}
\end{equation}
The representations therefore agree, so $C_3$ does not enforce a
zero of the comparison section at $\Gamma$. The explicit matrix of overlaps is also nonsingular:
\begin{equation}
S(\Gamma)
=
\braket{
v_{\Gamma}^\alpha
|
v_{\Gamma}^\sigma
}
\,\mathbb 1_{2},
\end{equation}
and hence
\begin{equation}
\left|\det S(\Gamma)\right|
=
\frac12
\left[
1+
\frac{
9t^2+\lambda_v^\alpha\lambda_v^\sigma
}{
\sqrt{9t^2+(\lambda_v^\alpha)^2}
\sqrt{9t^2+(\lambda_v^\sigma)^2}
}
\right].
\end{equation}
For the parameters of
Eq.~\eqref{eq:KM_relative_parameters},
\begin{equation}
\left|\det S(\Gamma)\right|
\simeq 0.987,
\end{equation}
so no zero of the comparison section occurs at $\Gamma$. The three constructions are summarized in
Table~\ref{tab:kane_mele_C3_summary}.
\begin{table}[t]
\centering
\begin{tabular}{c|c|c|c}
\hline
section
&
$K$
&
$K'$
&
$\Gamma$
\\
\hline
$\det O$
&
$0$ ($C_3$ enforced)
&
$0$ ($C_3$ enforced)
&
$\neq0$
\\
$p$
&
$0$ ($C_3$ enforced)
&
$0$ ($C_3$ enforced)
&
$\neq0$
\\
$\det S$
&
$0$ ($C_3$ enforced)
&
$0$ ($C_3$ enforced)
&
$\neq0$
\\
\hline
\end{tabular}
\caption{
Threefold-rotation constraints on the three Kane-Mele sections used
in Fig.~\ref{fig:kane_mele_z2_overlap}. For the reference and relative
determinants, the valley zeros follow from inequivalent $C_3$ sewing
matrices for the two spaces entering the overlap construction.
For the Pfaffian, the
zeros follow from the one-dimensional $C_3$ sectors that make up the occupied
space; the determinant of the sewing matrix alone does not decide this.
At $\Gamma$ no crystalline constraint forces a zero,
and all three coefficients are nonzero for the parameters considered.
}
\label{tab:kane_mele_C3_summary}
\end{table}

These results show that the positions of the valley zeros appearing in
Fig.~\ref{fig:kane_mele_z2_overlap} are not fixed by the class AII
invariant alone. For the Kane-Mele parameters used here, $C_3$
symmetry enforces their presence at $K$ and $K'$, while time reversal
relates the two valleys. This does not exclude additional zeros at
generic momenta. If such zeros are present, crystalline symmetry and
time reversal require them to occur in the corresponding
symmetry-related orbits.

\end{document}